\documentclass[a4paper,11pt]{article}
\usepackage{amsfonts}
\usepackage{mathrsfs}
\usepackage{amsmath,amssymb,amsthm}
\usepackage{bbm}
\usepackage{mathtools}
\usepackage{latexsym}
\usepackage{indentfirst}
\usepackage[top=1in, bottom=1in, left=1in, right=1in]{geometry}
\usepackage{color}
\usepackage{footmisc}
\usepackage{endnotes}
\usepackage{algpseudocode}
\usepackage{algorithm}
\usepackage{float}
\usepackage{graphicx}
\usepackage{caption}
\usepackage{subcaption}
\usepackage{epstopdf}
\usepackage{enumitem}
\usepackage{comment}
\usepackage{longtable}
\usepackage{multirow}
\usepackage[font={footnotesize}]{caption}
\usepackage{authblk}
\usepackage{bookmark}
\usepackage[authoryear]{natbib}
\usepackage[toc,page,titletoc]{appendix}
\usepackage[capitalise,nameinlink]{cleveref}
\usepackage{arydshln}

\let\footnote=\endnote

\input{./Definitions}

\newtheorem{theorem}{Theorem}
\newtheorem{lemma}{Lemma}

\newtheorem{assumption}{Assumption}

\allowdisplaybreaks

\DeclarePairedDelimiter\norm{\lVert}{\rVert}

\DeclareMathOperator*{\diag}{diag}

\DeclareMathOperator*{\dett}{det}
\DeclareMathOperator*{\ig}{InvGamma}
\DeclareMathOperator*{\TV}{TV}
\DeclareMathOperator*{\laz}{lazy}

\newcommand{\ud}{\mathrm{d}}

\newcommand{\ee}{\mathrm{e}}
\newcommand{\iid}{\mathrm{i.i.d.}}

\def\sfa{\mathsf{a}}
\def\sfb{\mathsf{b}}
\def\sfc{\mathsf{c}}

\def\sfg{\mathsf{g}}

\def\sfr{\mathsf{r}}
\def\sfs{\mathsf{s}}
\def\sfw{\mathsf{w}}
\def\sfx{\mathsf{x}}
\def\sfy{\mathsf{y}}
\def\sfz{\mathsf{z}}

\begin{document}

\title{\bf Pigeonhole Stochastic Gradient Langevin Dynamics for Large Crossed Mixed Effects Models}

\author[1]{Xinyu Zhang \thanks{zhang\_xinyu@u.nus.edu}}
\author[1]{Cheng Li \thanks{stalic@nus.edu.sg}}
\affil[1]{Department of Statistics and Data Science, National University of Singapore}

\date{}
\maketitle

\begin{abstract}
Large crossed mixed effects models with imbalanced structures and missing data pose major computational challenges for standard Bayesian posterior sampling algorithms, as the computational complexity is usually superlinear in the number of observations. We propose two efficient subset-based stochastic gradient MCMC algorithms for such crossed mixed effects models, which facilitate scalable inference on both the variance components and regression coefficients. The first algorithm is developed for balanced design without missing observations, where we leverage the closed-form expression of the precision matrix for the full data matrix. The second algorithm, which we call the pigeonhole stochastic gradient Langevin dynamics (PSGLD), is developed for both balanced and unbalanced designs with potentially a large proportion of missing observations. Our PSGLD algorithm imputes the latent crossed random effects by running short Markov chains and then samples the model parameters of variance components and regression coefficients at each MCMC iteration. We provide theoretical guarantees by showing the convergence of the output distribution from the proposed algorithms to the target non-log-concave posterior distribution. A variety of numerical experiments based on both synthetic and real data demonstrate that the proposed algorithms can significantly reduce the computational cost of the standard MCMC algorithms and better balance the approximation accuracy and computational efficiency.
\end{abstract}


\section{Introduction}
\label{sec:intro}
Datasets of massive sizes and complex dependence pose significant computational challenges to traditional statistical learning and inference. This paper studies one of such examples, the crossed mixed effects model, which is broadly applicable to e-commerce data and survey data with massive sizes. Such datasets are often routinely collected and have several typical features. First, the data consist of a large number of subjects and items. For example, the data from movie-reviewing websites (or e-commerce platforms) contain a large number of reviewer-IDs and movie-IDs (or customer-IDs and commodity-IDs). Second, the observed data are often at the intersections across subjects and items with no replication, such as the users' rating scores of movies or commodities. Third, the data often come with a high percentage of missingness and the observed ratings are sparse. As a result, it is often difficult and not of interest to predict the individual ratings for each subject and each item, but instead one can model the subjects and items as factors with crossed random effects and estimate the global variation across both subjects and items, after accounting for the fixed effects from certain predictors. This leads to the crossed mixed effects model studied in the literature (\citealt{gao2019estimation}, \citealt{ghosh2020backfitting}):
\begin{align}\label{eq:lme}
&Y_{ij}= x_{ij}^{\top}b +\alpha_i+ \beta_j+e_{ij}, ~~ x_{ij} \in \RR^p, ~~ i=1,\ldots,R,~~ j=1,\ldots,C,    \\
\text{where}~~ & \alpha_i \stackrel{\iid}{\sim} (0, \sigma^2_{\alpha}),~~ \beta_j \stackrel{\iid}{\sim} (0, \sigma^2_{\beta}), ~~ e_{ij} \stackrel{\iid}{\sim} (0, \sigma^2_e), ~~ b = (b_1,\cdots, b_p)^{\top} \in \RR^p,
\nonumber
\end{align}
where $x_{ij}$'s are known constants of predictors; $b= (b_1,\cdots, b_p)^{\top} \in \RR^p$ is the vector of fixed effects regression coefficients; the row random effects $\alpha_i$'s are independently and identically distributed (i.i.d.) with mean $0$ and variance $\sigma_{\alpha}^2$; the column random effects $\beta_j$'s are i.i.d. with mean $0$ and variance $\sigma_{\beta}^2$; and the random errors $e_{ij}$'s are i.i.d. with mean $0$ and variance $\sigma_e^2$. The two random effects have $R$ and $C$ different levels, respectively. The response $Y_{ij}$ is modeled as continuous in \eqref{eq:lme}, but it can be a categorical or ordinal observation in accordance with the properties of ratings and scores in applications.
The parameters of interest in the model \eqref{eq:lme} are $\theta = (b^{\top}, \sigma^2_{\alpha},\sigma^2_{\beta}, \sigma^2_e)^{\top}$, where $\sigma^2_{\alpha},\sigma^2_{\beta}, \sigma^2_e$ are often referred to as the variance components (\citealt{searle2009}). Observations of size $N$ can be laid out in an $R \times C$ matrix $\Yb$ $(1 \leqslant N \leqslant R \times C)$. The data design is balanced if there are no missing data in the matrix $\Yb$, i.e., it has the same number of observations at each level of each factor and the same number of observations at each combination of factor levels. As the numbers of subjects and items can become very large, and each subject only rates a small portion of all items, one should expect that the observations in the data matrix $\Yb$ are sparse. We are mainly concerned with the algorithms for unbalanced design where $R$ and $C$ are very large and $N$ is much smaller than $R \times C$ resulting from missing data.

Statistical inference for the model \eqref{eq:lme} is challenging due to the dependence structure in the data matrix $\Yb$; see Section \ref{subsec:balanced} for details on the covariance matrix. The lack of independence creates a complicated covariance structure when the data matrix $\Yb$ has a large size and contains missing data. Standard frequentist estimation methods that depend on optimization, such as the maximum likelihood estimation, typically incur a computational complexity of $O(N^{3/2})$ when $R$ and $C$ are both $O(N^{1/2})$ and require memory of size $O((R+C)^2)$ (\citealt{gao2019estimation}), which becomes computationally intractable for datasets with tens of thousands of subjects and items; see \citet{gao2017scalable} for a thorough review on the previous frequentist results on standard crossed mixed effects models. To reduce the computational cost, \cite{gao2017e} and \cite{gao2019estimation} proposed new moment-based estimators inspired by the Henderson I method. In particular, \cite{gao2017e} used the method of moments based on U-statistics to obtain unbiased estimates of the variance components in the model \eqref{eq:lme} with only an intercept term, and \cite{gao2019estimation} further developed an alternating algorithm to estimate the regression coefficients in the model \eqref{eq:lme} using generalized least squares. Their methods only require $O(N)$ computational time. With the same computational complexity, \cite{ghosh2020backfitting} improved the asymptotic efficiency for the estimation of regression coefficients by using a backfitting algorithm that takes account of the variance components of both random effects and the random error. They further extended the backfitting algorithm to the logistic regression in \cite{ghosh2021scalable}.

The goal of this paper is to develop scalable Bayesian algorithms for the crossed mixed effects model \eqref{eq:lme} when the numbers of rows and columns in $\Yb$ are large. Compared to frequentist methods, the Bayesian framework has the advantage of automatic uncertainty quantification for the model parameters via the posterior distribution. However, it is well known that standard Bayesian posterior sampling algorithms such as Gibbs samplers and Metropolis-Hastings (MH) algorithms suffer from high computational cost when the data have a large size, due to their sequential nature and that the updates of model parameters require to sweep over the entire data at each iteration. For the model \eqref{eq:lme} with large $N$, \cite{gao2017e} has shown that the standard Gibbs sampler is not scalable with an $O(N^{3/2})$ computational complexity for convergence to the stationary distribution. To address this problem, \citet{papa2020s} and \citet{papa2021} proposed to use the collapsed Gibbs sampler by integrating out the global mean and sampling each level of the remaining factors in blocks, and their method demonstrates the superior $O(N)$ computational complexity when $\Yb$ is balanced. Extension of their collapsed Gibbs sampler to the case of unbalanced $\Yb$ with missing data has shown some promising empirical results, though a theoretical justification will need a further extension of the multigrid decomposition techniques (\citealt{papa2021}). Alternatively, one may consider the variational Bayesian algorithm as developed in \citet{Menetal2022}, though this only provides an approximation to the posterior distribution and the theoretical property remains unclear.

On the other hand, there exists a rich literature on scaling the Bayesian posterior sampling algorithms using subsets of data, including: (i) the divide-and-conquer strategy for independent data (\citealt{scott2016bayes}, \citealt{Minetal17}, \citealt{srivastava2018scalable}, \citealt{XueLia19}), and some examples of dependent data such as hidden Markov models (\citealt{WanSri22}) and Gaussian processes (\citealt{Guhetal22}); (ii) MH algorithms using a subset of data for each acceptance-rejection step (\citealt{kora2014austerity}, \citealt{MacAda14}, \citealt{Baretal18}, \citealt{Quietal19}); (iii) nonreversible Markov chain Monte Carlo (MCMC) based on piecewise-deterministic Markov Processes for some special models with globally bounded log-posterior densities (\citealt{Feretal18}, \citealt{Bouetal18}, \citealt{Bieetal19}, \citealt{Senetal20}). All these methods become highly nontrivial for complex hierarchical models with many latent variables, including the crossed mixed effects model \eqref{eq:lme}. 

In this work, we focus on one of the most popular strategies for scalable Bayesian inference, the stochastic gradient MCMC (SGMCMC), which uses subsets of data to estimate the gradient of the log-posterior density inside a discretized version of continuous-time diffusion processes; see \citet{nemeth2021stochastic} for a review on SGMCMC and the references therein. In particular, the stochastic gradient Langevin dynamics (SGLD) algorithm proposed by \cite{welling2011bayesian} combines stochastic gradient optimization with Langevin dynamics, which has proved to be more efficient within a fixed computational budget than similar gradient-based posterior sampling algorithms such as the Metropolis-adjusted Langevin algorithm (MALA, \citealt{RobTwe96}) when the full data have a large size. Theoretical properties for the SGLD on models with independent data have been investigated in the literature, including the convergence to the target posterior distribution (\citealt{teh2016consistency}) and the upper bounds of the approximation error (\citealt{dalalyan2017}, \citealt{dalalyan2019user}). The SGLD algorithm has been further implemented for dependent data in state space models (\citealt{Maetal17}, \citealt{Aicetal19}).

Given the scalable performance of the SGLD in models with independent data, our main goal is to derive SGLD algorithms for the crossed mixed effects model \eqref{eq:lme} with massive and sparse observations. One can randomly select rows and columns from the full data matrix $\Yb$ and use the constructed subset matrix of data to estimate the gradients of the log-posterior density at each iteration. Nevertheless, this implementation of SGLD to the model \eqref{eq:lme} requires several special considerations. The first issue is that such a submatrix of $\Yb$ constructed from random subsets of rows and columns still consists of mutually dependent observations. Therefore, unlike the existing SGLD literature on independent or weakly dependent data, it is unclear how to construct unbiased estimators for the gradients of the log-likelihood and log-posterior density given a dependent subset of data. We show that when the data matrix $\Yb$ has no missing data and is fully balanced, there exist closed-form formulas to calculate the inverse covariance matrix for any submatrices of $\Yb$, which facilitates explicit unbiased estimation of the gradients of the log-likelihood in the SGLD algorithm.

The second issue is that the crossed mixed effects model \eqref{eq:lme} contains $R$ row random effects $\alpha_i$'s and $C$ column random effects $\beta_j$'s. They are not part of the model parameter $\theta$ but their numbers can be very large. When $\Yb$ contains missing data, it is not possible to integrate out these random effects to obtain the posterior distribution of $\theta$ in a closed form, because there is no explicit formula for the inverse covariance matrix of $\Yb$ or any subset of $\Yb$ in the presence of missing observations. To address this challenge from latent variables, we propose to adapt the extended SGLD algorithm in \cite{song2020extended} for our crossed mixed effects model \eqref{eq:lme}. In particular, for models with independent data and latent variables, at each SGLD iteration, \cite{song2020extended} approximates the gradient of the log-likelihood based on the subset of data by first sampling a short Markov chain of latent variables and then performing the Monte Carlo average of the gradient of log conditional posterior density of the model parameters given both the sampled latent variables and the subset of data. This extended version of the SGLD algorithm has been theoretically justified in  \cite{song2020extended} for models with independent data, and therefore motivates us to apply a similar procedure to the dependent data from the large crossed mixed effects model \eqref{eq:lme}. We treat the row and column random effects as latent variables and derive a scalable SGLD algorithm by sampling from their conditional posterior and performing the Monte Carlo average based on a subset of data at each iteration. We name our proposed algorithm as the ``pigeonhole SGLD'' (PSGLD) for crossed mixed effects models, imitating the name of the pigeonhole bootstrap method proposed by \cite{owen2007pigeonhole}, which resamples a subset of rows and columns of the full data matrix $\Yb$ independently and takes the intersections as the bootstrap sample. One difference is that \cite{owen2007pigeonhole} proposed to sample rows and columns with replacement, while we randomly select $r$ rows ($2\leqslant r<R$) and $c$ columns ($2\leqslant c<C$) from $\Yb$ without replacement at each iteration. Furthermore, different from the general Algorithm S1 in \cite{song2020extended}, our pigeonhole SGLD algorithm drops the step of importance resampling. For theoretical justification, we show its convergence to the target posterior distribution as both the sample size $N$ and the length of SGLD go to infinity.

The rest of the paper is organized as follows. In Section \ref{sec:notation}, we define necessary notations related to the crossed mixed effects model \eqref{eq:lme} and give the prior specification on the model parameters. Section \ref{sec:SGLD} presents our proposed SGLD algorithms for the large crossed mixed effects model in two cases: the balanced design without missing data, and the unbalanced model with missing data. Section \ref{sec:theorem} presents a theorem on the convergence of the pigeonhole SGLD algorithm. Section \ref{sec:experiments} includes the numerical results on two real data examples, which demonstrate the estimation accuracy and computation efficiency of our proposed algorithms. Section \ref{sec:discussion} includes some discussion on our SGLD algorithms and perspective on future research. Further technical details and simulation studies are provided in the Supplementary Material.

\section{Model Setup and Prior Specification}\label{sec:notation}

We first introduce some useful notations for the crossed mixed effects model \eqref{eq:lme}. Each response variable $Y_{ij}$ corresponds to a covariate vector $x_{ij}$ and two crossed random effects, the row effect $\alpha_i$ and the column effect $\beta_j$ for $i=1,\ldots,R$ and $j=1,\ldots,C$, where $R$ and $C$ denote the numbers of rows and columns. As such, the full dataset of $Y_{ij}$'s can be arranged in an $R\times C$  matrix $\Yb$, whose $(i,j)$-entry is $Y_{ij}$. Throughout the paper, we assume that there is at most one observation at each entry of the matrix $\Yb$.

In real applications such as customer ratings of movies or goods, it is rare that all the data in the matrix $\Yb$ are observable with $R$ and $C$ being very large, and we are liable to have $\Yb$ with a considerable amount of missing observations. We use another $R\times C$ matrix $\Zb$ consisting of 0s and 1s to describe the missingness of data in $\Yb$. The $(i,j)$-entry of $\Zb$, denoted by $Z_{ij}$, is equal to $1$ if $Y_{ij}$ is observed, and is equal to 0 if $Y_{ij}$ is missing. The total amount of observed data is therefore $N = \sum_{i=1}^R\sum_{j=1}^C Z_{ij}$. The numbers of observations in the $i$th row and in the $j$th column of $\Yb$ are denoted by $N_{i \bullet} = \sum_{j}Z_{ij}$ and $N_{\bullet j} = \sum_{i}Z_{ij}$, respectively. Without loss of generality, we remove all the rows and columns with no observations from $\Yb$, so there is at least one observation in each row and column. In other words, $R = \sum_i\mathbbm{1}(N_{i \bullet}>0)$ and $C = \sum_j\mathbbm{1}(N_{\bullet j}>0)$, where $\mathbbm{1}(\cdot)$ is the indicator function.

The algorithms we propose for large crossed mixed effects models are built upon the stochastic gradient MCMC algorithms, which process only a mini-batch of data at each iteration to obtain chains of approximate posterior distributions. Therefore, we also define some notations for the subset of data. Suppose that we select $r$ rows and $c$ columns ($2\leqslant r<R, 2\leqslant c<C$) from the full data matrix $\Yb$ randomly without replacement, which forms a submatrix $\Yb_n$ containing a subset of data. Without loss of generality, we will remove any rows and columns in $\Yb_n$ with no observations, until each row and column of $\Yb_n$ has at least one observation. Corresponding to $\Yb_n$, we also select the same $r$ rows and $c$ columns from $\Zb$ to construct a submatrix $\Zb_n$ of indicators, i.e., for the $(i,j)$-entry ($1 \leqslant i \leqslant r, 1 \leqslant j \leqslant c$), $(\Zb_{n})_{ij} =1$ if $(\Yb_{n})_{ij}$ is observed, and $(\Zb_{n})_{ij} =0$ otherwise. Let $n = \sum^r_{i=1} \sum^c_{j=1}(\Zb_{n})_{ij}$ denote the number of observations in the submatrix $\Yb_n$. The numbers of observations in row $i$ and column $j$ of the submatrix $\Yb_n$ are $n_{i \bullet} =  \sum^c_{j=1}(\Zb_{n})_{ij} $ and $n_{ \bullet j} =  \sum^r_{i=1}(\Zb_{n})_{ij} $, respectively. For the matrix $\Yb_n$, we use $s_1, \ldots, s_r \in \{1, \ldots, R\}$ and $q_1 \ldots, q_c \in \{1, \ldots, C \}$ to denote the original positions of the rows and columns of $\Yb_n$ in $\Yb$, for example, $(\Yb_n)_{ij}=\Yb_{s_iq_j}$ and similarly $(\Zb_n)_{ij} =\Zb_{s_iq_j}$.

We assume that both the row and column random effects as well as the errors follow normal distributions:
\begin{align}\label{eq:normalre}
& \alpha_i \mid \sigma^2_{\alpha} \stackrel{\iid}{\sim} N(0, \sigma^2_{\alpha}), \quad \text{for } i = 1, \ldots, R; ~~  \beta_j \mid \sigma^2_{\beta} \stackrel{\iid}{\sim} N(0, \sigma^2_\beta), \quad \text{for } j = 1, \ldots, C;  \nonumber \\
& e_{ij} \mid \sigma^2_e \stackrel{\iid}{\sim} N(0, \sigma^2_e), \quad \text{for } i = 1, \ldots, R,~ j =1 ,\ldots, C.
\end{align}
As such, the crossed mixed effects model \eqref{eq:lme} consists of finite dimensional parameters $\theta = \big(b^{\top}, \sigma^2_\alpha, \sigma^2_\beta, \sigma^2_e \big)^{\top}$. We assign the following priors:
\begin{align}\label{eq:prior}
& \pi(b) \propto 1, &   \sigma_\alpha^2 \mid \afrak_1, \bfrak_1 \sim \ig(\afrak_1, \bfrak_1), \nonumber \\
& \sigma_\beta^2 \mid \afrak_2, \bfrak_2 \sim \ig(\afrak_2, \bfrak_2), & \sigma_e^2 \mid \afrak_3, \bfrak_3 \sim \ig(\afrak_3, \bfrak_3),
\end{align}
where $\ig(c_1,c_2)$ stands for the inverse gamma distribution with the shape parameter $c_1$ and the rate parameter $c_2$.

Bayesian posterior sampling algorithms including stochastic gradient MCMC work best when the posterior chains can move freely on the entire real line for each parameter. Therefore, we reparameterize the variance components $(\sigma^2_\alpha, \sigma^2_\beta,\sigma^2_e)$ into the logarithm scale by letting $\eta_\alpha = \log \sigma^2_\alpha, ~\eta_\beta = \log \sigma^2_\beta$, and $\eta_e = \log \sigma^2_e$. By \eqref{eq:prior}, they have the prior densities $\pi(\eta_\alpha) \propto \exp\{-\afrak_1 \eta_\alpha - \bfrak_1\exp(-\eta_\alpha)\}$, $\pi(\eta_\beta) \propto \exp\{-\afrak_2 \eta_\beta - \bfrak_2\exp(-\eta_\beta)\} $, and $\pi(\eta_e) \propto \exp\{-\afrak_3 \eta_e - \bfrak_3\exp(-\eta_e)\}$, respectively.

\section{Stochastic Gradient MCMC for Crossed Mixed Effects Models}
\label{sec:SGLD}

When the sample size $N$ and the numbers of rows and columns $R,C$ become large in the crossed mixed effects model \eqref{eq:lme}, the standard MCMC algorithms based on the full data become computationally inefficient, as their computation cost can easily increase to superlinear in the number of observations $N$. For example, the Gibbs sampler requires theoretical $O(N^{3/2})$ iterations to converge to the stationary posterior distribution (\citealt{gao2017e}). To address this problem, we propose two scalable algorithms using stochastic gradient MCMC, which process only a subset of data at each iteration and therefore significantly speed up the posterior sampling.

We first briefly review the basic version of the stochastic gradient Langevin dynamics (SGLD) algorithm (\citealt{welling2011bayesian}, \citealt{teh2016consistency}) for $\iid$ data. For a large dataset $X_N$ consisting of $N$ $\iid$ samples $\{x_1,\ldots,x_N\}$ generated from the model $p(x \mid \theta)$ with the parameter $\theta \in \RR^D$, the full-data likelihood is $p(X_N \mid \theta) = \prod^N_{i=1}p(x_i \mid \theta)$. With the prior density $\pi(\theta)$, the log-posterior density of $\theta$ is $\log \pi(\theta\mid X_N) = \sum^N_{i=1} \log p(x_i \mid \theta) +\log\pi(\theta)$. To sample from $\pi(\theta\mid X_N)$, the SGLD algorithm finds the updated parameter $\theta^{(t+1)}$ from the last iteration $\theta^{(t)}$ using the following equation:
\begin{equation} 
\theta^{(t+1)} = \theta^{(t)} + \frac{\Ecal_t}{2}\left( \nabla_{\theta} \log \pi(\theta^{(t)})+ \frac{N}{n} \sum_{i\in \Scal} \nabla_{\theta} \log p(x_i \mid \theta^{(t)}) \right) + \psi^{(t)}, ~ \psi^{(t)} \sim N(0, \Ecal_t), \nonumber
\end{equation}
where $\Scal$ is a random subset of indexes in $\{1,\ldots,N\}$ with size $n$ and $\Ecal_t$ is a positive definite matrix for tuning the step sizes. Based on the subset of data $\{x_i:i\in \Scal\}$, the approximate gradient $(N/n) \sum_{i\in \Scal} \nabla_{\theta} \log p(x_i \mid \theta_t)$ is an unbiased estimator of $\sum^N_{i=1} \nabla_{\theta} \log p(x_i \mid \theta_t)$, the gradient of the log-likelihood based on the full data. SGLD only requires a computational cost of $O(n)$ at each iteration rather than $O(N)$ for the full-data-based MCMC algorithms. Furthermore, with properly tuned step sizes in $\Ecal_t$, SGLD is consistent for the target true posterior $\pi(\theta\mid X_N)$ (\citealt{teh2016consistency}). A potential issue of the SGLD algorithm is that the mixing rate can slow down if the different components of $\theta$ have very different scales or if they are highly correlated. Preconditioning adaptations of SGLD can partly alleviate this problem and improve the mixing (\citealt{ahn2012bayesian}, \citealt{patterson2013stochastic}).

In the following, we introduce two versions of SGLD algorithms for the crossed mixed effects models without missing data and with missing data in Sections \ref{subsec:balanced} and \ref{subsec:unbalanced}, respectively. Then we provide the detailed reasoning behind the construction of stochastic gradients used in the two algorithms in Section \ref{subsec:validity}.

\subsection{SGLD for the Balanced Model without Missing Data}
\label{subsec:balanced}
When the crossed mixed effects model is balanced with no missing data, there is exactly one observation $Y_{ij}$ in the intersection of each row and each column and $\Zb$ is a matrix of all $1$s. In this case, the inverse covariance matrix of $\Yb$ and the gradient of the log-likelihood function have analytically tractable closed forms. Therefore, we can derive the SGLD algorithm directly using data subsets, though we emphasize that this algorithm differs in essence from the original SGLD algorithm since the subset of data are not independent.

In particular, at each iteration, we randomly select $r$ rows and $c$ columns from the full data matrix $\Yb$ without replacement and obtain an $r\times c$ submatrix $\Yb_n$. To formulate the log-likelihood of data, we stack the rows of $\Yb_n$ into a column vector $Y_n \in \RR^n ~ (n = r \times c)$, and correspondingly stack the fixed effects $x_{ij}$s into a matrix $\Xb_n \in \RR^{n \times p}$. Let $\Ib_s$ be the $s\times s$ identity matrix and $\bm{1}_s \in \RR^s$ be the $s$-dimensional vector of all $1$s. Then the selected subset of data have the model
\begin{align}\label{SGLDcr}
& Y_n =\Xb_n b + \Zb_{\alpha n} \alphab_n + \Zb_{\beta n} \betab_n + e_n,
\end{align}
where $\Zb_{\alpha n}  = \Ib_r \otimes \bm{1}_c \in \{0,1\}^{n \times r}$, $\Zb_{\beta n} = \bm 1_r \otimes \Ib_c \in \{0,1\}^{n \times c}$, and $\otimes$ denotes the Kronecker product; $\alphab_n \in \RR^r$ and $\betab_n \in \RR^c$ are the selected vectors of row random effects and column random effects, and $e_n\in \RR^n$ is the vector consisting of all the random errors in $Y_n$.
As a result, after marginalizing out the random effects $\alphab_n$ and $\betab_n$ according to the model in \eqref{eq:normalre}, the covariance matrix of $Y_n$ can be written as $\bm{\Sigma}_n = \Zb_{\alpha n} \Zb_{\alpha n }^{\top}\sigma^2_\alpha + \Zb_{\beta n} \Zb_{\beta n}^{\top}\sigma^2_\beta + \Ib_n\sigma^2_e$, whose explicit form is
\begin{align}\label{eq:sig.inv0}
& \bm{\Sigma}_n={\left[ \begin{array}{cccc}
\Sigma_1 & \Sigma_2 & \ldots & \Sigma_2\\
\Sigma_2 & \Sigma_1 & \ldots & \Sigma_2\\
\vdots & \vdots & \ddots &\vdots \\
\Sigma_2 & \Sigma_2 & \ldots & \Sigma_1
\end{array}
\right ]}_{n \times n}, ~ \text{where}\\
& \Sigma_1={\left[ \begin{array}{cccc}
\sigma^2_\alpha+\sigma^2_\beta+\sigma^2_e & \sigma^2_\alpha & \ldots & \sigma^2_\alpha\\
\sigma_\alpha^2 & \sigma^2_\alpha+\sigma^2_\beta+\sigma^2_e & \ldots & \sigma^2_\alpha\\
\vdots & \vdots & \ddots &\vdots \\
\sigma_\alpha^2 & \sigma^2_\alpha & \ldots & \sigma^2_\alpha+\sigma^2_\beta+\sigma^2_e
\end{array}
\right ]}_{c \times c}, \text{ and } \Sigma_2 = \sigma^2_\beta \Ib_c . \nonumber
\end{align}
With the block matrix structure in \eqref{eq:sig.inv0}, the inverse covariance matrix $\bm{\Sigma}_n^{-1}$ can be explicitly derived:
\begin{align} \label{eq:sig.inv}
& \bm{\Sigma}_n^{-1}={\left[ \begin{array}{cccc}
\Sigma_3 & \Sigma_4 & \ldots & \Sigma_4\\
\Sigma_4 & \Sigma_3 & \ldots & \Sigma_4\\
\vdots & \vdots & \ddots &\vdots \\
\Sigma_4 & \Sigma_4 & \ldots & \Sigma_3
\end{array}
\right ]}_{n \times n},~\text{where } \\
 & \Sigma_3={\left[ \begin{array}{cccc}
\sfx & \sfy & \ldots & \sfy\\
\sfy & \sfx & \ldots & \sfy\\
\vdots & \vdots & \ddots &\vdots \\
\sfy & \sfy & \ldots & \sfx
\end{array}
\right ]}_{c \times c}, \quad  \Sigma_4={\left[ \begin{array}{cccc}
\sfw & \sfz & \ldots & \sfz\\
\sfz & \sfw & \ldots & \sfz\\
\vdots & \vdots & \ddots &\vdots \\
\sfz & \sfz & \ldots & \sfw
\end{array}
\right ]}_{c \times c}, ~ \text{and } \nonumber \\
& \sfz = \frac{\sigma^2_\alpha \sigma^2_\beta (2 \sigma^2_e + c\sigma^2_\alpha +r \sigma^2_\beta)}{\sigma^2_e(\sigma^2_e + r \sigma^2_\beta)(\sigma^2_e + c\sigma^2_\alpha)(\sigma^2_e + c \sigma^2_\alpha +r\sigma^2_\beta)},
\quad \sfy = \sfz-\frac{\sigma^2_\alpha}{\sigma^2_e(\sigma^2_e+c \sigma^2_\alpha)}, \nonumber \\
& \sfx = \sfy+\frac{\sigma^2_e +(r-1)\sigma^2_\beta}{\sigma^2_e(\sigma^2_e+r\sigma^2_\beta)},
\quad \sfw = \sfz-\frac{\sigma^2_\beta}{\sigma^2_e(\sigma^2_e + r \sigma^2_\beta)}. \nonumber
\end{align}

With the explicit formula of $\bm{\Sigma}^{-1}_n$ in \eqref{eq:sig.inv}, we can compute the log-likelihood function of the selected data subset and its gradient. The SGLD algorithm for balanced crossed mixed effects models without missing data is presented in Algorithm \ref{algo:SGLD}. The exact formulas for the gradients in \eqref{eq:SGLD} of Algorithm \ref{algo:SGLD} are presented in Section S1 of the Supplementary Material.

\begin{algorithm}[ht!]
\caption{Stochastic Gradient Langevin Dynamics for Balanced Crossed Mixed Effects Models without Missing Data} \label{algo:SGLD}
{\small
\begin{enumerate}[leftmargin=0mm]
\item[] {\bf Input}: Initial values of the model parameters $\theta^{(0)}=\big(b^{(0)\top}, \eta_\alpha^{(0)}, \eta_\beta^{(0)}, \eta_e^{(0)}\big)^{\top}$; the step size matrix of the vector $b$, $\Ecal_b^{(t)} = \diag\big\{\epsilon_{b_1}^{(t)}, \cdots, \epsilon_{b_p}^{(t)}\big\}$, and the full step size matrix $\Ecal^{(t)} = \diag \big\{\Ecal_b^{(t)}, \epsilon_{\eta_\alpha}^{(t)}, \epsilon_{\eta_\beta}^{(t)}, \epsilon_{\eta_e}^{(t)}\big\}$, for $t=0,\ldots,T-1$.
\item[] {\bf For} $t = 0, \ldots, T-1$ {\bf do}
\begin{enumerate}
\item {\bf (Sample the subset of data)} Select $r$ rows and $c$ columns ($2\leqslant r <R$, $2\leqslant c <C$) randomly without replacement from the full data matrix $\Yb$. Stack the selected submatrix $\Yb_n^{(t)}$ by rows and obtain the vector $Y_n^{(t)}$. Arrange the corresponding fixed effects in the subset matrix $\Xb_n^{(t)}$.

\item {\bf (Update parameters)} Update $\theta^{(t+1)} = \big(b^{(t+1)\top}, \eta_a^{(t+1)},\eta_\beta^{(t+1)}, \eta_e^{(t+1)} \big)^{\top}$ by the following equations:
\begin{align}\label{eq:SGLD}
b^{(t+1)} & = b^{(t)} +  \frac{\Ecal_{b}^{(t)}}{2} \bigg[ \frac{N}{n}\nabla_b \log p\left(Y_n^{(t)} \mid  b^{(t)}, \eta^{(t)}_\alpha, \eta^{(t)}_\beta, \eta^{(t)}_e \right) + \nabla_b \log \pi\left(b^{(t)}\right) \bigg ] + \psi_{b}^{(t)},  \nonumber\\
\eta_\alpha^{(t+1)} & = \eta_\alpha^{(t)} +  \frac{\epsilon_{\eta_{\alpha}}^{(t)}}{2} \bigg[ \frac{R}{r}\nabla_{\eta_\alpha} \log p\left(Y_n^{(t)} \mid  b^{(t)}, \eta^{(t)}_\alpha, \eta^{(t)}_\beta, \eta^{(t)}_e \right) + \nabla_{\eta_\alpha} \log \pi\left(\eta_\alpha^{(t)}\right) \bigg ] + \psi_{\eta_\alpha}^{(t)}, \nonumber\\
\eta_\beta^{(t+1)} & = \eta_\beta^{(t)} +  \frac{\epsilon_{\eta_{\beta}}^{(t)}}{2} \bigg[ \frac{C}{c}\nabla_{\eta_\beta} \log p\left(Y_n^{(t)} \mid b^{(t)}, \eta^{(t)}_\alpha, \eta^{(t)}_\beta, \eta^{(t)}_e \right) + \nabla_{\eta_\beta} \log \pi\left(\eta_\beta^{(t)}\right) \bigg ] + \psi_{\eta_\beta}^{(t)}, \nonumber \\
\eta_e^{(t+1)} & = \eta_e^{(t)} +  \frac{\epsilon_{\eta_e}^{(t)}}{2} \bigg[ \frac{N}{n}\nabla_{\eta_e} \log p\left(Y_n^{(t)} \mid  b^{(t)}, \eta^{(t)}_\alpha, \eta^{(t)}_\beta, \eta^{(t)}_e \right) + \nabla_{\eta_e} \log \pi\left(\eta_e^{(t)}\right) \bigg ] + \psi_{\eta_e}^{(t)},
\end{align}
where the exact formulas for the gradients are given in Section S1 of the Supplementary Material, and we sample $\psi_{b}^{(t)} \sim N(0,\Ecal_{b}^{(t)})$, $\psi_{\eta_{\alpha}}^{(t)} \sim N(0,\epsilon_{\eta_{\alpha}}^{(t)})$, $\psi_{\eta_{\beta}}^{(t)} \sim N(0,\epsilon_{\eta_{\beta}}^{(t)})$, $\psi_{\eta_e}^{(t)} \sim N(0,\epsilon_{\eta_e}^{(t)})$ independently.
\end{enumerate}
\item[] {\bf End for}
\item[] {\bf Return}: A sequence of parameters $\{\theta^{(t)}\}^{T}_{t=1} = \{b^{(t)\top}, \eta_\alpha^{(t)}, \eta_\beta^{(t)}, \eta_e^{(t)}\}^{T}_{t=1}$, whose empirical distribution is an approximation of the posterior distribution $\pi(\theta \mid \Yb)$.
\end{enumerate}
}
\end{algorithm}

In equation \eqref{eq:SGLD} of Algorithm \ref{algo:SGLD}, we have used different multiplicative factors for the four stochastic gradients of the log-likelihood function. In particular, we use $N/n, R/r, \\ C/c, N/n$ for the gradients with respect to the parameters $b,\eta_{\alpha},\eta_{\beta},\eta_e$, respectively. These choices are motivated by an expansion of the gradients by taking expectations with respect to the latent variables of row and column random effects. Detailed explanations on equation \eqref{eq:SGLD} of Algorithm \ref{algo:SGLD} are deferred to Section \ref{subsec:validity}. The step sizes $\epsilon_{b_1}^{(t)}, \ldots, \epsilon_{b_p}^{(t)}, \epsilon_{\eta_\alpha}^{(t)}, \epsilon_{\eta_\beta}^{(t)}, \epsilon_{\eta_e}^{(t)}$ in Algorithm \ref{algo:SGLD} are often chosen as constants in real applications. Depending on the model structure, especially on the design matrix of the fixed effects $\Xb$, we can also apply preconditioning methods to the step size matrix $\Ecal_b^{(t)}$. We will elaborate more on the choice of step sizes in the numerical experiments in Section \ref{sec:experiments} and Section S3 of the Supplementary Material.

\subsection{PSGLD for the Unbalanced Model with Missing Data}
\label{subsec:unbalanced}

When $\Yb$ contains missing data, the likelihood function for neither the full data $\Yb$ nor a subset of data $\Yb_n$ has a closed-form expression like in the balanced case of Section \ref{subsec:balanced}. This is because the form of the inverse covariance matrix of $\Yb$ or $\Yb_n$ that shows up in the likelihood function heavily depends on the specific missing pattern, and different subsets of $\Yb_n$ have vastly different missing patterns. As a result, the standard SGLD algorithm cannot be directly applied to the model \eqref{eq:lme} with missing data. To address this problem, we propose to include the row and column random effects as latent variables and apply the extended version of SGLD proposed in \citet{song2020extended}. The main purpose of \citet{song2020extended} is to apply the extended SGLD algorithm to Bayesian variable selection, where they utilized a special case of Fisher’s identity to provide a Monte Carlo estimator for the gradient of the log-posterior function (Lemma 1 of \cite{song2020extended}): For a large dataset $X_N$ consisting of $N$ $\iid$ samples $\{x_1,\ldots,x_N\}$ from the model $p(x \mid \theta, \vartheta)$ with parameter $\theta \in \RR^D$ and latent variable $\vartheta$, the gradients of log-posteriors $\log \pi(\theta \mid X_N)$ and $\log \pi(\theta \mid \vartheta, X_N)$ satisfy
\begin{align}  \label{eq:latent}
\nabla_{\theta} \log \pi(\theta \mid X_N) &= \int \left[\nabla_{\theta} \log \pi(\theta \mid \vartheta, X_N)\right] \pi(\vartheta \mid \theta, X_N) \ud \vartheta ,
\end{align}
where $\pi(\theta \mid \vartheta, X_N)$ and $\pi(\vartheta \mid \theta, X_N)$ are the conditional posterior densities of $\theta$ and $\vartheta$.

For $\iid$ data, the relation in \eqref{eq:latent} provides a way to approximate the intractable gradient $\nabla_{\theta} \log \pi(\theta \mid X_N)$ by computing the empirical expectation of samples of $\nabla_{\theta} \log \pi(\theta \mid \vartheta, X_N)$ with $\vartheta$ drawn from the conditional posterior $\pi(\vartheta \mid \theta, X_N)$. For a stochastic version, one can draw a subset data $X_n$ with size $n$ from $X_N$ randomly without replacement. Since $(N/n) \nabla_{\theta} \log p(X_n \mid \theta)$ is an unbiased estimator of $\nabla_{\theta}\log p(X_N\mid \theta)$ in the $\iid$ case, we have that $(N/n)\nabla_{\theta} \log p( X_n\mid \theta )  + \nabla \log \pi(\theta)$ is an unbiased estimator of $\nabla_{\theta} \log \pi(\theta \mid X_N)$. One can then use \eqref{eq:latent} and Monte Carlo samples of $\vartheta$ from $\pi(\vartheta\mid\theta,X_n)$ to further approximate the gradient, leading to the extended SGLD updating equation:
\begin{align} \label{exsgld}
&\theta^{(t+1)} = \theta^{(t)} + \frac{\Ecal_t}{2m}\sum^m_{k=1} \left[\frac{N}{n}\nabla_{\theta} \log p\big( X_n\mid \theta^{(t)}, \vartheta_k^{(t)} \big)  + \nabla_{\theta} \log \pi\big(\theta^{(t)}, \vartheta_k^{(t)}\big)\right]  + \psi^{(t)},
\end{align}
where $\psi^{(t)} \sim N(0, \Ecal_t)$, and $\big\{\vartheta_k^{(t)}:k=1,\ldots,m\big\}$ is a length-$m$ Markov chain drawn from $\pi(\vartheta\mid\theta^{(t)},X_n)$. The original extended SGLD Algorithm S1 in \citet{song2020extended} has included an extra importance resampling step and a correction term to \eqref{exsgld} to ensure that $\big\{\vartheta_k^{(t)}:k=1,\ldots,m\big\}$ are drawn from the posterior distribution of $\vartheta$ conditioning on $\theta^{(t)}$ and the expanded $N/n$-replicate of subset $X_n$. We find that this step can be skipped in implementation and will justify this in our theory in Section \ref{sec:theorem}.

To handle the missing data in crossed mixed effects models, we propose the pigeonhole SGLD algorithm (PSGLD), named in a similar fashion to the frequentist method of pigeonhole bootstrap in \citet{owen2007pigeonhole}. The pigeonhole SGLD algorithm adapts the extended SGLD in \citet{song2020extended} by treating the row and column effects of the selected subset of data at each iteration as the latent variables, i.e., $\vartheta=(\alphab_n^{\top}, \betab_n^{\top})^{\top}=(\alpha_{s_i},\ldots, \alpha_{s_r},\beta_{q_1},\ldots, \beta_{q_c})^{\top}$. In the $t$th iteration, the gradient of the subset log-posterior $\nabla_{\theta} \log \pi \big(\theta \mid \Yb_{n}^{(t)}\big)$ is approximated by first drawing $(\alphab_n, \betab_n)$ conditional on $\Yb_n^{(t)},\theta^{(t)}$ and then taking the empirical expectation similar to \eqref{exsgld}, which does not require the calculation of the intractable inverse covariance matrix for the subset of data $\Yb_n^{(t)}$. When the random effects and error terms are normally distributed as in \eqref{eq:normalre} and assigned the conjugate priors as in \eqref{eq:prior}, the conditional posterior distribution for each component of $(\alphab_n, \betab_n)$ is also normal, and hence we sample each row and column effect from its conditional posterior given the others and run a short Markov chain using the Gibbs sampler. Then we update $\theta$ by averaging over the gradient of log-posterior distributions of $\theta$ conditional on the latent variables $(\alphab_n, \betab_n)$ and the subset of data $\Yb_n$. We summarize the pigeonhole SGLD algorithm for large crossed mixed effects models in Algorithm \ref{algo:pigeonhole SGLD}.
The exact formulas of gradients in equation \eqref{eq:PSGLD} of Algorithm \ref{algo:pigeonhole SGLD} are provided in Section S2 of the Supplementary Material.

\subsection{Validity of Stochastic Gradients in Algorithms \ref{algo:SGLD} and \ref{algo:pigeonhole SGLD}}
\label{subsec:validity}

We now explain why we have constructed the stochastic gradients in the updating equations \eqref{eq:SGLD} of Algorithm \ref{algo:SGLD}, and in the equation \eqref{eq:PSGLD} of Algorithm \ref{algo:pigeonhole SGLD} using the Markov chain $\{\alphab_{n,k}, \betab_{n,k}\}_{k=1}^m$. In the derivation below, we omit the superscript $(t)$ for notational simplicity. Our argument proceeds in two steps of approximations.
\vspace{2mm}

\noindent \underline{\textsc{Step 1}: Markov chain based gradients approximate the subset gradients.}

In Step 1, we show that the Monte Carlo-based gradients in the equations \eqref{eq:PSGLD} are approximating some gradients given the subset of data $\Yb_n$.
\begin{itemize}[leftmargin=5mm]
\item For $\eta_e$, since the prior of latent variables $\{\alphab_n,\betab_n\}$ does not depend on $\eta_e$, we have
\begin{align} \label{eq:grad.eta.e.ex}
& \quad \frac{N}{n}\nabla_{\eta_e} \log p(\Yb_n| \theta) +  \nabla_{\eta_e} \log \pi(\eta_e) \nonumber \\
& = \int  \left[\frac{N}{n} \nabla_{\eta_e}\log p(\Yb_n|  b, \alphab_n, \betab_n, \eta_e) +  \nabla_{\eta_e} \log \pi(\eta_e)\right]\pi(\alphab_n,\betab_n | \theta, \Yb_n)\ud\alphab_n \ud \betab_n.
\end{align}
Therefore, with the Markov chain $\{\alphab_{n,k}, \betab_{n,k}\}_{k=1}^m$ drawn from their conditional posterior distributions based on the subset of data $\pi(\alphab_n, \betab_n | \theta, \Yb_n)$, we have that \\
$m^{-1} {\sum}^m_{k=1}\Big[ (N/n) \nabla_{\eta_e} \log p\big(\Yb_n | b, \alphab_{n,k}, \betab_{n,k},  \eta_e \big) + \nabla_{\eta_e} \log \pi\big(\eta_e \big) \Big]$ in equation \eqref{eq:PSGLD} of Algorithm \ref{algo:pigeonhole SGLD} is a Monte Carlo approximation of $(N/n)\nabla_{\eta_e} \log p(\Yb_n | \theta) +  \nabla_{\eta_e} \log \pi(\eta_e)$, as used in equation \eqref{eq:SGLD} of Algorithm \ref{algo:SGLD}. By the same argument, for the gradient with respect to $b$ in \eqref{eq:PSGLD}, $m^{-1} {\sum}^m_{k=1} \left[ (N/n) \nabla_{b} \log p\big(\Yb_n | b, \alphab_{n,k}, \betab_{n,k},  \eta_e \big) + \nabla_{b} \log \pi\big(b\big) \right]$ in equation \eqref{eq:PSGLD} of Algorithm \ref{algo:pigeonhole SGLD} is a Monte Carlo approximation of\\
$(N/n)\nabla_{b} \log p(\Yb_n | \theta) +  \nabla_{b} \log \pi(b)$, as used in equation \eqref{eq:SGLD} of Algorithm \ref{algo:SGLD}.

\item For $\eta_{\alpha}$, we have that
\begin{align} \label{eq:grad.eta.alpha.ex}
&\quad~ \frac{R}{r}\nabla_{\eta_\alpha}\log p(\Yb_n| \theta) + \nabla_{\eta_\alpha} \log \pi(\eta_\alpha) =   \frac{R}{r}\nabla_{\eta_\alpha}\log \pi(\theta| \Yb_n) - \frac{R-r}{r} \nabla_{\eta_\alpha} \log \pi(\eta_\alpha)   \nonumber\\
&= \frac{R}{r}\int  \left[\nabla_{\eta_\alpha}\log \pi(\theta|\Yb_n,\alphab_n,\betab_n)\right]\pi(\alphab_n,\betab_n | \theta, \Yb_n)\ud\alphab_n \ud \betab_n - \frac{R-r}{r} \nabla_{\eta_\alpha} \log \pi(\eta_\alpha)   \nonumber \\
&\overset{(i)}{=}  \int  \frac{R}{r}\left[\nabla_{\eta_\alpha}\log \pi( \alphab_n | \eta_\alpha) + \nabla_{\eta_\alpha} \log \pi(\eta_\alpha)\right]\pi(\alphab_n,\betab_n | \theta, \Yb_n)\ud\alphab_n \ud \betab_n  \nonumber \\
&\qquad - \frac{R-r}{r} \nabla_{\eta_\alpha} \log \pi(\eta_\alpha)   \nonumber \\
&=  \int \left[\frac{R}{r} \nabla_{\eta_\alpha}\log \pi( \alphab_n | \eta_\alpha) + \nabla_{\eta_\alpha} \log \pi(\eta_\alpha)\right]\pi(\alphab_n | \theta, \Yb_n)\ud\alphab_n ,
\end{align}

\begin{algorithm}[H]
\caption{Pigeonhole Stochastic Gradient Langevin Dynamics for Crossed Mixed Effects Models with Missing Data} \label{algo:pigeonhole SGLD}
{\footnotesize
\begin{enumerate}[leftmargin=0mm]
\item[]  {\bf Input}: Initial values of the model parameters $\theta^{(0)}=\big(b^{(0)\top}, \eta_\alpha^{(0)}, \eta_\beta^{(0)}, \eta_e^{(0)}\big)^{\top}$, row random effects $\alphab^{(0)}$, column random effects $\betab^{(0)}$; the step size matrix of the vector $b$, $\Ecal_b^{(t)} = \diag\big\{\epsilon_{b_1}^{(t)}, \cdots, \epsilon_{b_p}^{(t)}\big\}$, and the full step size matrix $\Ecal^{(t)} = \diag \big\{\Ecal_b^{(t)}, \epsilon_{\eta_\alpha}^{(t)}, \epsilon_{\eta_\beta}^{(t)}, \epsilon_{\eta_e}^{(t)}\big\}$, for $t=0,\ldots,T-1$.
\item[] {\bf For} $t = 0, \cdots, T -1$ {\bf do}
\begin{enumerate}
\item {\bf{(Sample the subset of data)}} Select $r$ rows and $c$ columns randomly without replacement from the matrix of the full data $\Yb$, and obtain the matrix of the subset of data $\Yb_n^{(t)}$.
\begin{enumerate}
\item[]{\bf While} $\big(\text{there is no observation in any rows (or columns) in } \Yb_n^{(t)}\big)$
\begin{enumerate}
\item[1.] Remove the rows (or columns) with no observations from $\Yb_n^{(t)}$;
\item[2.] Replace them with other rows (or columns) selected randomly without replacement from $\Yb$.
\end{enumerate}
\item[]{\bf End while}
\end{enumerate}
Based on the row indexes $\{s_1, \cdots, s_r\}$ and column indexes $\{q_1, \cdots, q_c\}$ in $\Yb_n^{(t)}$, collect the corresponding submatrices of fixed effects $\Xb_n^{(t)}$ and indicators $\Zb_n^{(t)}$. Let $n_{i \bullet}^{(t)} =  \sum^c_{j=1}(\Zb_{n}^{(t)})_{ij} $ and $n_{ \bullet j}^{(t)} =  \sum^r_{i=1}(\Zb_{n}^{(t)})_{ij}$.
\item {\bf{(Sample latent variables from $\pi\big(\vartheta \mid \theta^{(t)}, \Yb_n^{(t)}\big)$)}} Use the Gibbs sampler to generate a length-$m$ Markov chain of latent variables
$ \big\{ \alphab_{n,k}^{(t)} \big\}^m_{k=1} = \big\{\alpha^{(t)}_{s_1,k}, \ldots, \alpha^{(t)}_{s_r,k} \big\}^m_{k=1}$ and $\big\{ \betab_{n,k}^{(t)} \big\}^m_{k=1} = \big\{\beta^{(t)}_{q_1,k}, \ldots, \beta^{(t)}_{q_c,k} \big\}^m_{k=1}$ by iteratively sampling from the conditional posterior distributions 
\begin{align} 
& \alpha_{s_i} \mid \theta^{(t)},\betab^{(t)}_n, \Yb_n^{(t)} \sim N \left (\frac{\sum_{j=1}^{c}Z^{(t)}_{s_iq_j} (Y^{(t)}_{s_iq_j} - x_{s_iq_j}^{(t) \top}b^{(t)} -  \beta_{q_j}^{(t)} ) \ee^{\eta_\alpha^{(t)}}}{n^{(t)}_{i \bullet} \ee^{\eta_\alpha^{(t)}} + \ee^{\eta_e^{(t)}} }, ~~\frac{ \ee^{\eta_\alpha^{(t)} + \eta_e^{(t)}} }{n^{(t)}_{i \bullet} \ee^{\eta_\alpha^{(t)}} + \ee^{\eta_e^{(t)}} } \right), \nonumber \\
& \beta_{q_j} \mid \theta^{(t)},\alphab_n^{(t)}, \Yb_n^{(t)} \sim N \left (\frac{\sum_{i=1}^{r}Z^{(t)}_{s_iq_j} (Y^{(t)}_{s_iq_j} - x_{s_iq_j}^{(t)\top}b^{(t)} -   \alpha_{s_i}^{(t)}) \ee^{\eta_\beta^{(t)}}}{n^{(t)}_{ \bullet j} \ee^{\eta_\beta^{(t)}} + \ee^{\eta_e^{(t)}} }, ~~\frac{ \ee^{\eta_\beta^{(t)}+\eta_e^{(t)}} }{n^{(t)}_{ \bullet j} \ee^{\eta_\beta^{(t)}} + \ee^{\eta_e^{(t)}} } \right), \nonumber
\end{align}
where $i = 1, \ldots, r $ for $s_i$, $j = 1, \ldots, c $ for $q_j$, and $\theta^{(t)} = (b^{(t)\top}, \eta_\alpha^{(t)},\eta_\beta^{(t)}, \eta_e^{(t)})^{\top}$.
\item {\bf (Update $\theta$)} Update
$\theta^{(t+1)}= \left(b^{(t+1) \top}, \eta_\alpha^{(t+1)},\eta_\beta^{(t+1)}, \eta_e^{(t+1)} \right)^{\top}$ by the following equations:
\begin{align}\label{eq:PSGLD}
b^{(t+1)} & =  b^{(t)} + \frac{\Ecal_{b}^{(t)}}{2m} {\sum}^m_{k=1} \left[ \frac{N}{n^{(t)}} \nabla_b \log p\big(\Yb_n^{(t)} \mid b^{(t)}, \alphab_{n,k}^{(t)}, \betab_{n,k}^{(t)},  \eta^{(t)}_e \big) + \nabla_b \log \pi\big(b^{(t)}\big) \right] + \psi_{b}^{(t)}, \nonumber \\ 
\eta_\alpha^{(t+1)} & = \eta_\alpha^{(t)} + \frac{\epsilon_{\eta_\alpha}^{(t)}}{2m} {\sum}^m_{k=1} \left[  \frac{R}{r} \nabla_{\eta_\alpha} \log \pi\big(\alpha_{s_1,k}^{(t)}, \ldots, \alpha_{s_r,k}^{(t)} \mid \eta_\alpha^{(t)}\big) +\nabla_{\eta_\alpha} \log \pi\big(\eta_\alpha^{(t)}\big) \right] + \psi_{\eta_\alpha}^{(t)},  \nonumber\\ 
\eta_\beta^{(t+1)} & = \eta_\beta^{(t)} + \frac{\epsilon_{\eta_\beta}^{(t)}}{2m} {\sum}^m_{k=1} \left[ \frac{C}{c} \nabla_{\eta_\beta} \log \pi\big(\beta_{q_1,k}^{(t)}, \ldots, \beta_{q_c,k}^{(t)} \mid \eta_\beta^{(t)}\big) + \nabla_{\eta_\beta} \log \pi\big(\eta_\beta^{(t)}\big) \right] + \psi_{\eta_\beta}^{(t)}, \nonumber\\   
\eta_e^{(t+1)} & = \eta_e^{(t)} + \frac{\epsilon_{\eta_e}^{(t)}}{2m} {\sum}^m_{k=1} \left[ \frac{N}{n^{(t)}} \nabla_{\eta_e} \log p\big(\Yb_n^{(t)} \mid b^{(t)}, \alphab_{n,k}^{(t)}, \betab_{n,k}^{(t)}, \eta^{(t)}_e \big)  + \nabla_{\eta_e} \log \pi\big(\eta_e^{(t)}\big) \right]  + \psi_{\eta_e}^{(t)}, 
\end{align}
where the exact formulas for the gradients are given in Section S2 of the Supplementary Material, and we sample $\psi_{b}^{(t)} \sim N(0,\Ecal_{b}^{(t)})$, $\psi_{\eta_{\alpha}}^{(t)} \sim N(0,\epsilon_{\eta_{\alpha}}^{(t)})$, $\psi_{\eta_{\beta}}^{(t)} \sim N(0,\epsilon_{\eta_{\beta}}^{(t)})$, $\psi_{\eta_e}^{(t)} \sim N(0,\epsilon_{\eta_e}^{(t)})$ independently.
\end{enumerate}
\item[] {\bf End for}
\item[] {\bf Return}: A sequence of $\{\theta^{(t)}\}^{T}_{t=1} = \{b^{(t)\top}, \eta_\alpha^{(t)}, \eta_\beta^{(t)}, \eta_e^{(t)}\}^{T}_{t=1}$, whose empirical distribution is an approximation of the posterior distribution $\pi(\theta \mid \Yb)$.
\end{enumerate}
}
\end{algorithm}
where $(i)$ follows from the posterior decomposition $\pi(\theta| \Yb_n,\alphab_n,\betab_n) \propto p(\Yb_n |\theta,\alphab_n,\betab_n)\cdot \pi(\alphab_n|\eta_{\alpha}) \cdot \pi(\betab_n|\eta_{\beta}) \cdot \pi(\theta) $.
Therefore, with the Markov chain $\{\alphab_{n,k}, \betab_{n,k}\}_{k=1}^m$ drawn from their conditional posterior distributions based on the subset of data $\pi(\alphab_n, \betab_n | \theta, \Yb_n)$, we have that $m^{-1} {\sum}^m_{k=1} \left[(R/r) \nabla_{\eta_\alpha} \log \pi\big(\alpha_{s_1,k}, \ldots, \alpha_{s_r,k} | \eta_\alpha\big) +\nabla_{\eta_\alpha} \log \pi\big(\eta_\alpha\big) \right]$ in equation \eqref{eq:PSGLD} of Algorithm \ref{algo:pigeonhole SGLD} is a Monte Carlo approximation of $ (R/r)\nabla_{\eta_\alpha}\log p(\Yb_n | \theta) + \nabla_{\eta_\alpha} \log \pi(\eta_\alpha)$. By a similar argument, for the gradient of $\eta_{\beta}$ in \eqref{eq:PSGLD}, \\ $m^{-1} {\sum}^m_{k=1} \Big[(C/c) \nabla_{\eta_\beta} \log \pi\big(\beta_{q_1,k}, \ldots, \beta_{q_c,k} | \eta_\beta \big)$ $+\nabla_{\eta_\beta} \log \pi\big(\eta_\beta\big) \Big]$ in equation \eqref{eq:PSGLD} of Algorithm \ref{algo:pigeonhole SGLD} is a Monte Carlo approximation of
$ (C/c)\nabla_{\eta_\beta}\log p(\Yb_n | \theta) + \nabla_{\eta_\beta} \log \pi(\eta_\beta)$.
\end{itemize}
Therefore, for the model with missing data, we can use the Markov chain Monte Carlo-based subset gradients for $b,\eta_{\alpha},\eta_{\beta},\eta_e$ in equation \eqref{eq:PSGLD} of Algorithm \ref{algo:pigeonhole SGLD} to approximate the subset gradients in equation \eqref{eq:SGLD} of Algorithm \ref{algo:SGLD}.
\vspace{2mm}

\noindent \underline{\textsc{Step 2}: Subset gradients approximate the full data gradients.}

As we can see from equation \eqref{eq:grad.eta.e.ex}, for the gradient of the subset log-likelihood with respect to $\eta_e$, that is $(N/n)\nabla_{\eta_e} \log p(\Yb_n\mid \theta)$, we can express it as an expectation of the gradient $(N/n) \nabla_{\eta_e}\log p(\Yb_n\mid  b, \alphab_n, \betab_n, \eta_e)$, where the expectation is taken with respect to the posterior distribution of the latent row and column random effects $\alphab_n$ and $\betab_n$. The same relation holds for the full data $\Yb$, that is, we can write
\begin{align} \label{eq:grad.eta.e.ex2}
& \nabla_{\eta_e} \log p(\Yb \mid \theta) = \int \nabla_{\eta_e}\log p(\Yb \mid  b, \alphab, \betab, \eta_e) \pi(\alphab,\betab \mid \theta, \Yb)\ud\alphab \ud \betab,
\end{align}
where $\{\alphab,\betab\}$ contains all the $R$ row effects and $C$ column effects. On the other hand, from the exact formulas of the gradients in Section S2 of the Supplementary Material, we can see that the subset gradient $\nabla_{\eta_e} \log p\big(\Yb_n | b, \alphab_{n}, \betab_{n},  \eta_e \big)$ in equation \eqref{eq:grad.eta.e.ex} is a summation of $n$ independent terms if the model parameters are all fixed at their true values. Similarly, the full data gradient $\nabla_{\eta_e} \log p(\Yb \mid b, \alphab, \betab, \eta_e)$ in \eqref{eq:grad.eta.e.ex2} can also be written as
\begin{align*}
\nabla_{\eta_e}\log p(\Yb \mid  b, \alphab, \betab, \eta_e) & = \sum^R_{i=1} \sum^C_{j=1} Z_{ij} \Big[ -1 + \big(Y_{ij} -x_{ij}^{\top}b - \alpha_{i} - \beta_{j} \big)^2 \ee^{-\eta_e}  \Big] /2,
\end{align*}
which is the summation of $N$ independent terms if the model parameters are all fixed at their true values. Since we randomly select the rows and columns in Algorithms \ref{algo:SGLD} and \ref{algo:pigeonhole SGLD}, we can see that $(N/n)\nabla_{\eta_e}\log p(\Yb_n\mid  b, \alphab_n, \betab_n, \eta_e)$ in equation \eqref{eq:grad.eta.e.ex} is an unbiased estimator of $\nabla_{\eta_e}\log p(\Yb \mid  b, \alphab, \betab, \eta_e) $ in equation \eqref{eq:grad.eta.e.ex2}. This explains why we have used the multiplicative factor $N/n$ for the gradient with respect to $\eta_e$ in the two algorithms. We can use the same argument to explain the multiplicative factor $N/n$ for the gradient with respect to $b$ in both algorithms.

Now for the gradient with respect to $\eta_{\alpha}$, that is $(R/r)\nabla_{\eta_\alpha}\log p(\Yb_n \mid \theta)$, we can see from equation \eqref{eq:grad.eta.alpha.ex} that it can be written as an expectation of the gradient \\$(R/r)\nabla_{\eta_\alpha}\log \pi( \alphab_n \mid \eta_\alpha)$, where the expectation is taken with respect to the posterior distribution of the latent row random effects $\alphab_n$. The same relation holds for the full data $\Yb$, that is, we can write
\begin{align} \label{eq:grad.eta.alpha.ex2}
& \nabla_{\eta_\alpha} \log p(\Yb \mid \theta) = \int \nabla_{\eta_\alpha}\log \pi( \alphab\mid \eta_\alpha) \pi(\alphab \mid \theta, \Yb)\ud\alphab ,
\end{align}
where $\alphab$ contains all the $R$ row effects. On the other hand, from the exact formulas of the gradients in Section S2 of the Supplementary Material, we can see that the subset gradient $\nabla_{\eta_\alpha} \log \pi\big(\alphab_{n}| \eta_\alpha \big)$ in equation \eqref{eq:grad.eta.alpha.ex} is a summation of $r$ independent terms if the model parameters are all fixed at their true values. Similarly, the full data gradient $\nabla_{\eta_\alpha} \log \pi(\alphab | \eta_\alpha)$ in \eqref{eq:grad.eta.alpha.ex2} can also be written as
\begin{align*}
\nabla_{\eta_\alpha}\log \pi(\alphab | \eta_\alpha) & = \sum^{R}_{i=1} \Big[-1+ \big\{\alpha_{i}\big\}^2 \ee^{-\eta_\alpha} \Big] /2,
\end{align*}
which is the summation of $R$ independent terms if the model parameters are all fixed at their true values. Since we randomly select the rows and columns in Algorithms \ref{algo:SGLD} and \ref{algo:pigeonhole SGLD}, we can see that $(R/r)\nabla_{\eta_\alpha}\log \pi(\alphab_n | \eta_\alpha)$ in equation \eqref{eq:grad.eta.alpha.ex} is an unbiased estimator of $\nabla_{\eta_e}\log \pi(\alphab | \eta_\alpha)$ in equation \eqref{eq:grad.eta.alpha.ex2}. This explains why we have used the multiplicative factor $R/r$ for the gradient with respect to $\eta_\alpha$ in both algorithms. We can use a similar argument to explain the multiplicative factor $C/c$ for the gradient with respect to $\eta_\beta$ in both algorithms.
Finally, the validity of our Algorithms \ref{algo:SGLD} and \ref{algo:pigeonhole SGLD} follows by combining the approximations in Step 1 and Step 2 above.

There are two main differences from the pigeonhole SGLD Algorithm \ref{algo:pigeonhole SGLD} to the extended SGLD algorithm (Algorithm S1) in \citet{song2020extended}. First, Algorithm \ref{algo:pigeonhole SGLD} is designed for fitting the crossed mixed effects model in which the data $\Yb$ are dependent, while Algorithm S1 in \citet{song2020extended} is designed solely for independent data. Second, Algorithm S1 in \citet{song2020extended} contains an importance resampling step, which originates from the argument that the subset of data need to be augmented to the size of full data such that the posterior variation of both $\theta$ and $\vartheta$ can be correctly quantified. However, our SGLD algorithms drop this step based on an alternative perspective. We treat the gradients in \eqref{eq:PSGLD} merely as subset-based stochastic approximation estimators of the various gradients of log-posteriors. Hence it becomes unnecessary to justify that they are compatible or come from some well-defined adjusted posteriors conditional on augmented data. This is also the same perspective as in the originally proposed SGLD algorithm in \citet{welling2011bayesian}. Furthermore, the impact from the prior on such stochastic approximations is minimal in practice. We provide theoretical justification for the convergence of the pigeonhole SGLD Algorithm \ref{algo:pigeonhole SGLD} in Section \ref{sec:theorem}.

\section{Convergence Analysis of Pigeonhole SGLD}
\label{sec:theorem}

We derive the convergence of the proposed pigeonhole SGLD Algorithm \ref{algo:pigeonhole SGLD} applied to the crossed mixed effects model defined by \eqref{eq:lme}, \eqref{eq:normalre} and \eqref{eq:prior}. The theory can be derived similarly for the SGLD algorithm in the simpler case of balanced design without missing data in Algorithm \ref{algo:SGLD}. The convergence and approximation error analysis of SGLD for models with $\iid$ data has been studied in the literature for log-concave posterior densities (\citealt{dalalyan2017}, \citealt{dalalyan2019user}) and non-log-concave posterior densities (\citealt{zou2020faster}, \citealt{Chauetal2021}). For our crossed mixed effects model, establishing a similar convergence theory poses several challenges. First, the posterior density of $\theta$ in our model is clearly not log-concave for the three variance components $\sigma_{\alpha}^2,\sigma_{\beta}^2,\sigma_e^2$ or their logarithms. Second, most of the previous theoretical works on SGLD require a global Lipschitz condition on the gradient of the log density function, such as Equation (1) in both \citet{dalalyan2017} and \citet{dalalyan2019user}, Assumption 4.4 in \citet{zou2020faster}, and Assumption H2 in \citet{Chauetal2021}. This global Lipschitz condition is too strong and not satisfied by almost any statistical model that contains a variance parameter in the range $(0,+\infty)$. In fact, neither the log-concavity condition nor the global Lipschitz condition holds even for the posterior distribution of the simplest possible statistical model with $\iid$ data from $N(\mu,\sigma^2)$, where both $(\mu,\sigma^2)\in \RR\times (0,+\infty)$ are unknown parameters. For our crossed mixed effects model specified by \eqref{eq:lme}, \eqref{eq:normalre} and \eqref{eq:prior}, it is straightforward to see that the gradient of log-posterior density can grow unbounded as $\sigma^2_{\alpha},\sigma^2_{\beta},\sigma^2_e$ approach zero, and therefore, is not globally Lipschitz with respect to $\sigma^2_{\alpha},\sigma^2_{\beta},\sigma^2_e$ or their logarithms.

To overcome the issues of non-log-concave posterior density and unbounded gradient, we consider a constrained version of the posterior distribution and an adapted version of the pigeonhole SGLD algorithm. For positive constants $B_0,A_1,B_1,E_1$, we define the sieve parameter set
\begin{align} \label{eq:Theta}
\Theta_N &:= \Theta_N(B_0,A_1,B_1,E_1) \nonumber \\
&= \Big\{\theta=(b^\top,\eta_{\alpha},\eta_{\beta},\eta_e)^\top \in \RR^{p+3}:~ \|b\|_{\infty}\leqslant B_0\log N, |\eta_{\alpha}| \leqslant A_1 \log\log N, \nonumber \\
&\qquad |\eta_{\beta}| \leqslant B_1 \log\log N, ~ |\eta_e| \leqslant E_1 \log\log N \Big\} ,
\end{align}
where $\|\cdot\|_{\infty}$ denotes the $\ell_{\infty}$-norm. The size of the sieve $\Theta_N$ increases with $N$ and will eventually cover the entire space of $\RR^{p+3}$ as $N\to\infty$. The increasing rates along the components of $\theta$ are set in the way such that the radius increases at the $\log N$ rate for each parameter of $b,\sigma_{\alpha}^2,\sigma_{\beta}^2,\sigma_e^2$. On the sieve $\Theta_N$, the first and second derivatives of the log posterior density satisfy the Lipschitz condition with the Lipschitz constants growing polynomially in $N$. As a consequence, the global Lipschitz condition holds on the bounded set $\Theta_N$, and we can choose $T$ and the step sizes dependent on $N$ to establish the convergence of PSGLD using the techniques in \citet{zou2020faster}.

Let $\Pi_N^*(\ud \theta) \propto \Pi(\ud \theta\mid \Yb) \cdot \mathbbm{1}(\theta\in \Theta_N)$ be the truncated version of the posterior distribution $\Pi(\ud \theta\mid \Yb)$ to the sieve $\Theta_N$, where we have suppressed the conditional on $\Yb$ to simplify the notation. Correspondingly, we also consider an adapted version of Algorithm \ref{algo:pigeonhole SGLD} inside the sieve $\Theta_N$. At the end of Step (c), we add another checking step: if $\theta^{(t+1)} \in \Theta_N$, then we accept it; otherwise, we redo the normal proposal of $\psi_{b}^{(t)},\psi_{\eta_{\alpha}}^{(t)},\psi_{\eta_{\beta}}^{(t)},\psi_{\eta_e}^{(t)}$ until $\theta^{(t+1)} \in \Theta_N$ is satisfied. This additional step is equivalent to modifying the proposal distribution from an unconstrained normal on $\RR^{p+3}$ to a truncated normal with the support $\Theta_N$. We still call this algorithm the pigeonhole SGLD algorithm in the following theorem, and this additional step is only for the theory development within this section. In practice, since $N$ is typically large and $B_0,A_1,B_1,E_1$ in the definition of $\Theta_N$ in \eqref{eq:Theta} can be arbitrarily large, this does not affect the practical performance of Algorithm \ref{algo:pigeonhole SGLD} with the unconstrained normal proposal.

We are mainly concerned about the convergence from the empirical distribution of the posterior sample of parameters $\{\theta^{(t)}:t=1,\ldots,T\}$ from the adapted version of Algorithm \ref{algo:pigeonhole SGLD}, denoted by $\Pi_T$, to the target posterior distribution $\Pi_N^*$ under the Bayesian model specified by \eqref{eq:lme}, \eqref{eq:normalre}, and \eqref{eq:prior}. The convergence is in the asymptotic regime under which both the amount of observed data $N$ and the number of SGLD iterations $T$ go to infinity, which is the same asymptotic regime adopted by the theory for the extended SGMCMC algorithm in \cite{song2020extended}. There are two reasons why we consider this asymptotic regime. First, the asymptotics of $N\to\infty$ for the posterior distribution based on the full data under the crossed mixed effects model \eqref{eq:lme} is not very well understood and possibly nonstandard. To the best of our knowledge, standard Bayesian asymptotic theory, such as the posterior consistency and Bernstein-von Mises theorem, has never been established for $\theta=(b^\top,\eta_{\alpha},\eta_{\beta},\eta_e)^\top$ in the crossed mixed effects model \eqref{eq:lme} before, possibly due to the technical challenge from the complex crossed dependence in the model \eqref{eq:lme}. Thus theoretically, one cannot simply claim or assume that the posterior distribution of $\theta$ is asymptotically normal as $N\to\infty$. Second, we will allow the model \eqref{eq:lme} to be misspecified for the observed data, as can be seen clearly from Assumption \ref{assump:bound} below. 
Therefore, it is meaningful to discuss how the PSGLD algorithm can recover the posterior distribution of $\theta$ as $N,T\to\infty$.

We make a series of assumptions on the data, the model, and the algorithm. For the data, we consider the case where the amount of missing data increases proportionally to the total size of the data matrix $\Yb$.
\begin{assumption} \label{assump:asymp}
There exist two constants $0<\underline c< \overline c\leqslant 1$, such that $\underline c \leqslant N/(RC)\leqslant \overline c$, where $N=\sum_{i=1}^R \sum_{j=1}^C Z_{ij}$. The number of rows $r$ and the number of columns $c$ in each subset are kept as constants.
\end{assumption}
For the response variable $y$ and the predictors $x$, we impose the following regularity conditions as in many regression literature.
\begin{assumption} \label{assump:bound}
The $R\times C$ full data matrix $\Yb$ consists of random variables $Y_{ij}$, such that $\PP(|Y_{ij}|\geqslant C_y \log N) \leq \exp\{-(1/2)\log^2 N\}$ for a constant $C_y>0$, for all $i=1,\ldots,R$ and $j=1,\ldots,C$. The covariates $x_{ij}$ are known constants and satisfy $\max_{1\leqslant i\leqslant R, 1\leqslant j\leqslant C}|x_{ij}| \leqslant C_x$ for a constant $C_x>0$.
\end{assumption}

\begin{assumption} \label{assump:eigen}
There exist two constants $0< \underline \lambda_x < \overline \lambda_x < \infty$, such that for any row index set $\{s_1,\ldots,s_r\}\subseteq \{1,\ldots,R\}$ and any column index set $\{q_1,\ldots,q_c\}\subseteq \{1,\ldots,C\}$, the positive definite matrix $n^{-1}\sum_{i=1}^r \sum_{j=1}^c Z_{s_iq_j}x_{s_iq_j}x_{s_iq_j}^\top$ has its eigenvalues lower bounded by $\underline \lambda_x$ and upper bounded by $\overline \lambda_x$, for all $R,C,N \in \ZZ_+$, where $n=\sum_{i=1}^r\sum_{j=1}^c Z_{s_iq_j}$.
\end{assumption}
Assumption \ref{assump:bound} is very general and does not require the true data generating process of the response variable $Y_{ij}$ to strictly follow the crossed mixed effects model specified by \eqref{eq:lme} and \eqref{eq:normalre}. In other words, our convergence theory even works when the model in \eqref{eq:lme} and \eqref{eq:normalre} is misspecified. Such misspecification is common in real applications. For example, one may have missed some important covariates in $x$ for modeling $y$. By Assumption \ref{assump:bound}, we essentially do not assume the existence of a ``true model'' with ``true parameters''. As a result, our convergence theory below is essentially different from the standard frequentist evaluation of Bayesian procedures which typically requires a true model with true parameters; see for example, Chapter 10 in \citet{Van98}, and Chapters 6-9 in \citet{GhoVan17}. We notice that the inequality in Assumption \ref{assump:bound} only requires the distribution of $Y_{ij}$ to have small tail probabilities, which is trivially satisfied by any sub-Gaussian distribution. Assumption \ref{assump:eigen} imposes some restrictions on the eigenvalues of the predictor variables $x_{ij}$ and implicitly on the missing mechanism. Our convergence analysis will treat all $x_{ij}$'s and $Z_{ij}$'s as known constants rather than random variables, and our theory works for all missing mechanisms that satisfy Assumptions \ref{assump:asymp} and \ref{assump:eigen}.

The next assumption is on the step size and the initial values.
\begin{assumption} \label{assump:initial}
In both Algorithms \ref{algo:SGLD} and \ref{algo:pigeonhole SGLD}, the step size matrix $\Ecal=\Ecal^{(t)}$ is a constant diagonal matrix, with $\epsilon_{\min}$ and $\epsilon_{\max}$ being its minimum and maximum diagonal entries and satisfying $\epsilon_{\max}/\epsilon_{\min} \leqslant \overline c_{\epsilon} <\infty$ for a constant $\overline c_{\epsilon}$. The initial value $\theta^{(0)}$ is drawn from a distribution $\nu_0$ whose support is inside $\Theta_N$. Furthermore, $\nu_0$ is a $\lambda$-warm start with respect to $\Pi^*_N$ for some constant $\lambda>0$, i.e., $\sup_{\Acal \subseteq \Theta_N} \nu_0(\Acal)/\Pi_N^*(\Acal) \leqslant \lambda$.
\end{assumption}
Constant step sizes are commonly used in real applications of SGLD. The initial distribution $\nu_0$ from which the initial value $\theta^{(0)}$ is drawn is a reasonably good proxy to the true posterior $\Pi_N^*$. This condition has been commonly adopted by the theory on SGLD such as \cite{zou2020faster}, etc.

For stating our theory, we need the following definition of the Cheeger constant. For a probability measure $\nu$ on $\Theta_N$, we say that $\nu$ satisfies the isoperimetric inequality with Cheeger constant $\rho$ if for any $\Acal \subseteq \Theta_N$,
\[ \liminf_{d \rightarrow 0+} \frac{\nu(\Acal_d) - \nu(\Acal)}{d} \geqslant \rho \min\{\nu(\Acal), 1-\nu(\Acal) \},   \]
where $\Acal_d = \{x \in \Theta_N: \exists y \in \Acal, \|x - y\|_2 \leqslant d \}$ and $\|\cdot\|_2$ is the Euclidean norm.

For two positive sequences $\sfa_n$ and $\sfb_n$, we use $\sfa_n\prec \sfb_n$ and $\sfb_n\succ \sfa_n$ to denote the relation $\lim_{n\to\infty} \sfa_n/\sfb_n=0$. We use $\sfa_n\preceq \sfb_n$, $\sfb_n\succeq \sfa_n$, and $\sfa_n=O(\sfb_n)$ to denote the relation $\limsup_{n\to\infty} \sfa_n/\sfb_n<+\infty$, and $\sfa_n\asymp \sfb_n$ to denote the relation $\sfa_n\preceq \sfb_n$ and $\sfa_n\succeq \sfb_n$. For two probability measures $P_1,P_2$, let $\|P_1-P_2\|_{\TV}=\sup_{\Acal}|P_1(\Acal)-P_2(\Acal)|$ be the total variation distance between $P_1$ and $P_2$, where the supremum is taken over all measurable sets $\Acal$.

The following theorem states the convergence of the pigeonhole SGLD algorithm.
\begin{theorem} \label{th1}
Suppose that Assumptions \ref{assump:asymp}, \ref{assump:bound}, \ref{assump:eigen} and \ref{assump:initial} hold. Suppose that $\log T \asymp \log N$ as $N,T\to\infty$. Suppose that for a constant $\zeta>0$, the maximal step size $\epsilon_{\max}$ satisfies
\begin{align} \label{eq:emax.cond0}
& {\epsilon}_{\max} \asymp \min(\rho^2,1) N^{-4(1+\zeta)} ,
\end{align}
where $\rho$ is the Cheeger constant of the posterior distribution $\Pi_N^*$.
\begin{enumerate}[label=(\roman*),leftmargin=5mm]
\item[(i)] The total variation distance between the empirical distribution of the output from the pigeonhole SGLD $\Pi_T$ and the target posterior distribution $\Pi_N^*$ satisfies that with probability at least $1 - (Tmr + Tmc + \underline c^{-1} N) \exp\left\{-(1/2)\log^2 N\right\}  - 4\exp(-\sqrt{T}N^{-\zeta}/8)$, as $N,T \to\infty$,
\begin{align}\label{th1f1}
\left\|\Pi_T - \Pi_N^*\right\|_{\TV} \leqslant \lambda  \left(1- C_1 \rho^2 \epsilon_{\max}\right)^T + C_2 N^{-\zeta},
\end{align}
for some positive constants $C_1,C_2$.

\item[(ii)] Furthermore, if $T = C_T \zeta \rho^{-4}N^{4(1+\zeta)}\log N$, $m\leqslant N^{\varsigma}$, and $\rho\succeq N^{-c_{\nu}}$ for some positive constants $C_T, \varsigma, c_{\nu}$, then $\|\Pi_T-\Pi^*_N\|_{\TV} = O\left(N^{-\zeta}\right)$  almost surely as $N \to\infty$.

\item[(iii)] Following (ii), for any continuous function $f(\cdot)$ defined on $\Theta_N$ that satisfies $|f(\theta)|\leqslant C_f$ for all $\theta\in \Theta_N$ and a finite constant $C_f$,
$$\left|T^{-1}\sum^T_{t=1}f\big(\theta^{(t)}\big) - \int_{\Theta_N} f(\theta) \Pi_N^*(\ud \theta)\right| \rightarrow 0, $$
in probability as $N,T \rightarrow \infty$.
\end{enumerate}
\end{theorem}

Theorem \ref{th1} shows that as $N, T\to \infty$, the empirical distribution of the posterior samples from the pigeonhole SGLD algorithm is close in total variation distance to the true posterior distribution truncated to $\Theta_N$. The asymptotics of $N,T\to\infty$ are the same as Theorem 1 of \citet{song2020extended} for the extended SGMCMC algorithm. The convergence in total variation distance on the bounded parameter space $\Theta_N$ is stronger than the convergence in Wasserstein-2 distance in \citet{dalalyan2017}, \citet{dalalyan2019user}, and \citet{Chauetal2021}. In Part (i), we provide an upper bound on the total variation distance between these two distributions, similar to Theorem 4.5 of \citet{zou2020faster}. In fact, we have adapted the proof techniques and used the same auxiliary sequence of Metropolized SGLD as \citet{zou2020faster}. As a result, the first term on the right-hand side of \eqref{th1f1} is the sampling error of the auxiliary sequence generated by the Metropolized SGLD, and the second term accounts for the distance between the outputs from the pigeonhole SGLD and the Metropolized SGLD; see Section S4 of the Supplementary Material for details of the technical proof. If we further specify the polynomial order of $T$ in $N$, then Part (ii) shows that this total variation distance converges to zero as $N\to\infty$. Part (iii) is a consequence of Part (ii) for the convergence of sample average of bounded functions to the true posterior mean.

In Theorem \ref{th1}, there are various quantities dependent on the Cheeger constant $\rho$ of the target posterior distribution $\Pi_N^*$. When the crossed mixed effects model \eqref{eq:lme} is correctly specified for the data matrix $\Yb$, one would expect that most of the posterior probability mass of $\Pi_N^*$ concentrates on a small neighborhood around the ``true'' model parameters as $N\to\infty$. In regular parametric models, this neighborhood typically has a radius of order $O(N^{-1/2})$. Therefore, the Cheeger constant $\rho$ of $\Pi_N^*$ can be of some polynomial order of $N$; see the discussion in Remark 4.6 of \citet{zou2020faster} on various existing results on $\rho$. While it is desirable to derive an explicit lower bound for $\rho$, to the best of our knowledge, the Bayesian posterior contraction theory of the model parameters $(b,\sigma_{\alpha}^2,\sigma_{\beta}^2,\sigma_e^2)$ in the crossed mixed effects model or even the general linear mixed effects model has remained an open problem and requires further investigation.

We emphasize that Theorem \ref{th1} provides the convergence guarantee for the pigeonhole SGLD Algorithm \ref{algo:pigeonhole SGLD} when the sample size $N$ and the number of iterations $T$ become large. In the most ideal case, when the Cheeger constant $\rho$ is of constant order, we can see that the order of $T$ is at least $O(N^{4(1+\zeta)}\log N)$ from Part (ii). This excessively large order of $T$ is mainly the result of our general assumptions and current proof techniques. Theorem \ref{th1} does not necessarily imply that the pigeonhole SGLD algorithm requires more than $O(N^4)$ iterations to converge, for the following reasons. First, our Assumptions \ref{assump:asymp}-\ref{assump:initial} are very general. We do not require the correct specification of model \eqref{eq:lme} and we have almost no assumption on the missing mechanism in the data matrix $\Yb$ other than the sub-Gaussian tails and the proportional size of missing data. Second, as already explained above, if the crossed mixed effects model is correctly specified, one can expect that the posterior distribution $\Pi_N^*$ concentrates on an $O(N^{-1/2})$ neighborhood of the true parameters. In such cases, many of our upper bounds for the gradient of log-posterior densities used in the proofs can be significantly improved. On the other hand, such improvement is only possible when a rigorous Bayesian posterior asymptotic theory for $\theta$, such as the Bernstein-von Mises theorem, is established in the first place. Third, instead of adapting the proof techniques of \citet{zou2020faster} for non-log-concave posteriors, one can also consider other proof ideas such as bounding the difference to the Langevin dynamics in the continuous-time setting as in \citet{Chauetal2021}. However, a close examination reveals that the theory in \citet{Chauetal2021}  will lead to an even worse result that $T$ has to increase at least at an exponential order of $N$. Finally, our empirical results on many experiments in Section \ref{sec:experiments} show that the pigeonhole SGLD converges much faster to the true posterior distribution than the Gibbs sampler based on the full data, and is therefore a promising method for real applications.

\section{Numerical Experiments} \label{sec:experiments}
We apply the proposed Algorithms \ref{algo:SGLD} and \ref{algo:pigeonhole SGLD} for Bayesian inference on the coefficients of fixed effects $b=(b_1, \ldots, b_p)^{\top}$ and the variance components $\sigma^2_{\alpha}, \sigma^2_\beta, \sigma^2_e$ using both simulated data and two real datasets. We present extensive simulation studies on Algorithms \ref{algo:SGLD} and \ref{algo:pigeonhole SGLD} in Section S3 of the Supplementary Material. In the real data examples, we compare the proposed pigeonhole SGLD algorithm with the full-data Gibbs sampler and the restricted maximum likelihood estimator (REML) computed from the R package \texttt{lme4} (\citealt{Batetal15}). All experiments were run on a Windows machine with Intel(R) core(TM) i$7$-$9700$ CPU with $3.0$GHz $8$ core compute nodes and $32$GB memory. All the SGLD algorithms and Gibbs samplers were implemented in R version $4.0.5$. The method of moments in \cite{gao2019estimation} was implemented in Python version $3.8.5$.

For each Bayesian algorithm, we drop the initial $10^4$ iterations as burn-in, and then run the posterior chain for another $T=10^4$ iterations with a thinning step of every $10$th sample. The posterior samples from the full-data Gibbs sampler are used as the benchmark in all our comparisons.

For the marginal posterior distribution of each component of $\theta$, the approximation error from the proposed SGLD algorithms to the true posterior distribution is evaluated by the Wasserstein-$2$ ($W_2$) distance between the empirical distributions of the samples from the proposed algorithms and the one from the Gibbs sampler. In particular, for two generic univariate distribution functions $F_1,F_2$, their $W_2$ distance is given by $W_2(F_1, F_2) = \left[\int_0^1 \{F^{-1}_1(u) - F^{-1}_2(u)\}^2\ud u \right]^{1/2}$, where $F^{-1}(u)=\inf\{x:F(x)\geqslant u\}$ is the quantile function of $F(x)$. This approximation error in $W_2$ distance can be accurately evaluated based on empirical quantiles (\citealt{li2017simple}), which are readily available from the posterior samples from the Gibbs sampler and our SGLD algorithms.

\subsection{MovieLens Data} \label{realdata:1}

We illustrate the application of the pigeonhole SGLD algorithm through two real data examples. The main purpose is to show both the performance in the posterior estimation of parameters and the computational efficiency of the PSGLD relative to the full-data Gibbs sampler.

In this section, we analyze a MovieLens dataset containing evaluations of movies, which is freely available at \url{https://grouplens.org/datasets/movielens/} as a zip archive \texttt{ml-1m.zip}. The dataset contains $1,000,209$ anonymous ratings of around $3,900$ movies made by $6,040$ MovieLens users who joined MovieLens in 2000. Each user has at least $20$ ratings. We remove all the movies with ratings fewer than $20$ from the dataset and obtain the full data matrix $\Yb$ with $R = 6,040$ rows of users, $C = 3,043$ columns of movies, and in total $N = 995,492$ ratings. The ratings are in the $1$ to $5$ scales in the increment of $1$. In the dataset of ratings, each observation consists of a user-ID, a movie-ID, a rating, and the time of rating; in the dataset of movies, there are $19$ genres of movies and each movie is classified into at least one of them. We fit the dataset with the crossed mixed effects model described in \eqref{eq:lme} and \eqref{eq:normalre}, using the movie ratings as responses, the user-IDs and movie-IDs as random effects $\alpha_i$s and $\beta_j$s, and some user- and movie-specific information as fixed effects. Obviously, a generalized linear version of \eqref{eq:lme} fits better for a dataset with categorical responses; nonetheless, \eqref{eq:lme} is still a reasonable model to fit for the MovieLens dataset to some extent.

Following \cite{srivastava2018scalable} and \cite{song2020extended}, we generate three new fixed effects for accurate modeling of ratings. They are the \textit{Genera} predictor, the \textit{Popularity} predictor, and the \textit{Positive} predictor.

\begin{itemize}[leftmargin=5mm]
\item \textit{Genera} predictor, a categorical variable to reduce $19$ genres of movies into $4$ categories, namely `Action', `Children', `Comedy', and `Drama'. \textit{Action} category consists of Action, Adventure, Fantasy, Horror, Sci-Fi, and Thriller genres; \textit{Children} category consists of Animation and Children genres;  \textit{Comedy} category consists of Comedy genre; and \textit{Drama} category consists of Crime, Documentary, Drama, Film-Noir, Musical, Mystery, Romance, War, and Western genres. We use the same coding as that of \cite{song2020extended} to represent each category, i.e., $(1,0,0), (0,1,0), (0,0,1), (-1,-1,-1)$ representing Children, Comedy, Drama and Action. If a movie is classified into several genres, the \textit{Genera} predictor of the movie would be the summation of fractions proportional to the number of all categories to which the genres of the movie belong.

\item \textit{Popularity} predictor, defined as logit$\{(l_j+0.5)/(L_j+1.0)\}$ for the rating $Y_{ij}$, where $L_j$ is the number of recent ratings of movie $j$, and $l_j$ is the number of recent ratings of movie $j$ with the score higher than $3$. Here ``recent'' means 30 or fewer most recent ratings.

\item \textit{Positive} predictor, a dummy variable for the rating $Y_{ij}$, which is defined as $1$ if user $i$ rates more than half of the movies which he/she has rated with scores higher than $3$, and $0$ otherwise. This variable shows whether user $i$ is liable to give a positive review to a movie.
\end{itemize}

We choose $6$ coefficients for fixed effects with $b = (b_0, b_1, b_2, b_3, b_4, b_5)^{\top}$, where $b_0$ is the intercept; $b_1$ is the coefficient of \textit{Positive} predictor; $b_2, b_3, b_4$ are the coefficients of \textit{Genera} predictor; $b_5$ is the coefficient of \textit{Popularity} predictor. We also construct an indicator matrix $\Zb$ with the same dimension of the full data matrix $\Yb$, where $Z_{ij}=1$ if the score of user $i$ giving to movie $j$ is recorded, and $Z_{ij}=0$ otherwise. As described in \eqref{eq:prior}, we assign the following priors on the model parameters: $\pi(b) \propto 1$, $\sigma^2_\alpha \sim \ig(1,1)$, $\sigma^2_\beta \sim \ig(1,1)$ and $\sigma^2_e \sim \ig(0.01,0.01)$. We fit the dataset by the pigeonhole SGLD in Algorithm \ref{algo:pigeonhole SGLD}, and compare the estimated model parameters with the full-data Gibbs sampler and the frequentist REML computed from the R package \texttt{lme4}. For the PSGLD and the Gibbs sampler, we set the initial values to be equal to $1$ for all the parameters $b,\sigma_{\alpha}^2,\sigma_{\beta}^2,\sigma_e^2$.

At each iteration of the PSGLD, we randomly select $r = 200$ rows and $c = 200$ columns from the full data matrix $\Yb$ and construct the submatrix of data $\Yb_n$ with the number of observations $n = \sum^r_{i=1}\sum^c_{j=1}(Z_n)_{s_iq_j}$. A short Markov chain with the length $m=50$ for the latent variables $\{\alpha_{s_1,k}, \cdots, \alpha_{s_r, k}\}^m_{k=1}$, and $\{\beta_{q_1,k}, \cdots, \beta_{q_c, k}\}^m_{k=1}$ is generated by the Gibbs sampler following the conditional distributions in Section S2 of the Supplementary Material before updating the model parameters. We select the step sizes in the PSGLD by a grid search and adopt the combination of step sizes with the lowest $W_2$ distances between the empirical distribution of the PSGLD and that of the full-data Gibbs sampler. We set constant step sizes for the coefficients of fixed effects and the variance components in $O(N^{-1})$.

\begin{figure}[htb]
\caption{Boxplots of posterior samples for the coefficients of fixed effects $b = (b_0, b_1, b_2, b_3, b_4, b_5)^{\top}$ and the variance components $\sigma^2_\alpha, \sigma^2_\beta, \sigma^2_e$ for the crossed mixed effects model for the MovieLens dataset. The results are averaged over $10$ independent runs with each algorithm. PSGLD, pigeonhole stochastic gradient Langevin dynamics; Gibbs, Gibbs sampler.}
\label{boxplot_MovieLens}
\centering
\includegraphics[width=\textwidth]{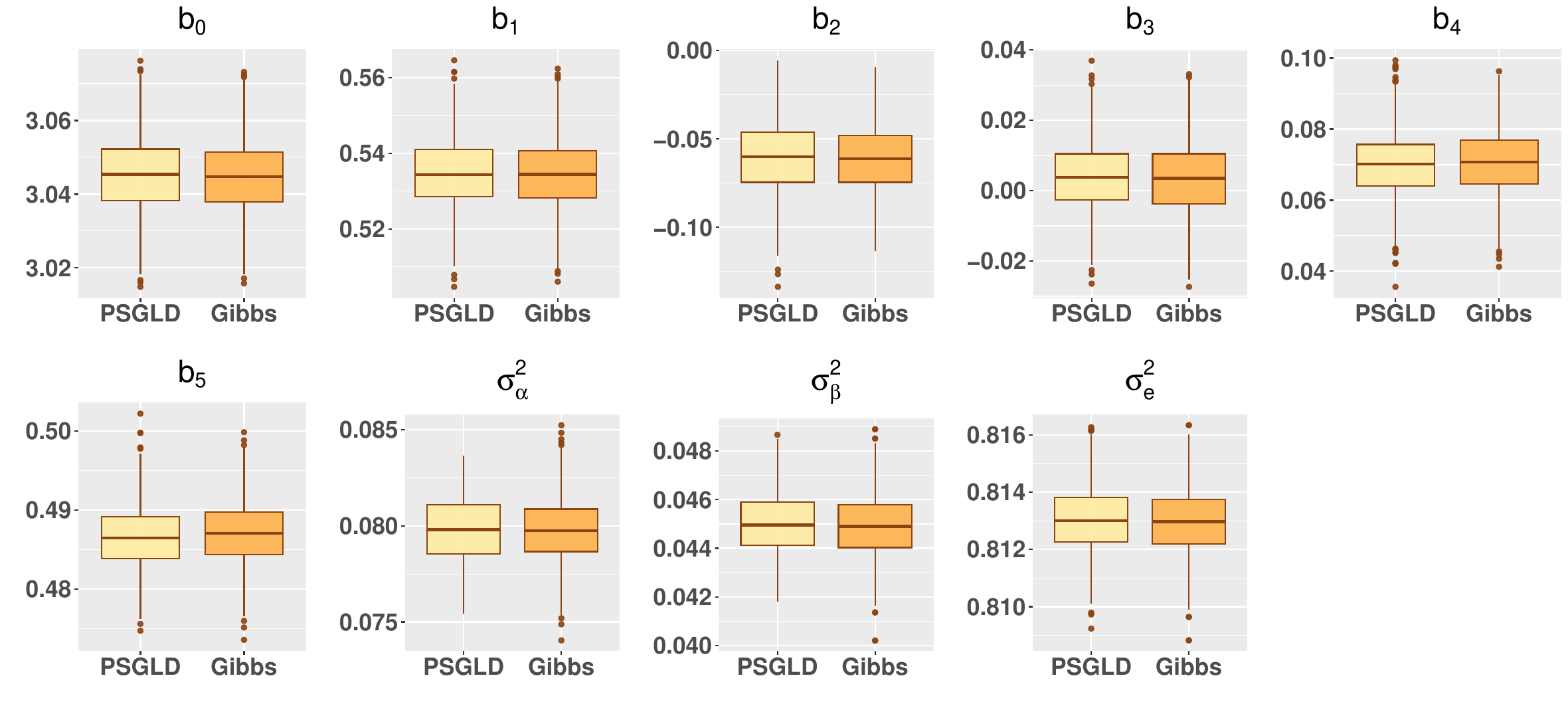}
\end{figure}

Figure \ref{boxplot_MovieLens} shows the boxplots of posterior samples from the pigeonhole SGLD and the Gibbs sampler. We run each algorithm for a chain of length $2\times 10^4$ iterations with the first $10^4$ samples discarded as burn-in. The marginal posterior distributions from the PSGLD are close to those from the Gibbs sampler for all the $9$ parameters including the three variance components, which demonstrates the approximation accuracy of the PSGLD. Furthermore, the posterior means of both the PSGLD and the Gibbs sampler are consistent with the REML computed from the R package \texttt{lme4} (\citealt{Batetal15}), which are $(\widehat{b}_0, \widehat{b}_1, \widehat{b}_2, \widehat{b}_3, \widehat{b}_4, \widehat{b}_5, \widehat{\sigma}^2_{\alpha}, \widehat{\sigma}^2_{\beta}, \widehat{\sigma}^2_e) = (3.0446, 0.5347, -0.0624, 0.0040, 0.0709, 0.4873,$\\ 
$0.0794, 0.0441, 0.8130)^{\top}$.

\begin{figure}[htb]
\caption{$W_2$ distances of the coefficients of fixed effects $b = (b_0, b_1, b_2, b_3, b_4, b_5)^{\top}$ and the variance components $\sigma^2_\alpha, \sigma^2_\beta, \sigma^2_e$ against CPU time (seconds) for the MovieLens dataset, where the brown line is for the pigeonhole stochastic gradient Langevin dynamics algorithm and the yellow line is for the Gibbs sampler. PSGLD, pigeonhole stochastic gradient Langevin dynamics; Gibbs, Gibbs sampler.}
\label{W2_distance_MovieLens}
\centering
\includegraphics[width=\textwidth]{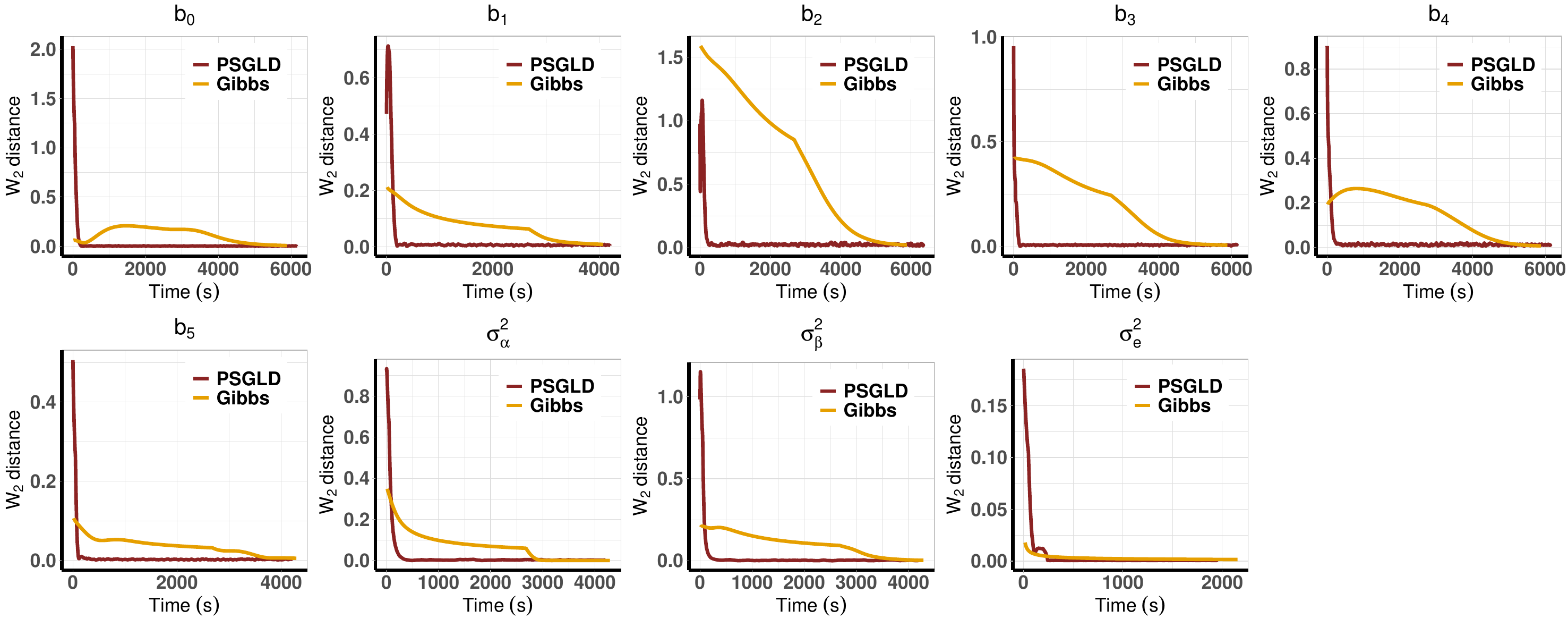}
\end{figure}

We evaluate the computational efficiency of the PSGLD and the Gibbs sampler approaching the target posterior distribution in Figure \ref{W2_distance_MovieLens}. We take the first $500$ samples after $10^4$ burn-in iterations from the Gibbs sampler as the stationary distribution regarded as the true posterior distribution, and iteratively compute the $W_2$ distance between the samples from each algorithm and this benchmark. Starting from iteration $t=3$ (for $t \geqslant 3)$, we record the elapsed CPU time (in seconds) and the $W_2$ distance between the latest $500$ samples ($\{\theta^{(i)}\}_{i = t-499}^{i=t}$ in the case of $t \geqslant 500$, or $\{\theta^{(i)}\}_{i = 1}^{i=t}$ if $t<500$) and the benchmark of the stationary distribution. We plot how the $W_2$ distance decreases with the elapsed CPU time for each of the $9$ model parameters in Figure \ref{W2_distance_MovieLens}, as a lower $W_2$ distance represents better convergence performance. It is clear that for most of the parameters, the posterior samples from the PSGLD have converged to the target true posterior distributions significantly faster than those from the full-data Gibbs sampler. The $W_2$ distance between samples of the PSGLD and the benchmark has dropped to a very low and stable level within $130$ seconds for all the parameters of interest. For all of the coefficients of fixed effects $(b_0, b_1, b_2, b_3, b_4, b_5)^{\top}$ and the variance components of random effects $\sigma^2_\alpha, \sigma^2_\beta$ to reach the same level in the $W_2$ distance, the full-data Gibbs sampler has required much more CPU time, approximately $3,000$ to $6,000$ seconds more than the PSGLD. The only exception is the error variance $\sigma^2_e$, for which the $W_2$ distance from the Gibbs sampler is much smaller than that from the PSGLD at the beginning of iterations, for the trajectory of $\sigma^2_e$ from the Gibbs sampler has converged in $10$ iterations, faster than the pigeonhole SGLD does. For the frequentist REML, it takes $450$ seconds for the R package \texttt{lme4} to fit the MovieLens dataset, which is also slower than the PSGLD.

\subsection{ETH Lecturer Evaluation Data} \label{realdata:2}

Our second real data example is for the dataset of evaluations of lecturers in ETH Zurich named \texttt{InstEval} freely available from the R package \texttt{lme4} (\citealt{Batetal15}), which consists of $73,421$ anonymous evaluations of $1,128$ lecturers made by $2,972$ students. We remove data of students giving fewer than $5$ evaluations and construct a full data matrix $\Yb$ with $R = 2,937$ rows of students, $C = 1,128$ columns of lecturers, and in total $N = 73,328$ evaluations. The evaluations are in the $1$ to $5$ scales in the increment of $1$. There are factors affecting the evaluations contained in the dataset, including \textit{studage}, denoting the number of semesters that the student has been enrolled; \textit{lectage}, measuring the number of semesters back the lecture rated had taken place; \textit{service}, a binary factor showing if a lecture is held for a different department from the lecturer's main one; and \textit{dept}, coding the department of the lecture. We take the factors \textit{studage}, \textit{lectage}, and \textit{service} as fixed effects, and use students and lecturers as the row and column random effects in the analysis. The coefficients for the intercept and fixed effects \textit{studage}, \textit{lectage}, \textit{service} are $b = (b_0, b_1, b_2, b_3)^{\top}$. An indicator matrix $\Zb$ with the same dimension as the full data matrix $\Yb$ is constructed to present the missingness of evaluations.

We implement the pigeonhole SGLD in Algorithm \ref{algo:pigeonhole SGLD} for the \texttt{InstEval} dataset and compare its performance with the full-data Gibbs sampler and the frequentist REML using the R package \texttt{lme4}. For the PSGLD and the Gibbs sampler, the prior distributions are the same as those in the analysis of the MovieLens dataset in Section \ref{realdata:1}. The initial values of the fixed effects coefficients $b$ are set to be $1$, and those of the variance components are set to be $2$. Similar to Section \ref{realdata:1}, for the PSGLD, we use the subset size $r=c=200$ and $m=50$ for the length of Markov chains for the latent variables of random effects. We set constant step sizes for the coefficients of fixed effects and the variance components in $O(N^{-1})$.

\begin{figure}[htb]
\caption{Boxplots of posterior samples for the coefficients of fixed effects $b = (b_0, b_1, b_2, b_3)^{\top}$ and the variance components $\sigma^2_\alpha, \sigma^2_\beta, \sigma^2_e$ for the crossed mixed effects model for the InstEval dataset. The results are averaged over $10$ independent runs with each algorithm. PSGLD, pigeonhole stochastic gradient Langevin dynamics; Gibbs, Gibbs sampler.}
\label{boxplot_InstEval}
\centering
\includegraphics[width=.9\textwidth]{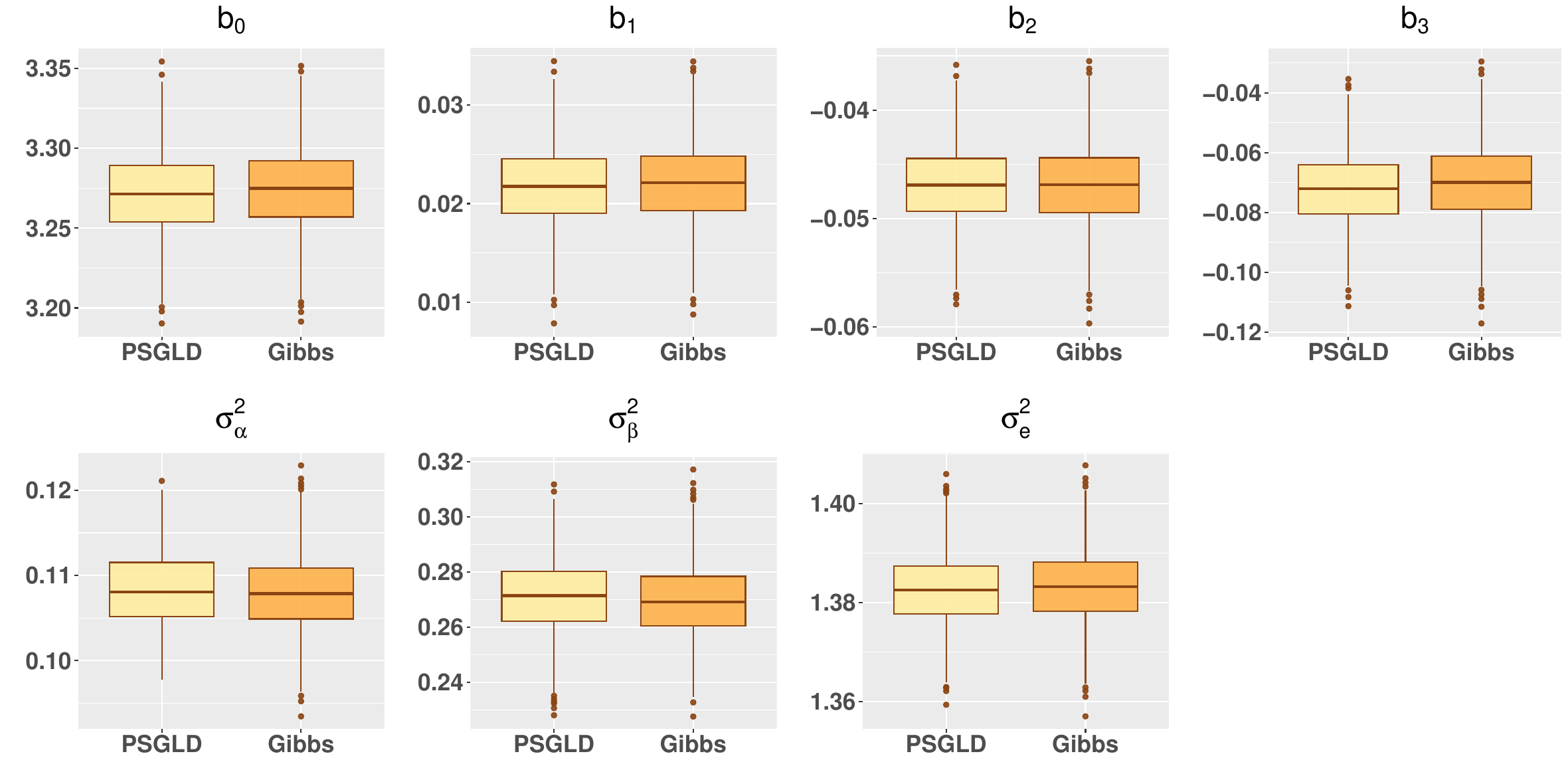}
\end{figure}

Figure \ref{boxplot_InstEval} shows the boxplots of posterior samples from the pigeonhole SGLD and the Gibbs sampler. We run each algorithm for a chain of length $2\times 10^4$ iterations with the first $5,000$ samples discarded as burn-in. The PSGLD provides an accurate approximation of the true posterior distributions from the full-data Gibbs sampler for all the $7$ parameters. Furthermore, the posterior means of both the PSGLD and the Gibbs sampler are consistent with the REML computed from the R package \texttt{lme4}, which are $(\widehat{b}_0, \widehat{b}_1, \widehat{b}_2, \widehat{b}_3, \widehat{\sigma}^2_{\alpha}, \widehat{\sigma}^2_{\beta}, \widehat{\sigma}^2_e)^{\top} = (3.2754, 0.0218, -0.0468, -0.0700, 0.1064, 0.2673, 1.3834)^{\top}$.

\begin{figure}[htb]
\caption{$W_2$ distances of the coefficients of fixed effects $b = (b_0, b_1, b_2, b_3, b_4, b_5)^{\top}$ and the variance components $\sigma^2_\alpha, \sigma^2_\beta, \sigma^2_e$ against CPU time (seconds) for the InstEval dataset, where the brown line is for the pigeonhole stochastic gradient Langevin dynamics algorithm and the yellow line is for the Gibbs sampler. PSGLD, pigeonhole stochastic gradient Langevin dynamics; Gibbs, Gibbs sampler.}
\label{W2_distance_InstEval}
\centering
\includegraphics[width=.9\textwidth]{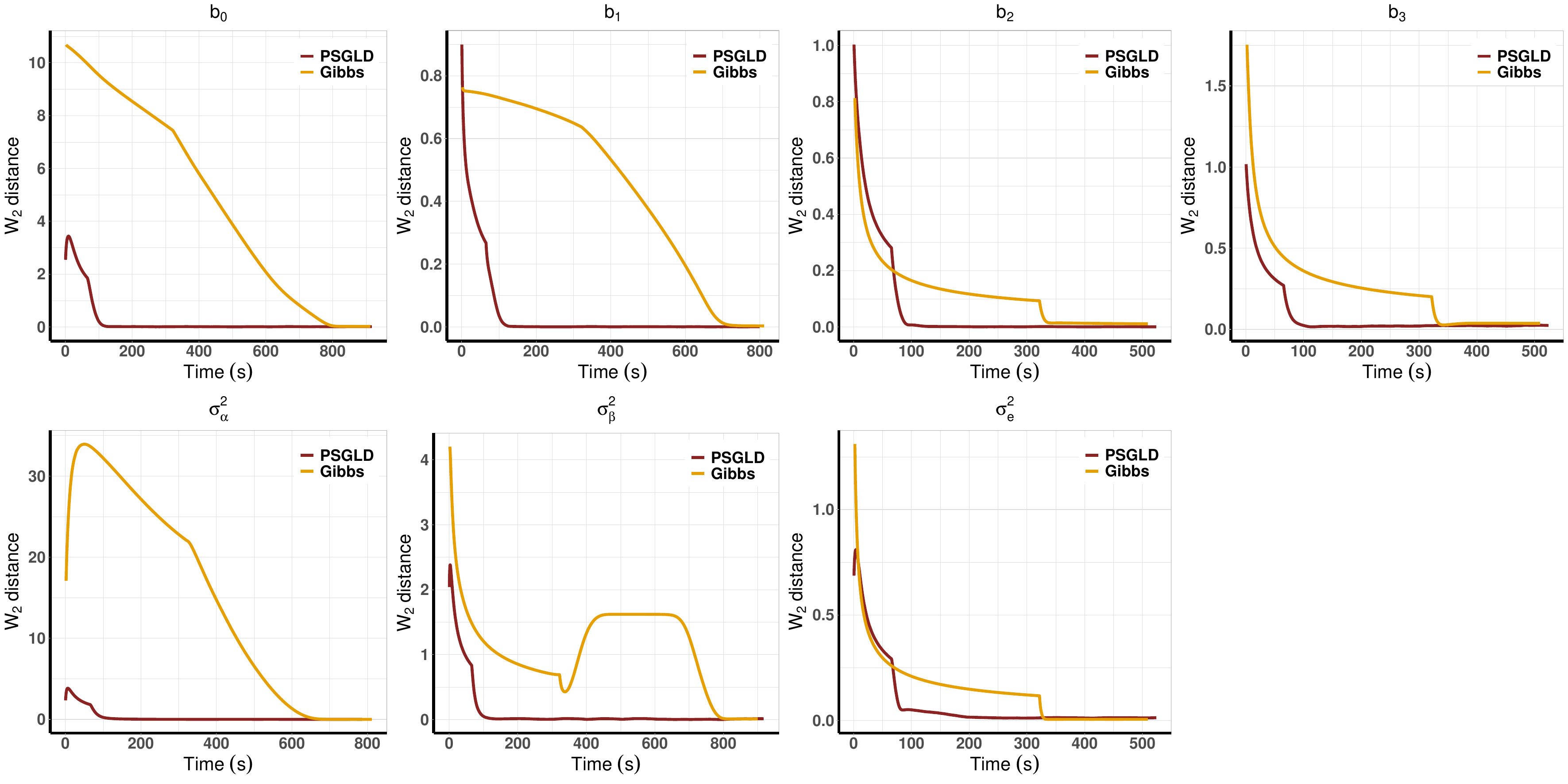}
\end{figure}

Similar to the MovieLens dataset in Section \ref{realdata:1}, we also plot the $W_2$ distance versus the elapsed CPU time for each parameter to compare the computational efficiency of the PSGLD and the Gibbs sampler in Figure \ref{W2_distance_InstEval}. Again, we take the first $500$ samples after $10^4$ burn-in iterations from the Gibbs sampler as the stationary distribution regarded as the true posterior distribution, and iteratively compute the $W_2$ distance between the samples from each algorithm and this benchmark. For all the $7$ parameters, the $W_2$ distances between the samples of the PSGLD and the benchmark have dropped quickly to a low and stable level very close to $0$ within around $100$ seconds. This is significantly faster than the full-data Gibbs sampler, which takes at least $800$ seconds to arrive at the same level of convergence. These plots again demonstrate the excellent computational efficiency of the PSGLD in approaching the target posterior distribution.

\section{Discussion} \label{sec:discussion}
Crossed mixed effects models are useful for analyzing massive datasets with missing data from e-commerce and large surveys, yet standard Bayesian posterior sampling algorithms often require prohibitively long computational time to reach convergence.
We have derived the stochastic gradient Langevin dynamics algorithms for large crossed mixed effects models. For the balanced design without missing observations, we leverage the closed-form formula for the inverse covariance matrix of subset data to efficiently estimate the gradient of log-posterior densities. For the unbalanced design with missing observations, we propose the pigeonhole SGLD algorithm that generates a short Markov chain of row and column effects and computes the gradients using Monte Carlo averages. We have shown the convergence of the output distribution from the pigeonhole SGLD to the true posterior distributions in total variation distance. The results of our numerical experiments demonstrate that the proposed SGLD algorithms can approximate the target posterior distribution accurately under various metrics, and meanwhile have much better computational efficiency than the standard Gibbs sampler based on the full data.

There are some important aspects of the proposed SGLD algorithms that require further investigation. First, while our convergence theory is derived under general model assumptions, it would be of interest to investigate the exact theoretical convergence rates when the crossed mixed effects model is correctly specified, which requires a new Bayesian posterior contraction theory for dependent data in the crossed mixed effects model. Second, it would be important to understand how different missing patterns of observations in the data matrix $\Yb$ will affect the convergence and computational efficiency of the pigeonhole SGLD, which may further provide guidance on the choice of step sizes. Third, besides the SGLD, it is worth exploring other more efficient SGMCMC algorithms for the crossed mixed effects model, such as the various versions of stochastic variance reduced gradient and stochastic gradient Hamiltonian Monte Carlo (\citealt{ma2015complete, Xuetal2018, Zouetal2019}). Finally, given that most e-commerce datasets consist of categorical ratings, it will be of interest to replace the continuous responses in the model \eqref{eq:lme} with a generalized linear model with either a logistic or probit link for categorical responses, and study the similar subset-based Bayesian algorithms for such models. We hope to explore these directions in future research.

\vspace{1cm}

\noindent {\large \bf Acknowledgement}
\vspace{3mm}

\noindent The authors would like to thank the anonymous referees, an Associate Editor and the Editor for their constructive comments that improved the quality of this paper. The research is supported by the Singapore Ministry of Education Academic Research Funds Tier 1 Grants A-0004822-00-00 and A-8002495-00-00.

\newpage

\noindent {\bf \LARGE Supplementary Material}
\vspace{5mm}

\setcounter{section}{0}
\setcounter{equation}{0}
\setcounter{lemma}{0}
\setcounter{proposition}{0}
\renewcommand{\theequation}{S.\arabic{equation}}
\renewcommand{\theproposition}{S.\arabic{proposition}}
\renewcommand{\thelemma}{S.\arabic{lemma}}
\renewcommand\thesection{S\arabic{section}}

\makeatletter
\renewcommand{\thefigure}{S\@arabic\c@figure}
\makeatother

We provide exact formulas for the gradients in Algorithms \ref{algo:SGLD} and \ref{algo:pigeonhole SGLD}, the numerical results of simulation studies, and the technical proof of Theorem \ref{th1}.

\section{Formulas of SGLD for Balanced Crossed Mixed Effects Models} \label{sec:formula_SGLD}

We provide formulas regarding the SGLD for balanced crossed mixed effects models in Section \ref{subsec:balanced} of the main paper. At each iteration, we randomly select $r$ rows and $c$ columns from the full data matrix $\Yb$ without replacement and formulate the subset of data as a vector $Y_n \in \RR^n ~ (n = r \times c)$: 
\begin{align*}
& Y_n =\Xb_n b + \Zb_{\alpha n} \alphab_n + \Zb_{\beta n} \betab_n + e_n,
\end{align*}
where $\Xb_n \in \RR^{n \times p}$ is the matrix of fixed effects stacked in the order of rows; $\Zb_{\alpha n}  = \Ib_r \otimes \bm{1}_c \in \{0,1\}^{n \times r}$, $\Zb_{\beta n} = \bm 1_r \otimes \Ib_c \in \{0,1\}^{n \times c}$, $\otimes$ denotes the Kronecker product; $\alphab_n \in \RR^r$ and $\betab_n \in \RR^c$ are the selected vectors of row random effects and column random effects, and $e_n\in \RR^n$ is the vector of random errors. The ${n \times n}$ covariance matrix of $Y_n$ can be written as $\bm{\Sigma}_n = \Zb_{\alpha n} \Zb_{\alpha n }^{\top}\sigma^2_\alpha + \Zb_{\beta n} \Zb_{\beta n}^{\top}\sigma^2_\beta + \Ib_n\sigma^2_e$, whose explicit form is
\begin{align}\label{eq:sig.inv0.dup}
& \bm{\Sigma}_n={\left[ \begin{array}{cccc}
\Sigma_1 & \Sigma_2 & \ldots & \Sigma_2\\
\Sigma_2 & \Sigma_1 & \ldots & \Sigma_2\\
\vdots & \vdots & \ddots &\vdots \\
\Sigma_2 & \Sigma_2 & \ldots & \Sigma_1
\end{array}
\right ]}_{n \times n}, ~ \text{where}\\
& \Sigma_1={\left[ \begin{array}{cccc}
\sigma^2_\alpha+\sigma^2_\beta+\sigma^2_e & \sigma^2_\alpha & \ldots & \sigma^2_\alpha\\
\sigma_\alpha^2 & \sigma^2_\alpha+\sigma^2_\beta+\sigma^2_e & \ldots & \sigma^2_\alpha\\
\vdots & \vdots & \ddots &\vdots \\
\sigma_\alpha^2 & \sigma^2_\alpha & \ldots & \sigma^2_\alpha+\sigma^2_\beta+\sigma^2_e
\end{array}
\right ]}_{c \times c}, \text{ and } \Sigma_2 = \sigma^2_\beta \Ib_c . \nonumber
\end{align}
With the block matrix structure in \eqref{eq:sig.inv0.dup}, the inverse covariance matrix $\bm{\Sigma}_n^{-1}$ can be explicitly derived:
\begin{align} \label{eq:sig.inv.dup}
& \bm{\Sigma}_n^{-1}={\left[ \begin{array}{cccc}
\Sigma_3 & \Sigma_4 & \ldots & \Sigma_4\\
\Sigma_4 & \Sigma_3 & \ldots & \Sigma_4\\
\vdots & \vdots & \ddots &\vdots \\
\Sigma_4 & \Sigma_4 & \ldots & \Sigma_3
\end{array}
\right ]}_{n \times n},~\text{where } \\
 & \Sigma_3={\left[ \begin{array}{cccc}
\sfx & \sfy & \ldots & \sfy\\
\sfy & \sfx & \ldots & \sfy\\
\vdots & \vdots & \ddots &\vdots \\
\sfy & \sfy & \ldots & \sfx
\end{array}
\right ]}_{c \times c}, \quad  \Sigma_4={\left[ \begin{array}{cccc}
\sfw & \sfz & \ldots & \sfz\\
\sfz & \sfw & \ldots & \sfz\\
\vdots & \vdots & \ddots &\vdots \\
\sfz & \sfz & \ldots & \sfw
\end{array}
\right ]}_{c \times c}, ~ \text{and } \nonumber \\
& \sfz = \frac{\sigma^2_\alpha \sigma^2_\beta (2 \sigma^2_e + c\sigma^2_\alpha +r \sigma^2_\beta)}{\sigma^2_e(\sigma^2_e + r \sigma^2_\beta)(\sigma^2_e + c\sigma^2_\alpha)(\sigma^2_e + c \sigma^2_\alpha +r\sigma^2_\beta)},
\quad \sfy = \sfz-\frac{\sigma^2_\alpha}{\sigma^2_e(\sigma^2_e+c \sigma^2_\alpha)}, \nonumber \\
& \sfx = \sfy+\frac{\sigma^2_e +(r-1)\sigma^2_\beta}{\sigma^2_e(\sigma^2_e+r\sigma^2_\beta)},
\quad \sfw = \sfz-\frac{\sigma^2_\beta}{\sigma^2_e(\sigma^2_e + r \sigma^2_\beta)}. \nonumber
\end{align}
With the explicit formula of $\bm{\Sigma}^{-1}_n$ in \eqref{eq:sig.inv.dup}, we can compute the log-likelihood function of the selected data subset and its gradient. 

The exact formulas for the gradients in \eqref{eq:SGLD} of Algorithm \ref{algo:SGLD} are as follows:
\begin{align} \label{eq:SGLD.8grads}
& \nabla_b \log p \left(Y_n^{(t)} \mid  b^{(t)}, \eta^{(t)}_\alpha, \eta^{(t)}_\beta, \eta^{(t)}_e \right) = \Xb_n^{(t)\top} \bm{\Sigma}_n^{-1}\big[Y_n^{(t)} - \Xb_n^{(t)}b^{(t)}\big];  \nonumber \\
& \nabla_{\eta_\alpha} \log p\left(Y_n^{(t)} \mid  b^{(t)}, \eta^{(t)}_\alpha, \eta^{(t)}_\beta, \eta^{(t)}_e \right) = -\frac{1}{2} \big[\sfx^{(t)} + (c-1)\sfy^{(t)}\big]n\exp\big(\eta_a^{(t)} \big)\nonumber \\
&\qquad + (1/2){\sum}^r_{i=1}{\sum}^c_{j=1}{\sum}^r_{g=1}{\sum}^c_{h=1}\big[Y^{(t)}_{s_iq_j}-x_{s_iq_j}^{(t)\top}b^{(t)}\big]
\big[Y^{(t)}_{s_gq_h}-x_{s_gq_h}^{(t)\top}b^{(t)}\big] \exp\big(\eta_a^{(t)}\big)  \nonumber \\
&\qquad \times \left(v_1\mathbbm{1}(i=g)+v_2\mathbbm{1}(i \ne g)\right),  \nonumber \\
&\qquad \text{ with }
 v_1 = \big[\sfx^{(t)}+(c-1)\sfy^{(t)}\big]^2 +(r-1)\big[\sfw^{(t)}+(c-1)\sfz^{(t)}\big]^2, \nonumber \\
& \qquad \quad v_2 = 2\big[\sfx^{(t)} +(c-1)\sfy^{(t)}\big]\big[\sfw^{(t)} + (c-1)\sfz^{(t)}\big] + (r-2)\big[\sfw^{(t)}+(c-1)\sfz^{(t)}\big]^2; \nonumber \\
& \nabla_{\eta_\alpha} \log  \pi\big(\eta_\alpha^{(t)}\big) = -\afrak_1+ \bfrak_1 \exp\big(-\eta_\alpha^{(t)}\big) ;\nonumber \\
&\nabla_{\eta_\beta} \log p\left(Y_n^{(t)} \mid  b^{(t)}, \eta^{(t)}_\alpha, \eta^{(t)}_\beta, \eta^{(t)}_e \right) =-\frac{1}{2} \big[\sfx^{(t)} + (r-1)\sfw^{(t)}\big] n\exp\big(\eta_\beta^{(t)}\big) + \nonumber \\
&\qquad  (1/2){\sum}^r_{i=1}{\sum}^c_{j=1}{\sum}^r_{g=1}{\sum}^c_{h=1} \big[Y^{(t)}_{s_iq_j}-x^{(t)\top}_{s_iq_j}b^{(t)}\big]\big[Y^{(t)}_{s_gq_h}- x^{(t)\top}_{s_gq_h}b^{(t)}\big] \exp\big(\eta_\beta^{(t)}\big) \nonumber \\
&\qquad \times \big[v_3\mathbbm{1}(j=h)+v_4\mathbbm{1}(j \ne h)\big], \nonumber \\
&\qquad \text{ with } v_3 = \big[\sfx^{(t)} +(r-1)\sfw^{(t)}\big]^2 + (c-1)\big[\sfy^{(t)}+(r-1)\sfz^{(t)}\big]^2, \nonumber \\
&\qquad \quad v_4 = 2\big[\sfx^{(t)}+(r-1)\sfw^{(t)}\big]\big[\sfy^{(t)}+(r-1)\sfz^{(t)}\big] +(c-2)\big[\sfy^{(t)}+(r-1)\sfz^{(t)}\big]^2; \nonumber \\
& \nabla_{\eta_\beta} \log  \pi\big(\eta_\beta^{(t)}\big) = -\afrak_2+ \bfrak_2 \exp\big(-\eta_\beta^{(t)}\big) ; \nonumber \\
&\nabla_{\eta_e} \log p\left(Y_n^{(t)} \mid b^{(t)}, \eta^{(t)}_\alpha, \eta^{(t)}_\beta, \eta^{(t)}_e \right) = -\frac{1}{2} n \sfx^{(t)}\exp\big(\eta_e^{(t)}\big)  \nonumber \\
&\qquad + (1/2){\sum}^r_{i=1}{\sum}^c_{j=1}{\sum}^r_{g=1}{\sum}^c_{h=1}\big[Y^{(t)}_{s_iq_j}-x_{s_iq_j}^{(t)\top}b^{(t)}\big]
\big[Y^{(t)}_{s_gq_h}-x_{s_gq_h}^{(t)\top}b^{(t)}\big]\exp\big(\eta_e^{(t)}\big) \nonumber \\
&\qquad \times \big[v_5\mathbbm{1}(i=g,j=h)+v_6\mathbbm{1}(i=g,j \ne h)+ v_7\mathbbm{1}(i \ne g, j=h)+v_8\mathbbm{1}(i \ne g, j \ne h)\big],  \nonumber \\
&\qquad \text{ with } v_5 = {\sfx^{(t)}}^2 +(c-1){\sfy^{(t)}}^2+(r-1)\big[{\sfw^{(t)}}^2+(c-1){\sfz^{(t)}}^2\big], \nonumber \\
&\qquad \quad  v_6 = 2\sfx^{(t)}\sfy^{(t)}+(c-2){\sfy^{(t)}}^2 +(r-1)\big[2\sfw^{(t)}\sfz^{(t)}+(c-2){\sfz^{(t)}}^2\big], \nonumber \\
&\qquad \quad  v_7 = 2\sfx^{(t)}\sfw^{(t)}+2(c-1)\sfy^{(t)}\sfz^{(t)}+(r-2)\big[{\sfw^{(t)}}^2+(c-1){\sfz^{(t)}}^2\big], \nonumber \\
&\qquad \quad  v_8 = 2\sfx^{(t)}\sfz^{(t)}+2\sfy^{(t)}\sfw^{(t)}+2(c-2)\sfy^{(t)}\sfz^{(t)}+(r-2)\big[2\sfw^{(t)}\sfz^{(t)}+ (c-2){\sfz^{(t)}}^2\big]; \nonumber \\
& \nabla_{\eta_e} \log  \pi\big(\eta_e^{(t)}\big) = -\afrak_3+ \bfrak_3 \exp\big(-\eta_e^{(t)}\big),
\end{align}
where $\sfx^{(t)}, \sfy^{(t)},\sfz^{(t)},\sfw^{(t)}$ are defined the same as $\sfx, \sfy, \sfz, \sfw$ in \eqref{eq:sig.inv} by replacing the variance parameters $\sigma_{\alpha}^2=\exp(\eta_{\alpha}), \sigma_{\beta}^2=\exp(\eta_{\beta}), \sigma_e^2=\exp(\eta_e)$ with their values at the $t$th iteration.

\section{Formulas of Pigeonhole SGLD for Crossed Mixed Effects Models with Missing Data}
\label{sec:formula_PSGLD}

The pigeonhole SGLD algorithm treats the row and column effects of the selected subset of data at each iteration as the latent variables. 
In the $t$th iteration of Algorithm \ref{algo:pigeonhole SGLD}, after randomly selecting a subset of data $\Yb_n^{(t)}$, we use the Gibbs sampler to generate a length-$m$ Markov chain of latent variables $\{\vartheta^{(t)}_k\}^m_{k=1}=\big\{\alphab_{n,k}^{(t)}, \betab_{n,k}^{(t)}\big\}^m_{k=1}=\big \{\alpha_{s_1,k}^{(t)}, \cdots, \alpha_{s_r,k}^{(t)}, \beta_{q_1,k}^{(t)}, \cdots, \beta_{q_c,k}^{(t)} \big\}^m_{k=1}$ by iteratively sampling from the conditional posterior distributions. The conditional posterior distributions of $\alphab_n^{(t)}$ and $\betab_n^{(t)}$ in Algorithm \ref{algo:pigeonhole SGLD} are as follows:
\begin{align} \label{eq:conditional_d}
& \alpha_{s_i} \mid \theta^{(t)},\betab^{(t)}_n, \Yb_n^{(t)} \sim N \left (\frac{\sum_{j=1}^{c}Z^{(t)}_{s_iq_j} (Y^{(t)}_{s_iq_j} - x_{s_iq_j}^{(t) \top}b^{(t)} -  \beta_{q_j}^{(t)} ) \ee^{\eta_\alpha^{(t)}}}{n^{(t)}_{i \bullet} \ee^{\eta_\alpha^{(t)}} + \ee^{\eta_e^{(t)}} }, ~~\frac{ \ee^{\eta_\alpha^{(t)} + \eta_e^{(t)}} }{n^{(t)}_{i \bullet} \ee^{\eta_\alpha^{(t)}} + \ee^{\eta_e^{(t)}} } \right), \nonumber \\
& \beta_{q_j} \mid \theta^{(t)},\alphab_n^{(t)}, \Yb_n^{(t)} \sim N \left (\frac{\sum_{i=1}^{r}Z^{(t)}_{s_iq_j} (Y^{(t)}_{s_iq_j} - x_{s_iq_j}^{(t)\top}b^{(t)} -   \alpha_{s_i}^{(t)}) \ee^{\eta_\beta^{(t)}}}{n^{(t)}_{ \bullet j} \ee^{\eta_\beta^{(t)}} + \ee^{\eta_e^{(t)}} }, ~~\frac{ \ee^{\eta_\beta^{(t)}+\eta_e^{(t)}} }{n^{(t)}_{ \bullet j} \ee^{\eta_\beta^{(t)}} + \ee^{\eta_e^{(t)}} } \right),
\end{align}
where $i = 1, \ldots, r $ for $s_i$, $j = 1, \ldots, c $ for $q_j$, and $\theta^{(t)} = (b^{(t)\top}, \eta_\alpha^{(t)},\eta_\beta^{(t)}, \eta_e^{(t)})^{\top}$.
Then we update the model parameter $\theta$ by averaging over the gradients of log-posterior distributions of $\theta$ conditional on the latent variables $(\alphab_n, \betab_n)$ and the subset of data $\Yb_n$. The exact formulas of the gradients in equation \eqref{eq:PSGLD} of Algorithm \ref{algo:pigeonhole SGLD} are as follows: 
\begin{align} \label{eq:PSGLD.grad}
& \nabla_b \log p\big(\Yb_n^{(t)} \mid b^{(t)}, \alphab_{n,k}^{(t)}, \betab_{n,k}^{(t)},  \eta^{(t)}_e \big) \nonumber \\
& \quad =  \sum^r_{i=1}\sum^c_{j=1} Z_{s_iq_j}^{(t)}\left(Y^{(t)}_{s_iq_j}-x_{s_iq_j}^{(t)\top}b^{(t)} - \alpha^{(t)}_{s_i,k} - \beta_{q_j,k}^{(t)}\right)x_{s_iq_j}^{(t)} \ee^{-\eta_e(t)} ;  \nonumber \\
& \nabla_{\eta_\alpha} \log \pi\big(\alpha_{s_1,k}^{(t)}, \ldots, \alpha_{s_r,k}^{(t)} \mid \eta_\alpha^{(t)}\big) =   \sum^{r}_{i=1} \Big[-1+ \big\{\alpha_{s_i,k}^{(t)}\big\}^2 \ee^{-\eta_\alpha(t)} \Big] /2 ; \nonumber \\
& \nabla_{\eta_\beta} \log \pi \big(\beta_{q_1,k}^{(t)}, \ldots, \beta_{q_c,k}^{(t)} \mid \eta_\beta^{(t)}\big) = \sum^c_{j=1} \Big[ -1 + \big\{\beta_{q_j,k}^{(t)}\big\}^2\ee^{-\eta_\beta(t)} \Big] /2 ; \nonumber \\
& \nabla_{\eta_e} \log p\big(\Yb_n^{(t)} \mid b^{(t)}, \alphab_{n,k}^{(t)}, \betab_{n,k}^{(t)},  \eta^{(t)}_e \big)  \nonumber \\
& \quad =   \sum^r_{i=1} \sum^c_{j=1} Z^{(t)}_{s_iq_j} \Big[ -1 + \big(Y^{(t)}_{s_iq_j} -x_{s_iq_j}^{(t) \top}b^{(t)}- \alpha_{s_i,k}^{(t)} - \beta_{q_j,k}^{(t)} \big)^2 \ee^{-\eta_e {(t)}} \Big] /2 ,
\end{align}
and $\nabla_{\eta_\alpha} \log \pi\big(\eta_\alpha^{(t)}\big), \nabla_{\eta_\beta} \log \pi\big(\eta_\beta^{(t)}\big), \nabla_{\eta_e} \log \pi\big(\eta_e^{(t)}\big)$ are defined the same as in \eqref{eq:SGLD.8grads}.

\section{Simulations} \label{sec:simulations}

In this section, we present extensive simulation studies by applying the proposed SGLD Algorithms \ref{algo:SGLD} and \ref{algo:pigeonhole SGLD} for Bayesian inference on the coefficients of fixed effects $b=(b_1, \ldots, b_p)^{\top}$ and the variance components $\sigma^2_{\alpha}, \sigma^2_\beta, \sigma^2_e$ in the crossed mixed effects model. We compare the performance of Algorithms \ref{algo:SGLD} and \ref{algo:pigeonhole SGLD} with the Gibbs sampler and the method of moments estimator proposed by \cite{gao2019estimation}, for both the balanced model without missing data and the unbalanced model with missing data. We further implement Algorithm \ref{algo:pigeonhole SGLD} and the Gibbs sampler on datasets with different missing patterns to evaluate the effectiveness of the proposed algorithm under more challenging conditions.  

\subsection{Simulation for Balanced Design without Missing Data} \label{simulate:b}

We simulate the data following the model in \eqref{eq:lme} and \eqref{eq:normalre} with the numbers of row effects and column effects as $R = C =1000$, resulting in a full data matrix $\Yb$ of $10^6$ observations. We set the true coefficients of fixed effects as $b=(3,2,4,6,5)^{\top}$, and the true variance components as $(\sigma^2_{\alpha}, \sigma^2_\beta, \sigma^2_e) = (9,4,1)$. For the fixed effect $x_{ij} \in \RR^5$, all the elements in $x_{ij}=(x_{ij1}, \cdots, x_{ij5})^{\top}$ are generated independently from $N(0, 0.5)$.

We compare the performance of the pigeonhole SGLD (PSGLD) in Algorithm \ref{algo:pigeonhole SGLD}, the SGLD in Algorithm \ref{algo:SGLD}, the Gibbs sampler, and the method of moments (MoM) in \cite{gao2019estimation} for inference on the coefficients of fixed effects $b = (b_1,\cdots, b_p)^{\top}$ and the variance components $\sigma_\alpha^2, \sigma_\beta^2, \sigma^2_e$. We repeat the whole simulated datasets and estimation procedures for 100 macro replications and report the averaged results. For the pigeonhole SGLD and the SGLD algorithms, we randomly select $r = 20$ rows and $c = 20$ columns from the full data matrix $\Yb$ and obtain the submatrix of data $\Yb_n$ with the mini-batch size $n = 400$ at each iteration. For Step (b) of the pigeonhole SGLD proposed in Algorithm \ref{algo:pigeonhole SGLD}, we generate a  Markov chain of length $m=50$ for the latent variables $\{\alpha_{s_1,k}, \cdots, \alpha_{s_r, k}\}^m_{k=1}$, and $\{\beta_{q_1,k}, \cdots, \beta_{q_c, k}\}^m_{k=1}$ following the conditional posterior distributions in \eqref{eq:conditional_d}.

The step sizes for the parameters are selected by grid search, such that they produce the lowest $W_2$ distances between the samples from the two SGLD algorithms and those from the Gibbs sampler. For the SGLD and the pigeonhole SGLD algorithms, we used fixed step sizes, while adjusted them respectively after $1000$ iterations. In particular, for the SGLD, at the first $1000$ iterations, the step sizes $\epsilon_{b_1},\epsilon_{b_2},\epsilon_{b_3},\epsilon_{b_4},\epsilon_{b_5}$ are $O(10^{-8})$, $\epsilon_{\eta_\alpha} = 1.99 \times 10^{-7}$, $\epsilon_{\eta_\beta} = 1.11 \times 10^{-5}$, and $\epsilon_{\eta_e} = 4.96 \times 10^{-9}$. After $1000$ iterations, the step sizes $\epsilon_{b_1},\epsilon_{b_2},\epsilon_{b_3},\epsilon_{b_4},\epsilon_{b_5}$ are $O(10^{-9})$, $\epsilon_{\eta_\alpha} = 1.42 \times 10^{-7}$, $\epsilon_{\eta_\beta} = 9.09 \times 10^{-6}$, and $\epsilon_{\eta_e} = 3.97 \times 10^{-9}$. For the pigeonhole SGLD, at the first $1000$ iteration, the step sizes $\epsilon_{b_1},\epsilon_{b_2},\epsilon_{b_3} ,\epsilon_{b_4},\epsilon_{b_5}$ are $O(10^{-8})$, $\epsilon_{\eta_\alpha} = 9.97 \times 10^{-5}$, $\epsilon_{\eta_\beta} =3.02 \times 10^{-4}$, and $\epsilon_{\eta_e} = 2.48 \times 10^{-7}$. After $1000$ iterations, the step sizes $\epsilon_{b_1},\epsilon_{b_2},\epsilon_{b_3} ,\epsilon_{b_4},\epsilon_{b_5}$ are $O(10^{-9})$, $\epsilon_{\eta_\alpha} = 4.75 \times 10^{-6}$, $\epsilon_{\eta_\beta} =3.02 \times 10^{-4}$, and $\epsilon_{\eta_e} = 2.92 \times 10^{-9}$.

We report the posterior means with posterior standard deviations in parentheses of all Bayesian methods, together with the estimated parameters from MoM in Table \ref{tbl:postmeansd_b}. It is clear that the proposed two SGLD algorithms have accurate posterior mean and standard deviation estimates for all the parameters. All Bayesian methods give posterior means very close to the MoM estimates for the fixed effects coefficients $b$ and the variance of random error $\sigma^2_e$. However, for the variances of random effects $\sigma^2_\alpha$ and $\sigma^2_\beta$, the two SGLD algorithms output posterior means similar to the MoM estimates but slightly lower than those from the full-data Gibbs sampler. The SGLD and the PSGLD both generate posterior chains with similar standard deviations to the Gibbs sampler.

\begin{figure}[H]
\caption{Boxplots of posterior samples for the coefficients of fixed effects $b = (b_1, b_2, b_3, b_4, b_5)^{\top}$ and the variance components $\sigma^2_\alpha, \sigma^2_\beta, \sigma^2_e$ in the crossed mixed effects model with balanced design without missing data. All results are averaged over $100$ macro replications. PSGLD, pigeonhole stochastic gradient Langevin dynamics; SGLD, stochastic gradient Langevin dynamics; Gibbs, full-data Gibbs sampler; MoM, method of moments of \cite{gao2019estimation}.}
\label{boxplot_b}
\centering
\includegraphics[width=\textwidth]{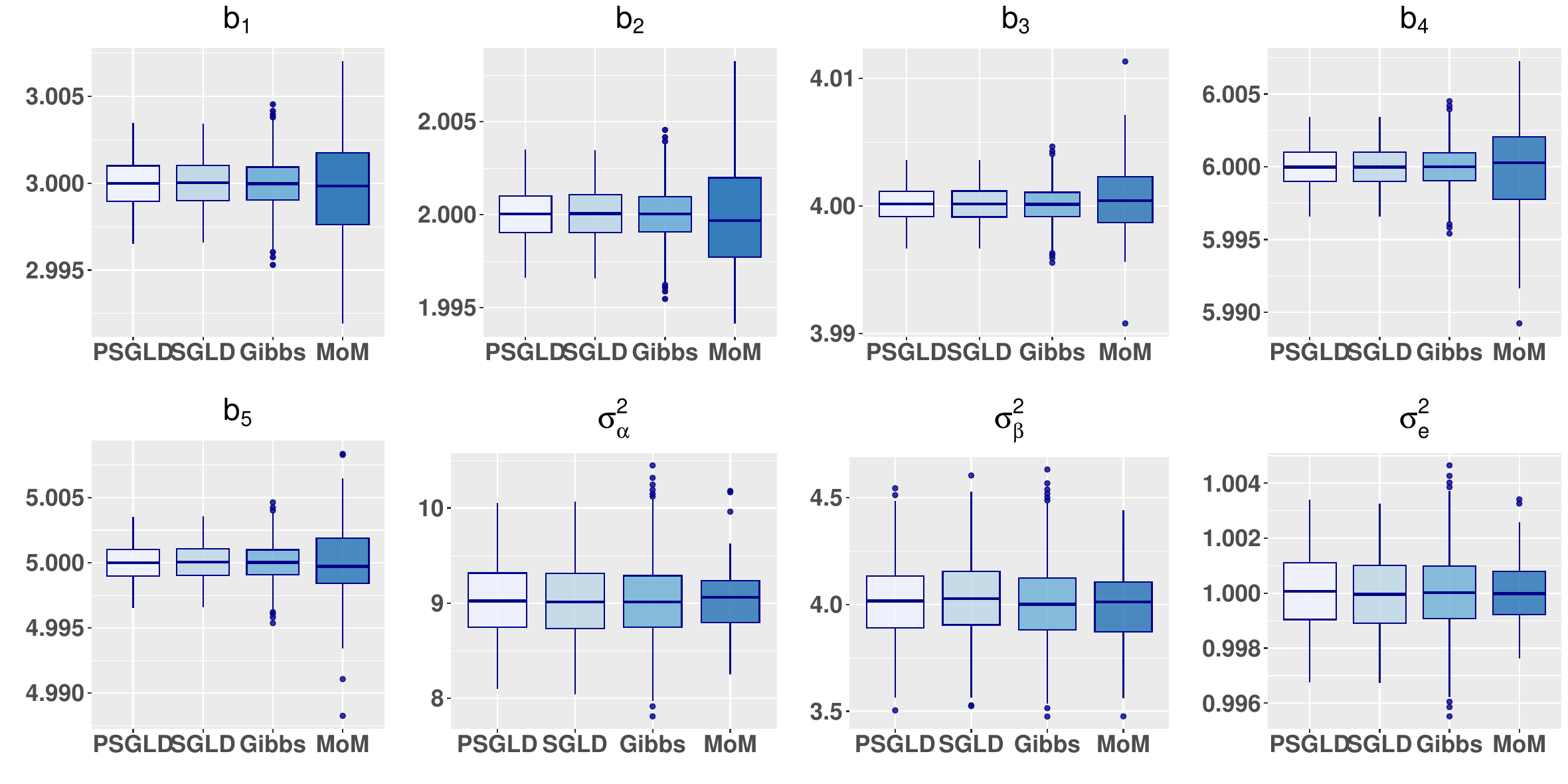} 
\end{figure}

Figure \ref{boxplot_b} shows the posterior distributions from the PSGLD, the SGLD, and the full-data Gibbs sampler based on the Wasserstein-2 barycenters of $100$ posterior distributions from $100$ simulated datasets (\citealt{li2017simple}). Regarding whether the Wasserstein-2 barycenter provides a representative summary of the individual SGLD chains, we provide some empirical results to quantify the uncertainty of individual chains in Section \ref{sec:UQ.PSGLD}. We also show the boxplots of the MoM estimators based on their empirical distributions from the same $100$ datasets. We can see that overall the two SGLD algorithms produce similar boxplots to the full-data Gibbs sampler for all the parameters. The variation of MoM estimators for the coefficients $b$ is significantly larger than that of all Bayesian posterior distributions, though MoM is a frequentist method and not directly comparable to the other Bayesian methods.

In Table \ref{tbl:w2_b}, we compute the $W_2$ distances from the marginal posterior distributions of the SGLD and the PSGLD to those of the full-data Gibbs sampler to quantify the approximation error of our proposed algorithms to the true target posteriors. The outputs from the full-data Gibbs sampler are used as the benchmarks here. For the fixed effects coefficients $b$ and the variance of random error $\sigma^2_e$, the marginal distributions from the SGLD and the PSGLD exhibit very small approximation errors in terms of the $W_2$ distance, while for the variances of random effects $\sigma^2_\alpha$ and $\sigma^2_\beta$, the SGLD and the PSGLD show higher but still small $W_2$ distances to the full data posterior. Meanwhile, the approximation errors from both the SGLD and the PSGLD are stable, as indicated by the low standard errors of the $W_2$ distances.

\begin{table}[H]
\caption{Posterior means and posterior standard deviations (in parentheses) for the coefficients of fixed effects $b = (b_1, b_2, b_3, b_4, b_5)^{\top}$ and the variance components $\sigma^2_\alpha, \sigma^2_\beta, \sigma^2_e$ in the crossed mixed effects model with balanced design without missing data. 
All results are averaged over $100$ macro replications. MoM, method of moments of \cite{gao2019estimation}; PSGLD, pigeonhole stochastic gradient Langevin dynamics; SGLD, stochastic gradient Langevin dynamics; Gibbs, full-data Gibbs sampler.
}
\label{tbl:postmeansd_b}
\centering
{\scriptsize
\begin{tabular}{rcccccccc}
\hline
&~ & $b_1$ & $b_2$ & $b_3$ & $b_4$ & $b_5$\\
\hline
& MoM & \multicolumn{1}{c}{2.9999 } & \multicolumn{1}{c} {2.0000 } & \multicolumn{1}{c} {4.0005} & \multicolumn{1}{c} {5.9999}  & \multicolumn{1}{c} {4.9998} \\
& PSGLD & \multicolumn{1}{c}{3.0000 (0.0014)} & \multicolumn{1}{c}{2.0000 (0.0014)} & \multicolumn{1}{c}{4.0002 (0.0014)} & \multicolumn{1}{c}{6.0000 (0.0014)} & \multicolumn{1}{c}{5.0000 (0.0014)}    \\
&  SGLD  & \multicolumn{1}{c}{3.0000 (0.0014)} & \multicolumn{1}{c}{2.0001 (0.0014)} & \multicolumn{1}{c}{4.0002 (0.0014)} & \multicolumn{1}{c}{6.0000 (0.0014)} & \multicolumn{1}{c}{5.0000 (0.0014)}\\
&  Gibbs & \multicolumn{1}{c}{3.0000 (0.0014)} & \multicolumn{1}{c}{2.0000 (0.0014)} & \multicolumn{1}{c}{4.0001 (0.0014)} & \multicolumn{1}{c}{6.0000 (0.0014)} & \multicolumn{1}{c}{5.0000 (0.0014)}\\
\hline
& ~  & \multicolumn{1}{c} {$\sigma^2_\alpha$} & \multicolumn{1}{c} {$\sigma^2_\beta$} & \multicolumn{1}{c} {$\sigma^2_e$}\\
\hline
& MoM & \multicolumn{1}{c}{ 9.0211} & \multicolumn{1}{c} {4.0012 } & \multicolumn{1}{c} {1.0000}   \\
& PSGLD & \multicolumn{1}{c}{9.0280 (0.4510)} & \multicolumn{1}{c}{4.0179 (0.2598)} & \multicolumn{1}{c}{1.0001 (0.0013)}    \\
& SGLD  & \multicolumn{1}{c}{9.0227 (0.4476)} & \multicolumn{1}{c}{4.026 (0.2594)} & \multicolumn{1}{c}{1.0000 (0.0014)} \\
& Gibbs & \multicolumn{1}{c}{9.0983 (0.4478)} & \multicolumn{1}{c}{4.0793 (0.2585)} & \multicolumn{1}{c}{1.0000 (0.0014)} \\
\hline
\end{tabular}
}
\end{table}

\begin{table}[H]
\caption{$W_2$ distances from the marginal distributions of the SGLD and the PSGLD to those of the full-data Gibbs sampler in the crossed mixed effects model with balanced design without missing data.  
The $W_2$ distances are averaged over $100$ macro replications. The standard errors are in parentheses. PSGLD, pigeonhole stochastic gradient Langevin dynamics; SGLD, stochastic gradient Langevin dynamics.
}
\label{tbl:w2_b}
\centering
{\scriptsize
\begin{tabular}{rcccccccc}
\hline
& ~ & $b_1$ & $b_2$ & $b_3$ & $b_4$  \\
\hline
& PSGLD & \multicolumn{1}{c}{0.00034 (0.00001)} & \multicolumn{1}{c}{0.00035 (0.00001)} & \multicolumn{1}{c}{0.00035 (0.00002)} & \multicolumn{1}{c}{0.00038 (0.00002)}    \\
& SGLD & \multicolumn{1}{c}{0.00035 (0.00002)} & \multicolumn{1}{c}{0.00038 (0.00002)} & \multicolumn{1}{c}{0.00036 (0.00002)} & \multicolumn{1}{c}{0.00036 (0.00001)}  \\
\hline
&~  & $b_5$ & \multicolumn{1}{c} {$\sigma^2_\alpha$} & \multicolumn{1}{c} {$\sigma^2_\beta$} & \multicolumn{1}{c} {$\sigma^2_e$}\\
\hline
& PSGLD  & \multicolumn{1}{c}{0.00037 (0.00002)} & \multicolumn{1}{c}{0.09817 (0.00398)} & \multicolumn{1}{c}{0.08213 (0.00119)} & \multicolumn{1}{c}{0.00037 (0.00002)}   \\
& SGLD & \multicolumn{1}{c}{0.00036 (0.00002)}  & \multicolumn{1}{c}{0.07295 (0.00259)} & \multicolumn{1}{c}{0.08512 (0.00186)} & \multicolumn{1}{c}{0.00035 (0.00002)} \\
\hline
\end{tabular}
}%
\end{table}

\subsection{Simulation for Unbalanced Design with Missing Data} \label{simulate:ub}

For the crossed mixed effects model with unbalanced design and missing data, we use exactly the same model setup as in Section \ref{simulate:b} and then censor part of the data. In particular, we consider two cases, with $50\%$ and $90\%$ of the data in the $1000\times 1000$ full data matrix $\Yb$ missing completely at random (MCAR). In this case, only the pigeonhole SGLD proposed in Algorithm \ref{algo:pigeonhole SGLD} is applicable, and we compare its performance with the full-data Gibbs sampler and the method of moments in \cite{gao2019estimation} for inference on the coefficients of fixed effects $b = (b_1,\cdots, b_p)^{\top}$ and the variance components $\sigma_\alpha^2, \sigma_\beta^2, \sigma^2_e$.

Similar to the experiments with no missing data in Section \ref{simulate:b}, at each iteration of the pigeonhole SGLD, we randomly select $r = 20$ rows and $c = 20$ columns from the matrix of full data $\Yb$ and construct the submatrix of data $\Yb_n$ with the number of observations $n = \sum^r_{i=1}\sum^c_{j=1}(Z_n)_{s_iq_j}$. For Step (b) of Algorithm \ref{algo:pigeonhole SGLD}, we generate a Markov chain of length $m=50$ for the latent variables $\{\alpha_{s_1,k}, \cdots, \alpha_{s_r, k}\}^m_{k=1}$, and $\{\beta_{q_1,k}, \cdots, \beta_{q_c, k}\}^m_{k=1}$. The step sizes are selected by a grid search similar to Section \ref{simulate:b}. For the datasets with $50\%$ observations, at the first $1000$ iterations, the step sizes
$\epsilon_{b_1}, \epsilon_{b_2}, \epsilon_{b_3}, \epsilon_{b_4}, \epsilon_{b_5}$ are $O(10^{-7})$, and $\epsilon_{\eta_\alpha} = 5.46 \times 10^{-6}, \epsilon_{\eta_\beta} = 1.91 \times 10^{-4} , \epsilon_{\eta_e} = 9.72 \times 10^{-9}$; after $1000$ iterations, the step sizes
$\epsilon_{b_1}, \epsilon_{b_2},\epsilon_{b_3},\epsilon_{b_4},\epsilon_{b_5}$ are $O(10^{-9})$ , and $\epsilon_{\eta_\alpha} = 4.04 \times 10^{-6}, \epsilon_{\eta_\beta} = 1.91 \times 10^{-4} , \epsilon_{\eta_e} = 7.85 \times 10^{-9}$. For the datasets with $90\%$ observations, at the first $1000$ iterations, the step sizes
$\epsilon_{b_1}, \epsilon_{b_2},\epsilon_{b_3},\epsilon_{b_4},\epsilon_{b_5}$ are $O(10^{-7})$ , and $\epsilon_{\eta_\alpha} = 5.46 \times 10^{-6}, \epsilon_{\eta_\beta} = 3.24 \times 10^{-5} , \epsilon_{\eta_e} = 4.08 \times 10^{-8}$; after $1000$ iterations, the step sizes
$\epsilon_{b_1}, \epsilon_{b_2},\epsilon_{b_3},\epsilon_{b_4},\epsilon_{b_5}$ are $O(10^{-8})$ , and $\epsilon_{\eta_\alpha} = 4.89 \times 10^{-6}, \epsilon_{\eta_\beta} = 1.16 \times 10^{-5} , \epsilon_{\eta_e} = 2.91 \times 10^{-8}$.

\begin{figure}[H]
\caption{Boxplots of posterior samples for the coefficients of fixed effects $b = (b_1, b_2, b_3, b_4, b_5)^{\top}$ and the variance components $\sigma^2_\alpha, \sigma^2_\beta, \sigma^2_e$ for the crossed mixed effects model with $50\%$ missing data in $100$ simulation replications. The results are averaged over $100$ simulation replications. PSGLD, pigeonhole stochastic gradient Langevin dynamics; Gibbs, Gibbs sampler; MoM, method of moments of \cite{gao2019estimation}.}
\label{boxplot_ub50}
\centering
\includegraphics[width=\textwidth]{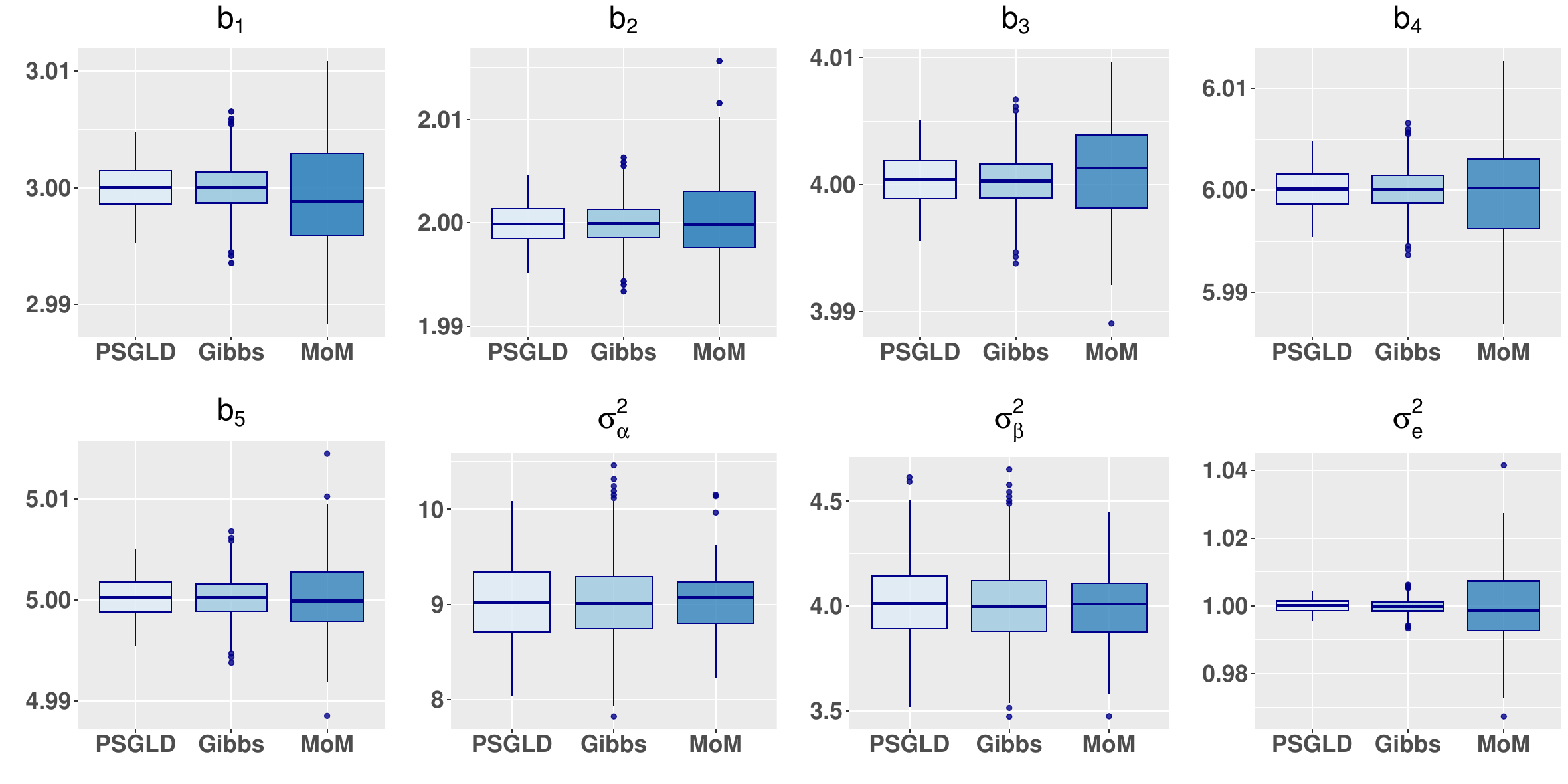} 
\end{figure}

\begin{figure}[htb]
\caption{Boxplots of posterior samples for the coefficients of fixed effects $b = (b_1, b_2, b_3, b_4, b_5)^{\top}$ and the variance components $\sigma^2_\alpha, \sigma^2_\beta, \sigma^2_e$ for the crossed mixed effects model with $90\%$ missing data in $100$ simulation replications. The results are averaged over $100$ simulation replications. PSGLD, pigeonhole stochastic gradient Langevin dynamics; Gibbs, Gibbs sampler; MoM, method of moments of \cite{gao2019estimation}.}
\label{boxplot_ub10}
\centering
\includegraphics[width=\textwidth]{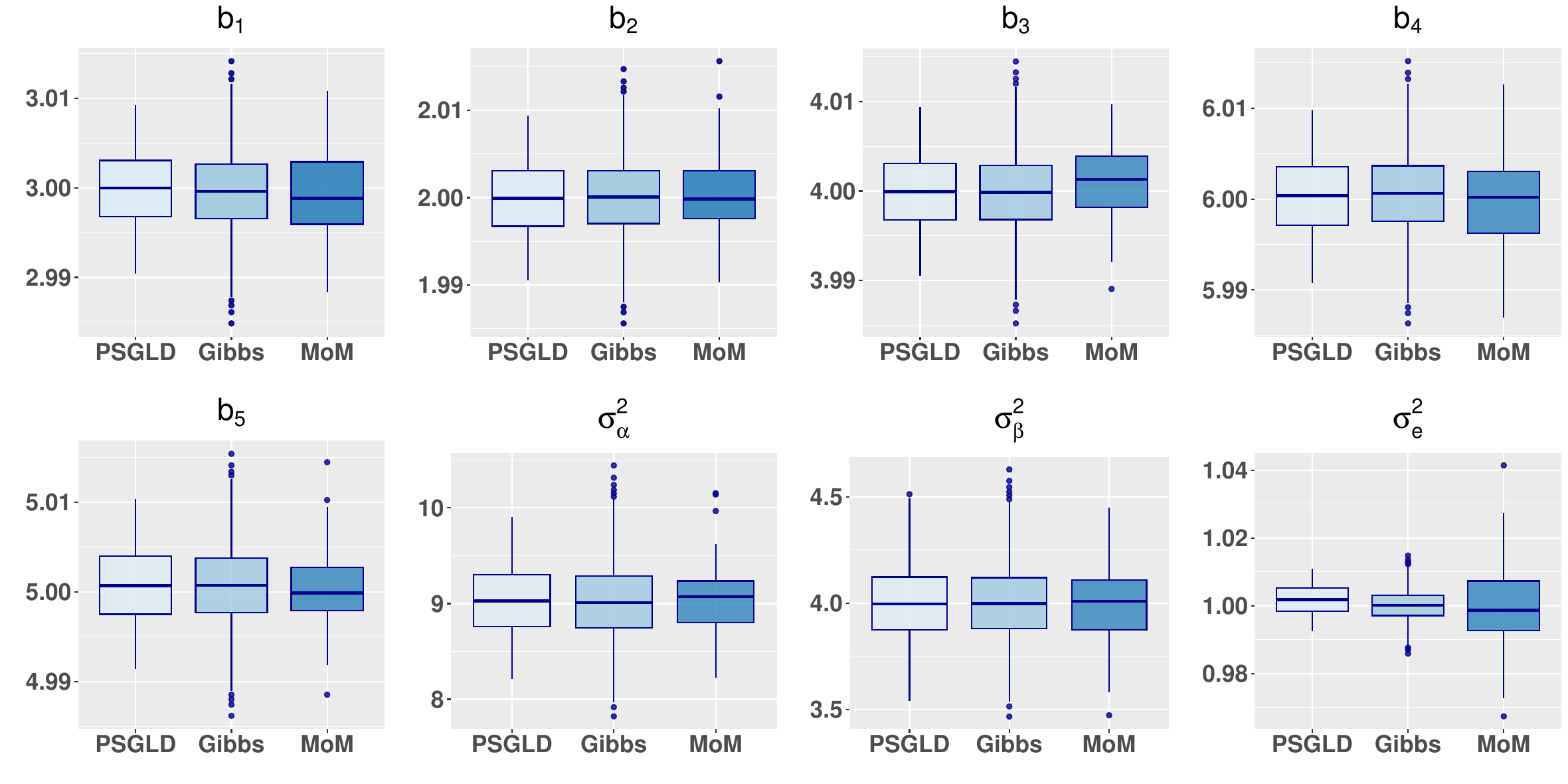} 
\end{figure}

Tables \ref{tbl:postmeansd_ub50} and \ref{tbl:postmeansd_ub10} report the posterior means and posterior standard deviations in parentheses of all the Bayesian methods as well as the estimated parameters from the MoM. Figures \ref{boxplot_ub50} and \ref{boxplot_ub10} show the boxplots of the marginal posterior distributions of all the parameters from the PSGLD and the Gibbs sampler as the Wasserstein-2 barycenters of $100$ posterior distributions, as well as the empirical distributions of the MoM estimators based on the same $100$ simulated datasets. From the tables and figures, we can see that the PSGLD still performs relatively well for the models with $50\%$ and $90\%$ missing data by providing posterior means of parameters similar to those from the full-data Gibbs sampler and the MoM. 
The posterior standard deviations from the PSGLD also closely resemble those from the Gibbs sampler. The boxplots in Figures \ref{boxplot_ub50} and \ref{boxplot_ub10} show that the marginal posterior distributions from the PSGLD are comparable to those from the full-data Gibbs sampler. In contrast, the empirical distributions of the MoM estimators tend to have very different variances from the Bayesian posterior variances for most of the parameters, especially for the variance of the error $\sigma^2_e$ where the two Bayesian methods have much lower uncertainty than the MoM.

To evaluate the approximation error from the PSGLD to the true target posteriors for the model with missing data, we compute the $W_2$ distances of marginal posterior distributions of the PSGLD to those of the full-data Gibbs sampler which are used as the benchmarks in Tables \ref{tbl:w2_ub50} and \ref{tbl:w2_ub10}. In both the cases of $50\%$ and $90\%$ observations, the PSGLD provides an accurate approximation to the target posteriors for the fixed effects coefficients $b=(b_1, b_2, b_3, b_4, b_5)^{\top}$ and the variance of random error $\sigma^2_e$ with very low $W_2$ distances. For random effects  $\sigma^2_\alpha$ and $\sigma^2_\beta$, the $W_2$ distances are higher than those of the other parameters, but overall they are still sufficiently small, demonstrating the good performance of the PSGLD in approaching the true posteriors even in the presence of high proportions of missing data. Low standard errors of these $W_2$ distances also show that the approximation is stable.

Meanwhile, we can also observe from comparing the results in Tables \ref{tbl:w2_b}, \ref{tbl:w2_ub50} and \ref{tbl:w2_ub10} that the increase in the proportion of missing observations in the data matrix $\Yb$ has deteriorated the approximation quality of the PSGLD to the target posterior distribution. 
For the fixed effects coefficients $b$ and the variance of random error $\sigma^2_e$, the balanced design with no missing data always has the lowest $W_2$ distances, and increasing the missing proportion from $50\%$ to $90\%$ has resulted in higher $W_2$ distances as well. This is expected because we have used the same minibatch size of $r=c=20$ for the subsets in the PSGLD for all three tables, and a missing proportion as high as $90\%$ will definitely result in significantly less amount of information and higher variation in the PSGLD algorithm, contributing the larger $W_2$ distances to the target posteriors.

\begin{table}[H]
\caption{Posterior means and posterior standard deviations for the coefficients of fixed effects $b = (b_1, b_2, b_3, b_4, b_5)^{\top}$ and the variance components $\sigma^2_\alpha, \sigma^2_\beta, \sigma^2_e$ in the crossed mixed effects model with $50\%$ missing data. The results are averaged over $100$ simulation replications. MoM, method of moments of \cite{gao2019estimation}; PSGLD, pigeonhole stochastic gradient Langevin dynamics; Gibbs, Gibbs sampler.
}
\label{tbl:postmeansd_ub50}
\centering
{\scriptsize
\begin{tabular}{rcccccccc}
\hline
&~& $b_1$ & $b_2$ & $b_3$ & $b_4$ & $b_5$\\
\hline
& MoM &  \multicolumn{1}{c}{2.9994 } & \multicolumn{1}{c} {2.0002 } & \multicolumn{1}{c} {4.0009} & \multicolumn{1}{c} {5.9998}  & \multicolumn{1}{c} {5.0004}   \\
& PSGLD & \multicolumn{1}{c}{3.0000 (0.0019)} & \multicolumn{1}{c}{1.9999 (0.0020)} & \multicolumn{1}{c}{4.0004 (0.0020)} & \multicolumn{1}{c}{6.0001 (0.0020)} & \multicolumn{1}{c}{5.0002 (0.0020)}    \\
& Gibbs & \multicolumn{1}{c}{3.0000 (0.0020)} & \multicolumn{1}{c}{1.9999 (0.0020)} & \multicolumn{1}{c}{4.0003 (0.0020)} & \multicolumn{1}{c}{6.0001 (0.0020)} & \multicolumn{1}{c}{5.0002 (0.0020)}\\
\hline
& ~  & \multicolumn{1}{c} {$\sigma^2_\alpha$} & \multicolumn{1}{c} {$\sigma^2_\beta$} & \multicolumn{1}{c} {$\sigma^2_e$}\\
\hline
&  MoM & \multicolumn{1}{c}{9.0216} & \multicolumn{1}{c} {4.0012} & \multicolumn{1}{c} {1.0004}   \\
& PSGLD & \multicolumn{1}{c}{9.0262 (0.4260)} & \multicolumn{1}{c}{4.0140 (0.2263)} & \multicolumn{1}{c}{1.0000 (0.0020)}    \\
& Gibbs & \multicolumn{1}{c}{9.0603 (0.4286)} & \multicolumn{1}{c}{4.0417 (0.2258)} & \multicolumn{1}{c}{0.9998 (0.0020)} \\
\hline
\end{tabular}
}%
\end{table}

\begin{table}[H]
\caption{$W_2$ distances between the marginal distributions of samples from the PSGLD and those from the Gibbs sampler in the crossed mixed effects model with $50\%$ missing data respectively.  
The $W_2$ distances are averaged over $100$ simulation replications. The standard errors of the average $W_2$ distances are in parentheses. PSGLD, pigeonhole stochastic gradient Langevin dynamics.
}
\label{tbl:w2_ub50}
\centering
{\scriptsize
\begin{tabular}{rcccccccc}
\hline
& $b_1$ & $b_2$ & $b_3$ & $b_4$  \\
\hline
PSGLD & \multicolumn{1}{c}{0.00063 (0.00003)} & \multicolumn{1}{c}{0.00060 (0.00003)} & \multicolumn{1}{c}{0.00058 (0.00003)} & \multicolumn{1}{c}{0.00068 (0.00004)}    \\
\hline
& $b_5$ & \multicolumn{1}{c} {$\sigma^2_\alpha$} & \multicolumn{1}{c} {$\sigma^2_\beta$} & \multicolumn{1}{c} {$\sigma^2_e$}\\
\hline
PSGLD & \multicolumn{1}{c}{0.00056 (0.00003)} & \multicolumn{1}{c}{0.09018 (0.00368)} & \multicolumn{1}{c}{0.04967 (0.00079)} & \multicolumn{1}{c}{0.00058 (0.00003)}    \\
\hline
\end{tabular}
}%
\end{table}

\begin{table}[H]
\caption{Posterior means and posterior standard deviations for the coefficients of fixed effects $b = (b_1, b_2, b_3, b_4, b_5)^{\top}$ and the variance components $\sigma^2_\alpha, \sigma^2_\beta, \sigma^2_e$ in the crossed mixed effects model with $90\%$ missing data. The results are averaged over $100$ simulation replications. MoM, method of moments of \cite{gao2019estimation}; PSGLD, pigeonhole stochastic gradient Langevin dynamics; Gibbs, Gibbs sampler.
}
\label{tbl:postmeansd_ub10}
\centering
{\scriptsize
\begin{tabular}{rcccccccc}
\hline
&~  & $b_1$ & $b_2$ & $b_3$ & $b_4$ & $b_5$\\
\hline
& MoM & \multicolumn{1}{c}{3.0011} & \multicolumn{1}{c} {1.9993} & \multicolumn{1}{c} {4.0003} & \multicolumn{1}{c} {5.9990}  & \multicolumn{1}{c} {4.9997}  \\
& PSGLD & \multicolumn{1}{c}{2.9998 (0.0046)} & \multicolumn{1}{c}{1.9999 (0.0045)} & \multicolumn{1}{c}{3.9999 (0.0044)} & \multicolumn{1}{c}{6.0004 (0.0045)} & \multicolumn{1}{c}{5.0009 (0.0045)}    \\
&  Gibbs & \multicolumn{1}{c}{2.9996 (0.0045)} & \multicolumn{1}{c}{2.0000 (0.0045)} & \multicolumn{1}{c}{3.9999 (0.0045)} & \multicolumn{1}{c}{6.0006 (0.0045)} & \multicolumn{1}{c}{5.0007 (0.0045)}\\
\hline
&~  & \multicolumn{1}{c} {$\sigma^2_\alpha$} & \multicolumn{1}{c} {$\sigma^2_\beta$} & \multicolumn{1}{c} {$\sigma^2_e$}\\
\hline
& MoM & \multicolumn{1}{c}{ 9.0156} & \multicolumn{1}{c} {3.9921 } & \multicolumn{1}{c} {1.0062}   \\
& PSGLD & \multicolumn{1}{c}{9.0275 (0.4084)} & \multicolumn{1}{c}{4.0029 (0.1841)} & \multicolumn{1}{c}{1.0013 (0.0045)}    \\
& Gibbs & \multicolumn{1}{c}{9.0243 (0.4043)} & \multicolumn{1}{c}{4.0023 (0.1799)} & \multicolumn{1}{c}{1.0001 (0.0045)} \\
\hline
\end{tabular}
}%
\end{table}

\begin{table}[H]
\caption{$W_2$ distances between the marginal distributions of samples from the PSGLD and those from the Gibbs sampler in the crossed mixed effects model with $90\%$ missing data respectively.  
The $W_2$ distances are averaged over $100$ simulation replications. The standard errors of the average $W_2$ distances are in parentheses. PSGLD, pigeonhole stochastic gradient Langevin dynamics.
}
\label{tbl:w2_ub10}
\centering
{\scriptsize
\begin{tabular}{rcccccccc}
\hline
& $b_1$ & $b_2$ & $b_3$ & $b_4$  \\
\hline
PSGLD & \multicolumn{1}{c}{0.00305 (0.00018)} & \multicolumn{1}{c}{0.00298 (0.00019)} & \multicolumn{1}{c}{0.00329 (0.00023)} & \multicolumn{1}{c}{0.00300 (0.00019)}     \\
\hline
& $b_5$ & \multicolumn{1}{c} {$\sigma^2_\alpha$} & \multicolumn{1}{c} {$\sigma^2_\beta$} & \multicolumn{1}{c} {$\sigma^2_e$}\\
\hline
PSGLD  & \multicolumn{1}{c}{0.00271 (0.00017)} & \multicolumn{1}{c}{0.11002 (0.00477)} & \multicolumn{1}{c}{0.03956 (0.00175)} & \multicolumn{1}{c}{0.00276 (0.00019)}    \\
\hline
\end{tabular}
}%
\end{table}

\subsection{Simulation for Different Missing Patterns} \label{simulate:pattern}

We illustrate the application of the pigeonhole SGLD algorithm on datasets with different missing patterns through three examples of simulated datasets. The first dataset includes varying degrees of missing data across its rows, and the other two datasets are simulated similarly to the real data analyzed in Section \ref{sec:experiments}, incorporating dummy indicators in the fixed effects and discrete responses. 

In the first example with varying degrees of missing data, we simulate the data following the model setup in Section \ref{subsec:balanced} with the size as $R = C = 1000$ in the full data matrix $\Yb$. In the first $300$ rows, we select $99\%$ observations missing completely at random; in the next $300$ rows and last $400$ rows, $95\%$ and $90\%$ of the data are missing completely at random, respectively. We set the true coefficients of fixed effects as $b = (0.5, 3, -2, 1.5, -1)^{\top}$, and the true variance components as $(\sigma^2_\alpha, \sigma^2_\beta, \sigma^2_e) =(1, 2, 0.5)$. For the fixed effect $x_{ij} \in \RR^5$, all the elements in $x_{ij} = (x_{ij1}, \cdots, x_{ij5})^{\top}$ are generated independently from $N (0, 1)$. We compare the performance of the pigeonhole SGLD in Algorithm \ref{algo:pigeonhole SGLD}, the Gibbs sampler, and the method of moments in
\cite{gao2019estimation} for inference on the coefficients of fixed effects $b = (b_1, \cdots, b_p)^{\top}$ and the variance components $\sigma^2_\alpha, \sigma^2_\beta, \sigma^2_e$. 

We repeat the whole simulation and estimation procedures for $10$ macro replications and report the averaged results. For the PSGLD algorithm, we randomly select $r = 50$ rows and $c = 50$ columns from the full data matrix $\Yb$ to form the submatrix of data $\Yb_n$ with the mini-batch size $n = \sum^r_{i=1}\sum^c_{j=1}(Z_{n})_{s_iq_j}$ at each iteration. For Step (b) of Algorithm \ref{algo:pigeonhole SGLD}, we generate a Markov chain of length $m = 50$ for the latent variables $\{\alpha_{s_1,k}, \cdots, \alpha_{s_r, k}\}^m_{k=1}$, and $\{\beta_{q_1,k}, \cdots, \beta_{q_c, k}\}^m_{k=1}$. The step sizes are selected by a grid search similar to Section \ref{simulate:b}. At the first $1100$ iterations, the step sizes of the coefficients of fixed effects 
$\epsilon_{b_1}, \epsilon_{b_2}, \epsilon_{b_3}, \epsilon_{b_4}, \epsilon_{b_5}$ are $O(10^{-6})$, and those of the variance components are $\epsilon_{\eta_\alpha} = 2.50 \times 10^{-5}, \epsilon_{\eta_\beta} = 1.11 \times 10^{-5} , \epsilon_{\eta_e} = 1.00 \times 10^{-5}$; after $1100$ iterations, the step sizes
$\epsilon_{b_1}, \epsilon_{b_2},\epsilon_{b_3},\epsilon_{b_4},\epsilon_{b_5}$ are $O(10^{-8})$ , and $\epsilon_{\eta_\alpha} = 5.81 \times 10^{-6}, \epsilon_{\eta_\beta} = 6.17 \times 10^{-8} , \epsilon_{\eta_e} = 1.00 \times 10^{-5}$.

\begin{figure}[H]
\caption{Dataset with varying degrees of missing data: Boxplots of posterior samples for the coefficients of fixed effects $b = (b_1, b_2, b_3, b_4, b_5)^{\top}$ and the variance components $\sigma^2_\alpha, \sigma^2_\beta, \sigma^2_e$ in $10$ simulation replications. The results are averaged over $10$ simulation replications. PSGLD, pigeonhole stochastic gradient Langevin dynamics; Gibbs, Gibbs sampler; MoM, method of moments of \cite{gao2019estimation}.}
\label{boxplot_vm}
\centering
\includegraphics[width=\textwidth]{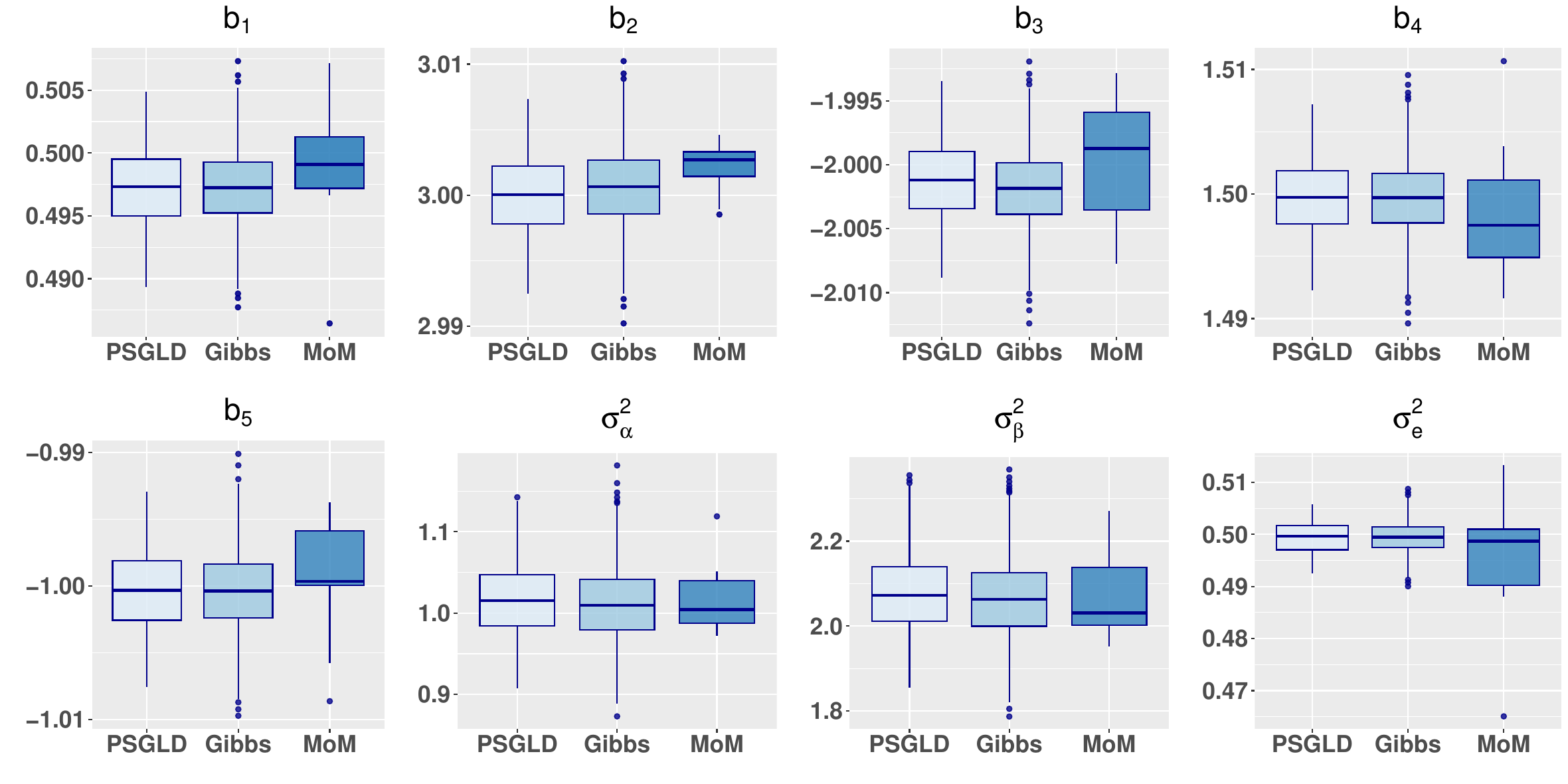} 
\end{figure}

Table \ref{tbl:postmeansd_vm} reports the posterior means and posterior standard deviations in parentheses for the two Bayesian methods, as well as the estimated parameters from the MoM. Figure \ref{boxplot_vm} shows the boxplots of the marginal posterior distributions of all the parameters from the PSGLD and the Gibbs sampler using the Wasserstein-2 barycenter of $10$ posterior distributions, as well as the empirical distributions of the MoM estimators based on the same $10$ simulated datasets. Similar posterior means and posterior standard deviations from the PSGLD and the Gibbs sampler demonstrate the estimation accuracy of the PSGLD in datasets with varying degrees of missing data and a greater missing proportion. 
The marginal posterior distributions from PSGLD are also comparable to those from the full-data Gibbs sampler as shown in the boxplots in Figure \ref{boxplot_vm}, indicating good approximation performance from PSGLD to the true posterior distribution. In contrast, the empirical distributions of the MoM estimators differ substantially from the Bayesian posterior distributions for most of the model parameters.

To assess the approximation error from the PSGLD to the true target posteriors for the dataset with varying degrees of missing data, we compute the $W_2$ distances between the marginal posterior distributions of samples from the PSGLD and those from the full-data Gibbs sampler regarded as the benchmarks in Table \ref{tbl:w2_vm}. Remarkably low $W_2$ distances associated with small standard errors indicate an accurate and stable approximation of the PSGLD to the true posteriors for all the parameters in datasets with high proportions and varying degrees of missing data. Additionally, we note that the $W_2$ distances of all the parameters in Table \ref{tbl:w2_vm} are lower than those observed for the dataset with $90\%$ missing data in Table \ref{tbl:w2_ub10}. Despite the dataset in Table \ref{tbl:w2_vm} having a proportion of missing data greater than $90\%$, we select $50$ rows and $50$ columns for the submatrix $\Yb_n$ in the PSGLD and process more data per iteration than the example of $90\%$ missing data with the mini-batch size $r=c=20$. This consequently yields a higher amount of information in PSGLD and improves the approximation quality to the target posterior distributions.
\vspace{2cm}

\begin{table}[H]
\caption{Dataset with varying degrees of missing data: Posterior means and posterior standard deviations for the coefficients of fixed effects $b = (b_1, b_2, b_3, b_4, b_5)^{\top}$ and the variance components $\sigma^2_\alpha, \sigma^2_\beta, \sigma^2_e$. The results are averaged over $10$ simulation replications. MoM, method of moments of \cite{gao2019estimation}; PSGLD, pigeonhole stochastic gradient Langevin dynamics; Gibbs, Gibbs sampler.
}
\label{tbl:postmeansd_vm}
\centering
{\scriptsize
\begin{tabular}{rcccccccc}
\hline
&~  & $b_1$ & $b_2$ & $b_3$ & $b_4$ & $b_5$\\
\hline
& MoM & \multicolumn{1}{c}{0.4989} & \multicolumn{1}{c} {3.0021} & \multicolumn{1}{c} {-1.9996} & \multicolumn{1}{c} {1.4985}  & \multicolumn{1}{c} {-0.9995}  \\
& PSGLD & \multicolumn{1}{c}{0.4972 (0.0030)} & \multicolumn{1}{c}{3.0002 (0.0029)} & \multicolumn{1}{c}{-2.0016 (0.0031)} & \multicolumn{1}{c}{1.4997 (0.0030)} & \multicolumn{1}{c}{-1.0003 (0.0031)}    \\
&  Gibbs & \multicolumn{1}{c}{0.4973 (0.0030)} & \multicolumn{1}{c}{3.0007 (0.0030)} & \multicolumn{1}{c}{-2.0020 (0.0030)} & \multicolumn{1}{c}{1.4998 (0.0030)} & \multicolumn{1}{c}{-1.0002 (0.0031)}\\
\hline
&~  & \multicolumn{1}{c} {$\sigma^2_\alpha$} & \multicolumn{1}{c} {$\sigma^2_\beta$} & \multicolumn{1}{c} {$\sigma^2_e$}\\
\hline
& MoM & \multicolumn{1}{c}{1.0176} & \multicolumn{1}{c} {2.0680} & \multicolumn{1}{c} {0.4951}   \\
& PSGLD & \multicolumn{1}{c}{1.0162 (0.0460)} & \multicolumn{1}{c}{2.0781 (0.0916)} & \multicolumn{1}{c}{0.4994 (0.0029)}    \\
& Gibbs & \multicolumn{1}{c}{1.0112 (0.0466)} & \multicolumn{1}{c}{2.0652 (0.0931)} & \multicolumn{1}{c}{0.4996 (0.0029)} \\
\hline
\end{tabular}
}%
\end{table}

\vspace{2cm}

\begin{table}[H]
\caption{Dataset with varying degrees of missing data: $W_2$ distances between the marginal distributions of samples from the PSGLD and those from the Gibbs sampler. 
The $W_2$ distances are averaged over $10$ simulation replications. The standard errors of the average $W_2$ distances are in parentheses. PSGLD, pigeonhole stochastic gradient Langevin dynamics.
}
\label{tbl:w2_vm}
\centering
{\scriptsize
\begin{tabular}{rcccccccc}
\hline
& $b_1$ & $b_2$ & $b_3$ & $b_4$  \\
\hline
PSGLD & \multicolumn{1}{c}{0.00154 (0.00030)} & \multicolumn{1}{c}{0.00151 (0.00030)} & \multicolumn{1}{c}{0.00113 (0.00010)} & \multicolumn{1}{c}{0.00108 (0.00019)}     \\
\hline
& $b_5$ & \multicolumn{1}{c} {$\sigma^2_\alpha$} & \multicolumn{1}{c} {$\sigma^2_\beta$} & \multicolumn{1}{c} {$\sigma^2_e$}\\
\hline
PSGLD  & \multicolumn{1}{c}{0.00137 (0.00021)} & \multicolumn{1}{c}{0.03620 (0.00403)} & \multicolumn{1}{c}{0.02843 (0.00443)} & \multicolumn{1}{c}{0.00223 (0.00040)}    \\
\hline
\end{tabular}
}%
\end{table}

We further validate the applicability of the PSGLD algorithm using two simulated datasets examples. Compared to previous simulations, these datasets are more challenging with greater proportions of missing data and truncated responses, closely resembling the real data characteristics in Section \ref{sec:experiments} of the main text. With the understanding of the ground truth regarding the model and parameters from which the data are generated, we can justify the convergence of the approximate posterior samples from PSGLD. The two challenging datasets are akin to the MovieLens dataset in Section \ref{sec:experiments} in terms of fixed effects features, the order of parameter magnitudes, dataset sizes, and discrete responses. The fixed effects $x_{ij}=(x_{ij0}, x_{ij1}, x_{ij2}, x_{ij3}, x_{ij4}, x_{ij5})^{\top}$ in both datasets are created according to the following mechanism: $x_{ij0}=1$ serves as the intercept; $x_{ij1}$ is randomly assigned a value of $1$ or $0$ with the equal probability $1/2$; $x_{ij2}, x_{ij3}, x_{ij4}$ mimic the \textit{Genera} predictor in the MovieLens dataset, each of which is generated independently from a Bernoulli distribution with equal probabilities of  $0$ or $1$, ensuring that their sum equals $1$. For example, if we draw values $(1,0,0)$ from the Bernoulli distribution, then $(x_{ij2},x_{ij3},x_{ij4})=(1,0,0)$, whereas if we generate $(1,1,1)$, then $(x_{ij2},x_{ij3},x_{ij4})=(1/3,1/3,1/3)$. $x_{ij5}$ is independently generated from $N(0,0.25)$. 

For the first challenging dataset, we generate data following the model in \eqref{eq:lme} and \eqref{eq:normalre} with the numbers of row effects and column effects as $R=6000,~ C=4000$. We select $99\%$ of data missing completely at random, resulting in $2.4 \times 10^5$ observations. We set the coefficients of fixed effects as $b=(2, 0.8, 0.2, -0.5, 0.07, 0.35)^{\top}$, and the true variance components as $(\sigma^2_\alpha, \sigma^2_\beta, \sigma^2_e) = (0.08, 0.2, 0.9)$. After generating responses $Y_{ij}$'s, we transform them to $5$ integer values based on the quantiles: responses lower than the $20\%$ quantile are assigned the value $0$, those between the $20\%$ and $40\%$ quantiles are assigned the value $1$, those between the $40\%$ and $60\%$ quantiles are assigned the value $2$, those between the $60\%$ and $80\%$ quantiles are assigned the value $3$, and those higher than the $80\%$ quantile are assigned the value $4$. This mimics the discrete observations in the real MovieLens dataset. Meanwhile, this allows us to assess the performance of the proposed algorithm in the presence of model misspecification. 

For the second dataset, we set the numbers of row effects and column effects as $R=3000,~ C=5000$. The proportion of missing observations and the method of generating the fixed effects are identical to those in the first dataset. The true values of parameters are $b = (1.5, 0.5, 1, -1.5, 0.1, -0.5)^{\top}$ for the coefficients of fixed effects and $(\sigma^2_\alpha, \sigma^2_\beta, \sigma^2_e) = (1.5, 2, 1)$ for the variances components. We also truncate the responses by sorting them in ascending order and categorizing them based on quantiles: data points in the lowest $5.6\%$ are assigned the value $-2$; those between the $5.6\%$ and $16.3\%$ quantiles are assigned the value $0$; those between the $16.3\%$ to $42.4\%$ quantiles are assigned the value $2$; those between the $42.3\%$ to $77.3\%$ quantiles are assigned the value $4$; those above $77.3\%$ are assigned the value $6$. 

We fit both datasets using the PSGLD in Algorithm \ref{algo:pigeonhole SGLD} and the full-data Gibbs sampler, evaluating both the posterior estimation accuracy and the computational efficiency of the PSGLD relative to the Gibbs sampler. For the first dataset, the initial values of all the parameters are set to be $1$ for the PSGLD and the Gibbs sampler. For the second dataset, the initial values of all fixed effects coefficients $b$ are set to be $2$, and those of the variance components $\sigma^2_\alpha, \sigma^2_\beta, \sigma^2_e$ are set to be $0.5, 1, 2$ for the PSGLD and the Gibbs sampler. 

While analyzing each of the two datasets, at each iteration of the PSGLD, we randomly select $r=200$ and $c=200$ columns from the full data matrix $\Yb$ and construct the submatrix of data $\Yb_n$ with the number of observations $n = \sum^r_{i=1}\sum^c_{j=1}(Z_n)_{s_iq_j}$. A short Markov chain with the length $m=50$ for the latent variables $\{\alpha_{s_1,k}, \cdots, \alpha_{s_r,k}\}^m_{k=1}$ and $\{\beta_{q_1,k}, \cdots, \beta_{q_c,k}\}^m_{k=1}$ is generated by the Gibbs sampler following the conditional posterior distributions \eqref{eq:conditional_d}. We select the step sizes in the PSGLD through a grid search and adopt the combination of step sizes that minimizes the $W_2$ distances between the empirical distribution of the PSGLD and that of the full-data Gibbs sampler. For the first dataset, at the first $1500$ iterations, the step sizes
$\epsilon_{b_0},\epsilon_{b_1}, \epsilon_{b_2}, \epsilon_{b_3}, \epsilon_{b_4}, \epsilon_{b_5}$ are $O(10^{-5})$, and $\epsilon_{\eta_\alpha} = 2.00 \times 10^{-5}, \epsilon_{\eta_\beta} = 3.33 \times 10^{-5} , \epsilon_{\eta_e} = 2.00 \times 10^{-5}$; after $1500$ iterations, the step sizes
$\epsilon_{b_0},\epsilon_{b_1}, \epsilon_{b_2},\epsilon_{b_3},\epsilon_{b_4},\epsilon_{b_5}$ are $O(10^{-7})$ , and $\epsilon_{\eta_\alpha} = 1.67 \times 10^{-6}, \epsilon_{\eta_\beta} = 8.33 \times 10^{-6} , \epsilon_{\eta_e} = 1.00 \times 10^{-7}$. For the second dataset, at the first $1000$ iterations, the step sizes
$\epsilon_{b_0},\epsilon_{b_1}, \epsilon_{b_2}, \epsilon_{b_3}, \epsilon_{b_4}, \epsilon_{b_5}$ are $O(10^{-6})$, and $\epsilon_{\eta_\alpha} = 2.00 \times 10^{-5}, \epsilon_{\eta_\beta} = 1.33 \times 10^{-5} , \epsilon_{\eta_e} = 1.00 \times 10^{-5}$; after $1000$ iterations, the step sizes
$\epsilon_{b_0},\epsilon_{b_1}, \epsilon_{b_2},\epsilon_{b_3},\epsilon_{b_4},\epsilon_{b_5}$ are $O(10^{-6})$ , and $\epsilon_{\eta_\alpha} = 1.33 \times 10^{-5}, \epsilon_{\eta_\beta} = 7.84 \times 10^{-6} , \epsilon_{\eta_e} = 1.43 \times 10^{-7}$.

\begin{figure}[H]
\caption{Challenging simulated dataset $1$: Boxplots of posterior samples for the coefficients of fixed effects $b = (b_0, b_1, b_2, b_3, b_4, b_5)^{\top}$ and the variance components $\sigma^2_\alpha, \sigma^2_\beta, \sigma^2_e$ for the crossed mixed effects model in $10$ simulation replications. The results are averaged over $10$ simulation replications. 
PSGLD, pigeonhole stochastic gradient Langevin dynamics; Gibbs, Gibbs sampler.}
\label{boxplot_c1}
\centering
\includegraphics[width=\textwidth]{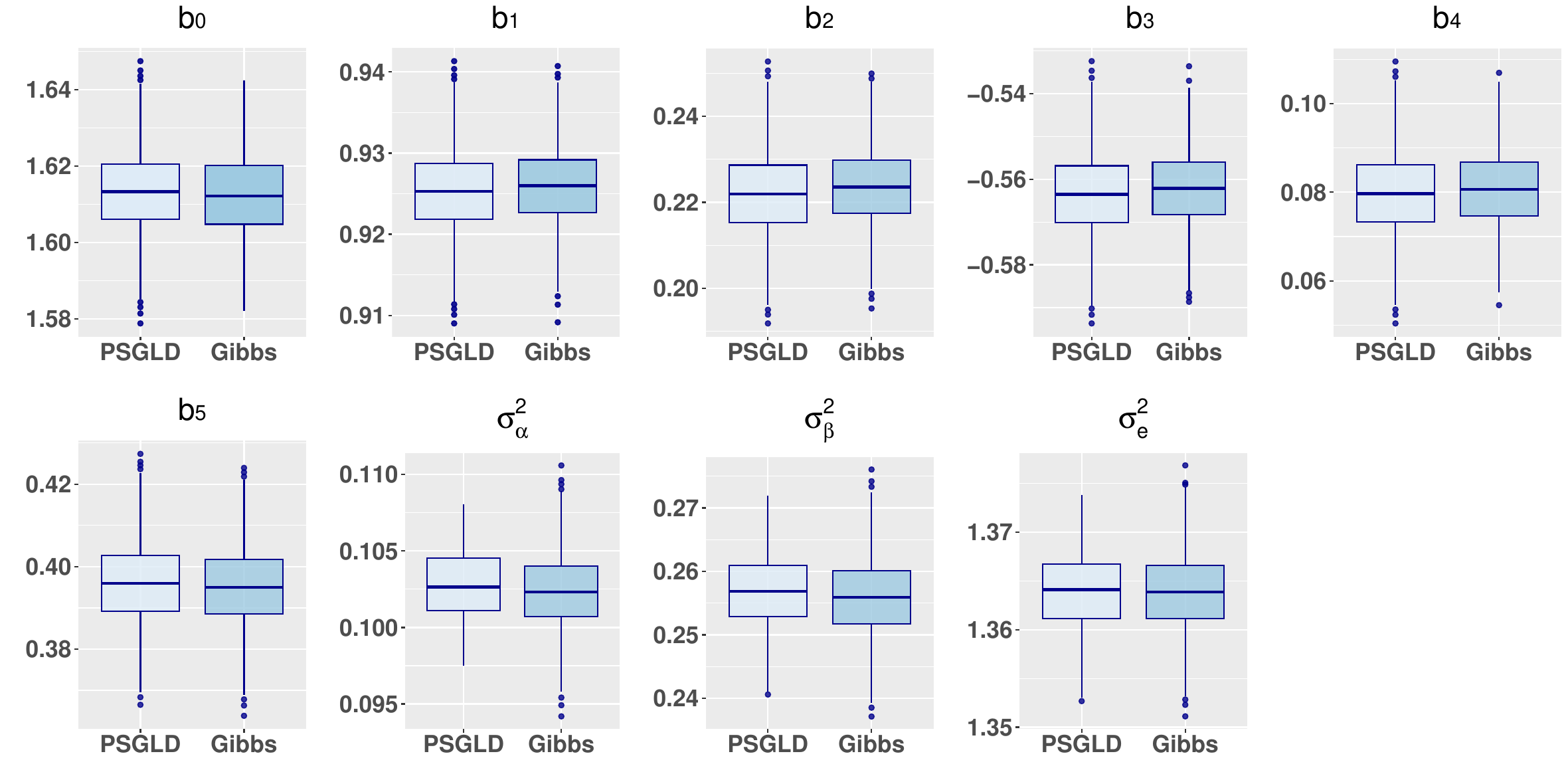}
\end{figure}

\begin{figure}[H]
\caption{Challenging simulated dataset $2$: Boxplots of posterior samples for the coefficients of fixed effects $b = (b_0, b_1, b_2, b_3, b_4, b_5)^{\top}$ and the variance components $\sigma^2_\alpha, \sigma^2_\beta, \sigma^2_e$ for the crossed mixed effects model in $10$ simulation replications. The results are averaged over $10$ simulation replications. 
PSGLD, pigeonhole stochastic gradient Langevin dynamics; Gibbs, Gibbs sampler.}
\label{boxplot_c2}
\centering
\includegraphics[width=\textwidth]{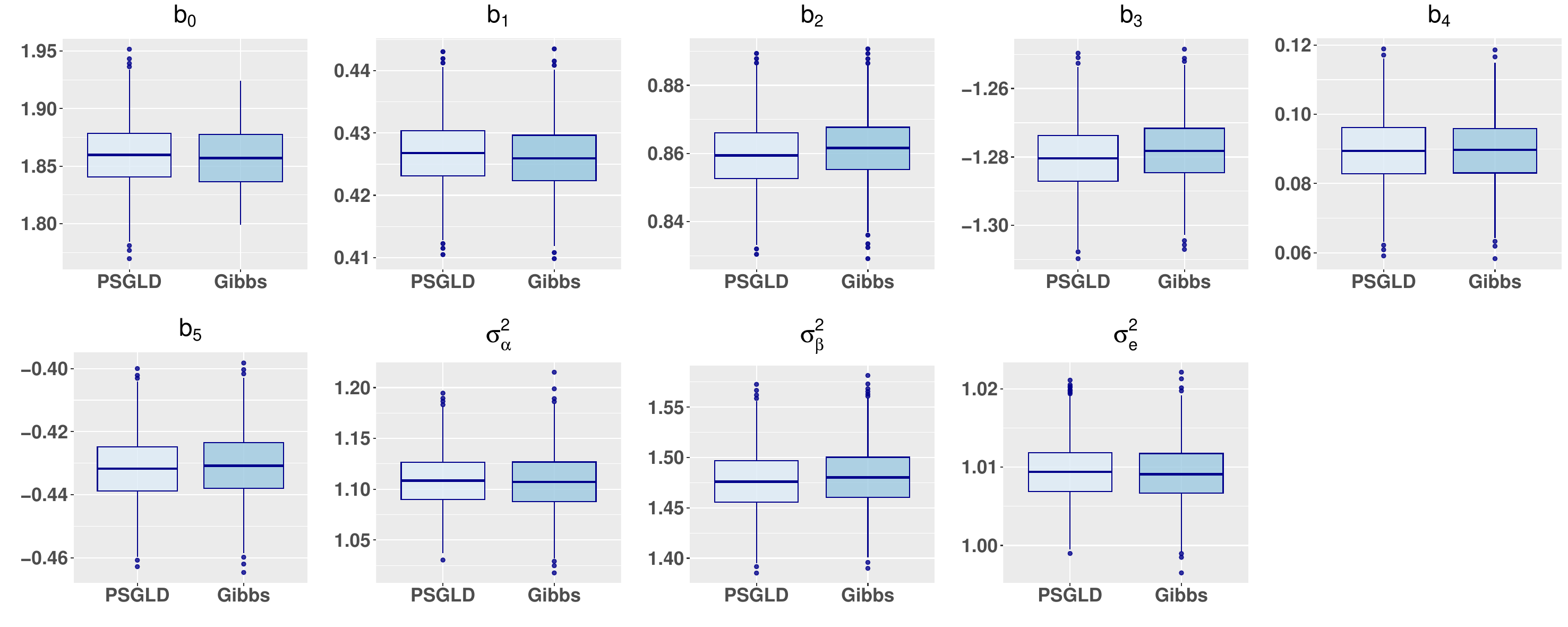}
\end{figure}

Figures \ref{boxplot_c1} and \ref{boxplot_c2} display the boxplots of the marginal posterior samples from the PSGLD and the Gibbs sampler for the two challenging datasets, respectively. We implement each algorithm for a chain of length $4 \times 10^4$ iterations with the first $10^4$ samples discarded as burn-in. The marginal posterior distributions from the PSGLD quickly approach those from the Gibbs sampler for both the fixed effect coefficients and the variance components in both datasets. The PSGLD provides an accurate approximation to the true posterior distributions from the full-data Gibbs sampler even with high proportions of missing data. Furthermore, the posterior means from both the PSGLD and the Gibbs sampler show deviations from the true values of model parameters for the two datasets, yet they still approximate the ground truth to some extent, which demonstrates the effectiveness and applicability of the PSGLD in crossed mixed effects models under slight model misspecification. 

\begin{figure}[H]
\caption{Challenging simulated dataset $1$: $W_2$ distances of the coefficients of fixed effects $b = (b_0, b_1, b_2, b_3, b_4, b_5)^{\top}$ and the variance components $\sigma^2_\alpha, \sigma^2_\beta, \sigma^2_e$ against CPU time (seconds), where the brown line is for the pigeonhole stochastic gradient Langevin dynamics algorithm and the yellow line is for the Gibbs sampler. PSGLD, pigeonhole stochastic gradient Langevin dynamics; Gibbs, Gibbs sampler.}
\label{W2_distance_c1}
\centering
\includegraphics[width=\textwidth]{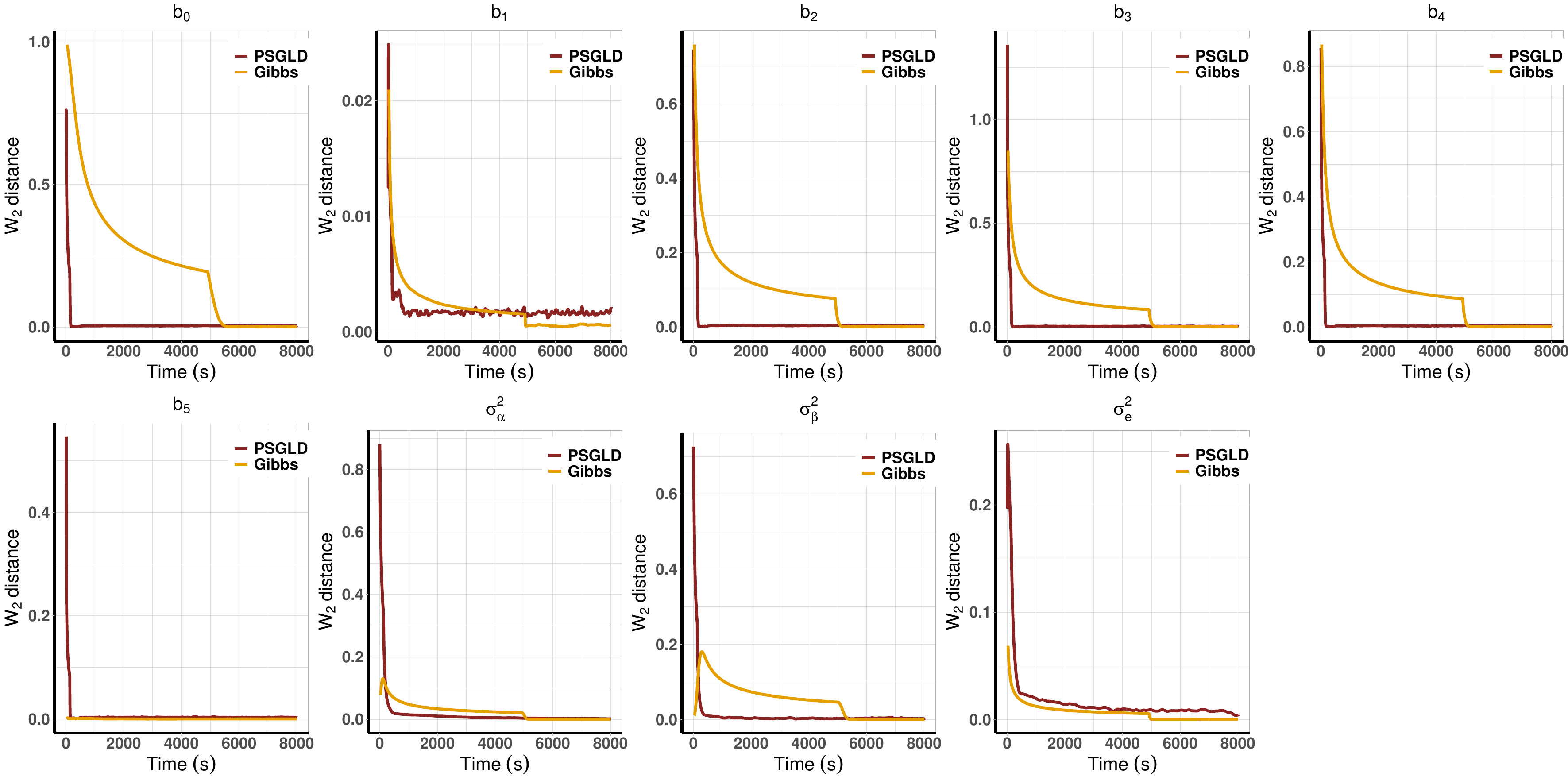}
\end{figure}

\begin{figure}[H]
\caption{Challenging simulated dataset $2$: $W_2$ distances of the coefficients of fixed effects $b = (b_0, b_1, b_2, b_3, b_4, b_5)^{\top}$ and the variance components $\sigma^2_\alpha, \sigma^2_\beta, \sigma^2_e$ against CPU time (seconds), where the brown line is for the pigeonhole stochastic gradient Langevin dynamics algorithm and the yellow line is for the Gibbs sampler. PSGLD, pigeonhole stochastic gradient Langevin dynamics; Gibbs, Gibbs sampler.}
\label{W2_distance_c2}
\centering
\includegraphics[width=\textwidth]{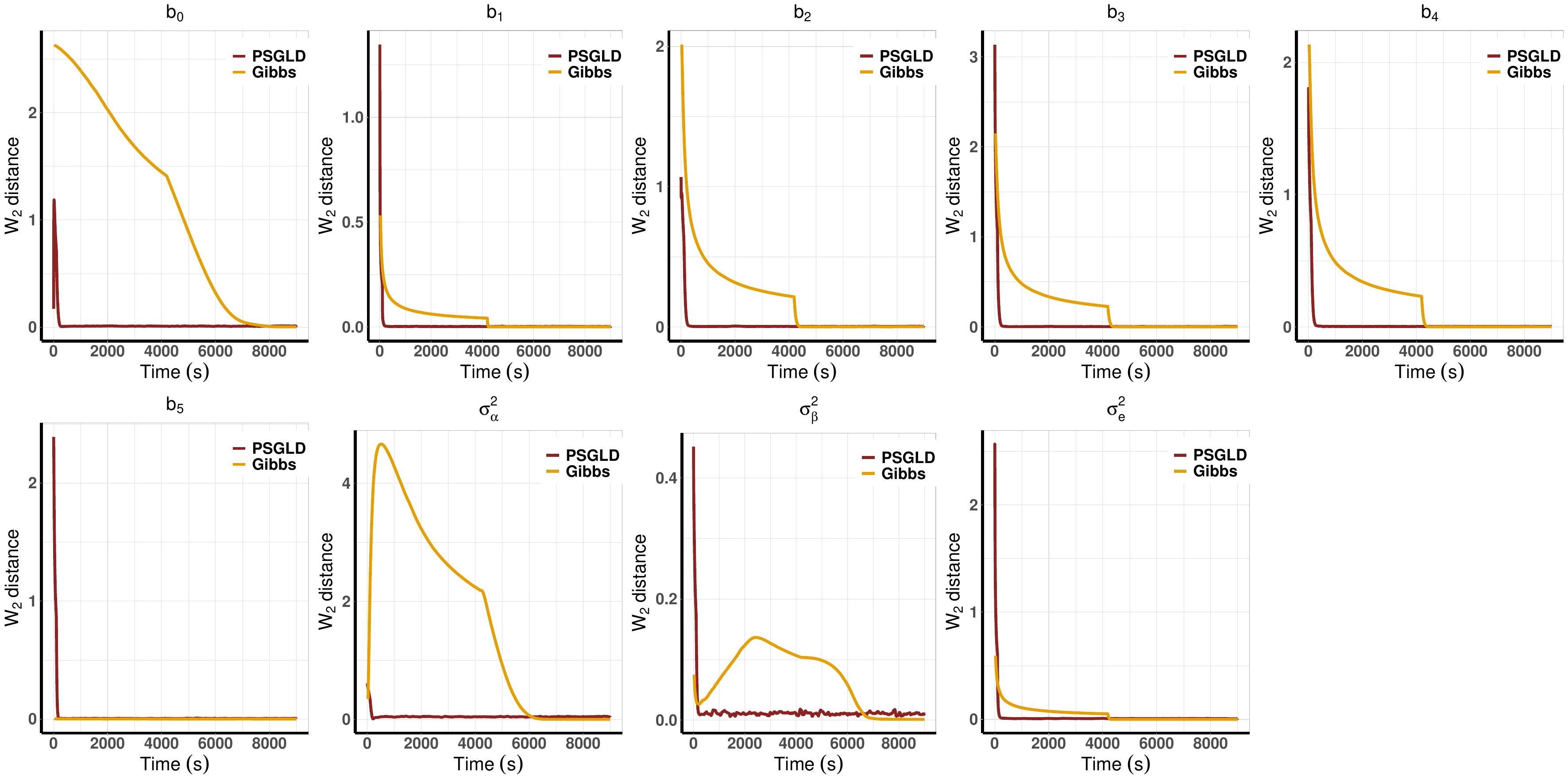}
\end{figure}

To assess the computational efficiency of the PSGLD and the Gibbs sampler in approximating the target posterior distribution, we follow the approach detailed in Section \ref{sec:experiments} and plot the $W_2$ distance against the elapsed CPU time for each model parameter across both datasets in Figures \ref{W2_distance_c1} and \ref{W2_distance_c2}. We consider the first $500$ samples after $10^4$ burn-in iterations from the Gibbs sampler as the stationary distribution serving as the true posterior distribution, and compute the $W_2$ distance iteratively between the samples from each algorithm and this benchmark. For the two datasets, the $W_2$ distances between the samples of each parameter from the PSGLD and the benchmark have decreased rapidly to a low and stable level very close to $0$ within $110$ seconds. However, convergence of the posterior samples of each parameter from the Gibbs sampler in the first dataset requires over $800$ seconds, and in the second dataset, it takes at least $2500$ seconds for samples of all the model parameters from the Gibbs sampler to reach convergence. Although the $W_2$ distances of posterior samples from the Gibbs sampler drop as quickly as the PSGLD for some parameters, such as $b_5, \sigma^2_e$ in the first dataset and $b_5$ in the second dataset, it is evident that the PSGLD achieves convergence to the true posterior distributions significantly faster for most parameters than the full-data Gibbs sampler. These plots again demonstrate the remarkable computational efficiency of the PSGLD in approximating the target posterior distribution with a high proportion of missing data and model misspecification. 

\newpage

\vspace{-4mm}

\vspace{-4mm}

\newpage

\newpage

\section{Uncertainty Quantification for PSGLD Algorithm} \label{sec:UQ.PSGLD}

In all the previous simulation studies and real data applications, we have used the Wasserstein-2 barycenter to summarize the posterior samples from multiple chains from different runs from the SGLD and PSGLD algorithms in Algorithms \ref{algo:SGLD} and \ref{algo:pigeonhole SGLD}. This raises the question of how representative the Wasserstein-2 barycenter is compared to the individual chains, and how much variation we observe for different runs of our SGLD algorithms. In this section, we provide an empirical study on the uncertainty quantification of the individual PSGLD chains by comparing them with the benchmark, the posterior distribution from full-data Gibbs sampler. 

Table \ref{tbl:Summary_w2} presents the mean, median, maximum, and standard deviation of the $W_2$ distances between 
individual PSGLD chains and the chain from full-data Gibbs sampler, as well as the $W_2$ distances between the $W_2$ barycenter of PSGLD chains and the chain from full-data Gibbs sampler for each parameter. 
We report the results on the two challenging datasets in Section \ref{simulate:pattern} and the two real data examples in Section \ref{sec:experiments}, both of which have high proportions of missing data ($94.6\%-99\%$) and either resemble or originate from real-world data.
For the challenging datasets in Section \ref{simulate:pattern}, $10$ macro-replicated datasets were generated under the same model setup, with $10$ PSGLD chains run on each dataset using different random seeds, yielding $100$ chains in total. For the real data examples in Section \ref{sec:experiments}, $10$ PSGLD chains were run on the same dataset using $10$ different random seeds. As shown in Table \ref{tbl:Summary_w2}, the means, medians, and maximums of these $W_2$ distances for individual PSGLD chains are all small and no more than the order $O(10^{-2})$. The $W_2$ distances between the $W_2$ barycenter and the chain from full-data Gibbs sampler are also in the same order. This indicates that the PSGLD algorithm is relatively stable and generates similar posterior samples to those from the Gibbs sampler in each run. In addition, the standard deviations are generally small, indicating that the individual PSGLD chains are not substantially different from one another. Given the low uncertainty and stability of the PSGLD chains, it is reasonable to summarize the individual PSGLD posterior chains using their Wasserstein-2 barycenter.

\newpage

\begin{table}[H]
\caption{Summary statistics of $W_2$ distances between individual PSGLD chains and the chain from Gibbs Sampler, as well as the $W_2$ distances between the $W_2$ barycenter of $10$ individual PSGLD chains and the chain from Gibbs sampler, for the coefficients of fixed effects $b$ and the variance components $\sigma^2_\alpha, \sigma^2_\beta, \sigma^2_e$ in the challenging simulated datasets $1$ and $2$ in Section \ref{simulate:pattern}, as well as the MovieLens dataset and ETH Lecturer Evaluation dataset in Section \ref{sec:experiments}. Each statistic is computed based on $10$ $W_2$ distances from $10$ PSGLD runs per dataset.  Max, maximum; SD, standard deviation.}
\vspace{2mm}
\label{tbl:Summary_w2}
\centering
{\scriptsize
\begin{tabular}{rcccccccccc}
\hline
\multicolumn{11}{c}{Challenging simulated dataset $1$} \\
\hline
& ~  & $b_0$ & $b_1$ & $b_2$ & $b_3$ & $b_4$ & $b_5$ & $\sigma^2_\alpha$ & $\sigma^2_\beta$ & $\sigma^2_e$ \\
\hline
& $W_2$ barycenter & \multicolumn{1}{c}{0.0027} & \multicolumn{1}{c} {0.0017} & \multicolumn{1}{c} {0.0036} & \multicolumn{1}{c} {0.0035}  & \multicolumn{1}{c} {0.0030} & \multicolumn{1}{c} {0.0028}& \multicolumn{1}{c} {0.0122} & \multicolumn{1}{c} {0.0130} & \multicolumn{1}{c}{0.0194}  \\
& Mean & \multicolumn{1}{c}{0.0045} & \multicolumn{1}{c} {0.0023} & \multicolumn{1}{c} {0.0049} & \multicolumn{1}{c} {0.0048}  & \multicolumn{1}{c} {0.0043} & \multicolumn{1}{c} {0.0046}& \multicolumn{1}{c} {0.0219} & \multicolumn{1}{c} {0.0142} & \multicolumn{1}{c}{0.0198}  \\
& Median & \multicolumn{1}{c}{0.0043} & \multicolumn{1}{c} {0.0023} & \multicolumn{1}{c} {0.0046} & \multicolumn{1}{c} {0.0043}  & \multicolumn{1}{c} {0.0040} & \multicolumn{1}{c} {0.0047}& \multicolumn{1}{c} {0.0217}& \multicolumn{1}{c} {0.0138} & \multicolumn{1}{c}{0.0195}   \\
&  Max & \multicolumn{1}{c}{0.0100} & \multicolumn{1}{c} {0.0045} & \multicolumn{1}{c} {0.0114} & \multicolumn{1}{c} {0.0113}  & \multicolumn{1}{c} {0.0082} & \multicolumn{1}{c} {0.0096}& \multicolumn{1}{c} {0.0477}& \multicolumn{1}{c} {0.0241} & \multicolumn{1}{c}{0.0414} \\
& SD & \multicolumn{1}{c}{0.0018} & \multicolumn{1}{c} {0.0008} & \multicolumn{1}{c} {0.0017} & \multicolumn{1}{c} {0.0019}  & \multicolumn{1}{c} {0.0014} & \multicolumn{1}{c} {0.0016}& \multicolumn{1}{c} {0.0088}& \multicolumn{1}{c} {0.0032} & \multicolumn{1}{c}{0.0064}\\
\hline
\multicolumn{11}{c}{Challenging simulated dataset $2$} \\
\hline
& ~  & $b_0$ & $b_1$ & $b_2$ & $b_3$ & $b_4$ & $b_5$ & $\sigma^2_\alpha$ & $\sigma^2_\beta$ & $\sigma^2_e$ \\
\hline
& $W_2$ barycenter & \multicolumn{1}{c}{0.0129} & \multicolumn{1}{c} {0.0040} & \multicolumn{1}{c} {0.0060} & \multicolumn{1}{c} {0.0063}  & \multicolumn{1}{c} {0.0059} & \multicolumn{1}{c} {0.0067}& \multicolumn{1}{c} {0.0247} & \multicolumn{1}{c} {0.0144} & \multicolumn{1}{c}{0.0088}  \\
& Mean & \multicolumn{1}{c}{0.0325} & \multicolumn{1}{c} {0.0048} & \multicolumn{1}{c} {0.0076} & \multicolumn{1}{c} {0.0088}  & \multicolumn{1}{c} {0.0087} & \multicolumn{1}{c} {0.0094}& \multicolumn{1}{c} {0.0497} & \multicolumn{1}{c} {0.0297} & \multicolumn{1}{c}{0.0091}  \\
& Median & \multicolumn{1}{c}{0.0274} & \multicolumn{1}{c} {0.0045} & \multicolumn{1}{c} {0.0072} & \multicolumn{1}{c} {0.0081}  & \multicolumn{1}{c} {0.0081} & \multicolumn{1}{c} {0.0083}& \multicolumn{1}{c} {0.0464}& \multicolumn{1}{c} {0.0292} & \multicolumn{1}{c}{0.0086}   \\
&  Max & \multicolumn{1}{c}{0.0695} & \multicolumn{1}{c} {0.0087} & \multicolumn{1}{c} {0.0162} & \multicolumn{1}{c} {0.0180}  & \multicolumn{1}{c} {0.0171} & \multicolumn{1}{c} {0.0208}& \multicolumn{1}{c} {0.0927}& \multicolumn{1}{c} {0.0467} & \multicolumn{1}{c}{0.0181} \\
& SD & \multicolumn{1}{c}{0.0192} & \multicolumn{1}{c} {0.0012} & \multicolumn{1}{c} {0.0023} & \multicolumn{1}{c} {0.0028}  & \multicolumn{1}{c} {0.0028} & \multicolumn{1}{c} {0.0037}& \multicolumn{1}{c} {0.0199}& \multicolumn{1}{c} {0.0077} & \multicolumn{1}{c}{0.0030}\\
\hline
\multicolumn{11}{c}{MovieLens Dataset} \\
\hline
& ~  & $b_0$ & $b_1$ & $b_2$ & $b_3$ & $b_4$ & $b_5$ & $\sigma^2_\alpha$ & $\sigma^2_\beta$ & $\sigma^2_e$ \\
\hline
& $W_2$ barycenter & \multicolumn{1}{c}{0.0031} & \multicolumn{1}{c} {0.0041} & \multicolumn{1}{c} {0.0018} & \multicolumn{1}{c} {0.0018}  & \multicolumn{1}{c} {0.0123} & \multicolumn{1}{c} {0.0041}& \multicolumn{1}{c} {0.0021} & \multicolumn{1}{c} {0.0034} & \multicolumn{1}{c}{0.0029}  \\
& Mean & \multicolumn{1}{c}{0.0081} & \multicolumn{1}{c} {0.0041} & \multicolumn{1}{c} {0.0061} & \multicolumn{1}{c} {0.0027}  & \multicolumn{1}{c} {0.0123} & \multicolumn{1}{c} {0.0064}& \multicolumn{1}{c} {0.0029} & \multicolumn{1}{c} {0.0052} & \multicolumn{1}{c}{0.0029}  \\
& Median & \multicolumn{1}{c}{0.0080} & \multicolumn{1}{c} {0.0043} & \multicolumn{1}{c} {0.0060} & \multicolumn{1}{c} {0.0028}  & \multicolumn{1}{c} {0.0123} & \multicolumn{1}{c} {0.0065}& \multicolumn{1}{c} {0.0029}& \multicolumn{1}{c} {0.0055} & \multicolumn{1}{c}{0.0029}   \\
&  Max & \multicolumn{1}{c}{0.0091} & \multicolumn{1}{c} {0.0046} & \multicolumn{1}{c} {0.0066} & \multicolumn{1}{c} {0.0034}  & \multicolumn{1}{c} {0.0126} & \multicolumn{1}{c} {0.0066}& \multicolumn{1}{c} {0.0033}& \multicolumn{1}{c} {0.0058} & \multicolumn{1}{c}{0.0033} \\
& SD & \multicolumn{1}{c}{0.0006} & \multicolumn{1}{c} {0.0005} & \multicolumn{1}{c} {0.0004} & \multicolumn{1}{c} {0.0005}  & \multicolumn{1}{c} {0.0002} & \multicolumn{1}{c} {0.0001}& \multicolumn{1}{c} {0.0003}& \multicolumn{1}{c} {0.0006} & \multicolumn{1}{c}{0.0002}\\
\hline
\multicolumn{11}{c}{~~~ETH Lecturer Evaluation Dataset} \\
\hline
& ~  & $b_0$ & $b_1$ & $b_2$ & $b_3$ & $\sigma^2_\alpha$ & $\sigma^2_\beta$ & $\sigma^2_e$ & ~  & ~  \\
\hline
& $W_2$ barycenter & \multicolumn{1}{c}{0.0043} & \multicolumn{1}{c} {0.0018} & \multicolumn{1}{c} {0.0008} & \multicolumn{1}{c} {0.0024}  & \multicolumn{1}{c} {0.0078} & \multicolumn{1}{c} {0.0142}& \multicolumn{1}{c} {0.0070}  \\
& Mean & \multicolumn{1}{c}{0.0087} & \multicolumn{1}{c} {0.0014} & \multicolumn{1}{c} {0.0011} & \multicolumn{1}{c} {0.0028}  & \multicolumn{1}{c} {0.0084} & \multicolumn{1}{c} {0.0144}& \multicolumn{1}{c} {0.0077} & ~ & ~  \\
& Median & \multicolumn{1}{c}{0.0088} & \multicolumn{1}{c} {0.0014} & \multicolumn{1}{c} {0.0011} & \multicolumn{1}{c} {0.0025}  & \multicolumn{1}{c} {0.0076} & \multicolumn{1}{c} {0.0145}& \multicolumn{1}{c} {0.0074}& ~ & ~  \\
&  Max & \multicolumn{1}{c}{0.0116} & \multicolumn{1}{c} {0.0017} & \multicolumn{1}{c} {0.0015} & \multicolumn{1}{c} {0.0038}  & \multicolumn{1}{c} {0.0124} & \multicolumn{1}{c} {0.0194}& \multicolumn{1}{c} {0.0103}& ~ & ~ \\
& SD & \multicolumn{1}{c}{0.0014} & \multicolumn{1}{c} {0.0002} & \multicolumn{1}{c} {0.0003} & \multicolumn{1}{c} {0.0007}  & \multicolumn{1}{c} {0.0021} & \multicolumn{1}{c} {0.0025}& \multicolumn{1}{c} {0.0014}& ~ & ~ \\
\hline
\end{tabular}
}

\end{table}

\vspace{2cm}

\newpage

\section{Proof of Theorem \ref{th1}} \label{sec:proof.thm1}

In the proof below, we use $\|x\|_2$ to denote the Euclidean norm of a vector $x$. For a generic matrix $A=(A_{ij})_{1\leqslant i\leqslant n_1, 1\leqslant j\leqslant n_2}$, let $\lambda_{\max}(A)$ and $\lambda_{\min}(A)$ denote the maximal and minimal eigenvalues of a generic square matrix $A$, $\|A\|_2 = \sqrt{\lambda_{\max}(A^\top A)}$ be the matrix operator norm, and $\|A\|_F=\sqrt{\sum_{i=1}^{n_1}\sum_{j=1}^{n_2}A_{ij}^2}$ be the Frobenius norm of $A$.

Let $\Bcal(\ub, \sfr)$ denote the Euclidean of radius $\sfr>0$ centered at $\ub \in \RR^{p+3}$. For any two generic probability measures $P_1,P_2$, we use $\|P_1 - P_2 \|_{\TV}$ and $D_\text{{KL}}(P_1,  P_2)$ to denote the total variation distance and the Kullback-Leibler divergence from $P_1$ and $P_2$.

\begin{lemma}\label{lem:upper.bound}
Suppose that Assumptions \ref{assump:asymp}, \ref{assump:bound} and \ref{assump:initial} hold. Let $\sfc_0 = C_y +  C_x B_0 p + 1$. Define the event
\begin{align} \label{eq:Gcal}
&\Gcal_N = \Bigg\{ \big|\alpha_{s_i,k}^{(t)}\big| \leqslant 2\sfc_0(r+c)(\log N)^{A_1+B_1+3E_1/2+1}, \text{ and } \nonumber \\
& \big|\beta_{q_j,k}^{(t)}\big| \leqslant 2\sfc_0(r+c)(\log N)^{A_1+B_1+3E_1/2+1}, \text{ for all } i=1,\ldots,r,~ j=1,\ldots,c , \nonumber \\
&  t=1,\ldots,T,~ k=1,\ldots,m, \text{and } |Y_{ij}|\leqslant C_y\log N, \text{ for all } 1\leqslant i\leqslant R, 1\leqslant j\leqslant C \Bigg\}.
\end{align}
Then for all sufficiently large $N$,
\begin{align} \label{eq:Gcal.tail}
& \PP (\Gcal_N) \geqslant 1 - (Tmr + Tmc + \underline c^{-1} N) \exp\left\{-(1/2)\log^2 N\right\}.
\end{align}
\end{lemma}

\begin{proof}[Proof of Lemma \ref{lem:upper.bound}]
Let $\Gcal_{1N}=\big\{|Y_{ij}|\leqslant C_y\log N, \text{ for all } 1\leqslant i\leqslant R, 1\leqslant j\leqslant C \big\}$. Then by Assumption \ref{assump:asymp} and Assumption \ref{assump:bound}, a simple union bound implies that as $N\to\infty$,
\begin{align} \label{eq:E1n.bound}
\PP\big(\Gcal_{1N}^c\big)\leqslant \underline c^{-1} N \exp\left\{-(1/2)\log^2 N\right\} \to 0 .
\end{align}

Next we turn to the random effects of $\big\{\alpha_{s_i,k}^{(t)},\beta_{q_j,k}^{(t)}:t=1,\ldots,T, k=1,\ldots,m\big\}$ in Algorithm \ref{algo:pigeonhole SGLD}. Define the numbers
\begin{align} \label{eq:abc.lem4}
& \sfc_{1N}  = \frac{1}{1+(\log N)^{-(A_1+E_1)}/c}, \quad \sfc_{2N}  = \frac{1}{1+(\log N)^{-(B_1+E_1)}/r}, \nonumber \\
& A_{3N} = \frac{\sfc_0\sfc_{1N} + \sfc_0}{1-\sfc_{1N}\sfc_{2N}}, \quad B_{3N} = \frac{\sfc_0\sfc_{2N} + \sfc_0}{1-\sfc_{1N}\sfc_{2N}} .
\end{align}
Notice that by Assumption \ref{assump:asymp}, $r$ and $c$ are constants, which implies that $\sfc_0>0, \sfc_{1N}\in (0,1) ,\sfc_{2N}\in (0,1)$, and $A_{3N}\to +\infty,B_{3N}\to +\infty$ as $N\to\infty$. Furthermore, by definition, it is straightforward to verify that $A_{3N}$ and $B_{3N}$ satisfy
\begin{align} \label{eq:A3B3}
& A_{3N} = \sfc_0 + \sfc_{1N} B_{3N}, \quad B_{3N} = \sfc_0 + \sfc_{2N} A_{3N}.
\end{align}
Our goal is to first show that on the event $\Gcal_{1N}$, for every $s_i$ ($i=1,\ldots,r$), every $q_j$ ($j=1,\ldots,c$), every $t=1,\ldots,T$, every $k=1,\ldots,m$, and for all sufficiently large $N$,
\begin{align}
& \PP \left( \big|\alpha_{s_i,k}^{(t)}\big| > A_{3N} (\log N)^{E_1/2+1} , \text{ and } \Gcal_{1N} \right) \leqslant \exp\left\{-(1/2)\log^2 N\right\} , \label{eq:Gtail.1}  \\
& \PP \left( \big|\beta_{q_j,k}^{(t)}\big| > B_{3N} (\log N)^{E_1/2+1} , \text{ and } \Gcal_{1N} \right) \leqslant \exp\left\{-(1/2)\log^2 N\right\} . \label{eq:Gtail.2}
\end{align}
We prove \eqref{eq:Gtail.1} and \eqref{eq:Gtail.2} by induction. The initial values of $\big\{\alpha_i^{(0)},\beta_j^{(0)}:~i=1,\ldots,R, ~j=1\ldots,C\big\}$ obviously satisfy \eqref{eq:Gtail.1} and \eqref{eq:Gtail.2} since they are finite numbers and they must be smaller than $\sfc_0 \log N + A_{3N}$ and $\sfc_0 \log N + B_{3N}$ in absolute value for all sufficiently large $N$. Now suppose that \eqref{eq:Gtail.1} and \eqref{eq:Gtail.2} hold true for all draws of $\alpha$'s and $\beta$'s before $\alpha_{s_i,k}^{(t+1)}$ (we assume without loss of generality that $\alpha$'s are drawn first and $\beta$'s are drawn second at each iteration of $k=1,\ldots,m$ and $t=1,\ldots,T$). Then according to the first updating equation in \eqref{eq:conditional_d}, we have that
\begin{align} \label{eq:alphat.1}
& \PP \Bigg( \left|\alpha_{s_i,k}^{(t+1)} - \frac{\sum_{j=1}^{c}Z^{(t)}_{s_iq_j} (Y^{(t)}_{s_iq_j} - x_{s_iq_j}^{(t) \top}b^{(t)} -  \beta_{q_j,k-1}^{(t)} ) \ee^{\eta_\alpha^{(t)}}}{n^{(t)}_{i \bullet} \ee^{\eta_\alpha^{(t)}} + \ee^{\eta_e^{(t)}} } \right| \nonumber \\
&\quad > \sqrt{\frac{ \ee^{\eta_\alpha^{(t)} + \eta_e^{(t)}} }{n^{(t)}_{i \bullet} \ee^{\eta_\alpha^{(t)}} + \ee^{\eta_e^{(t)}} }} \log N ~\Big|~ \theta^{(t)}, \betab_n^{(t)}, \Yb_n^{(t)} \Bigg) \nonumber \\
& \leqslant \exp\left\{-(1/2)\log^2 N\right\} .
\end{align}
We notice that on the event $\Gcal_{1N}$, on the parameter set $\Theta_N$, for all sufficiently large $N$,
\begin{align} \label{eq:alphat2}
&\quad~ \left| \frac{\sum_{j=1}^{c}Z^{(t)}_{s_iq_j} (Y^{(t)}_{s_iq_j} - x_{s_iq_j}^{(t) \top}b^{(t)} -  \beta_{q_j,k-1}^{(t)} ) \ee^{\eta_\alpha^{(t)}}}{n^{(t)}_{i \bullet} \ee^{\eta_\alpha^{(t)}} + \ee^{\eta_e^{(t)}} } \right| + \sqrt{\frac{ \ee^{\eta_\alpha^{(t)} + \eta_e^{(t)}} }{n^{(t)}_{i \bullet} \ee^{\eta_\alpha^{(t)}} + \ee^{\eta_e^{(t)}} }} \log N \nonumber \\
&\leqslant  \frac{\sum_{j=1}^{c}Z^{(t)}_{s_iq_j} \left(C_y \log N +  C_x B_0 p \log N +  \big|\beta_{q_j,k-1}^{(t)}\big| \right) \ee^{\eta_\alpha^{(t)}}}{n^{(t)}_{i \bullet} \ee^{\eta_\alpha^{(t)}} + \ee^{\eta_e^{(t)}} }  + \sqrt{\frac{ \ee^{\eta_\alpha^{(t)} + \eta_e^{(t)}} }{n^{(t)}_{i \bullet} \ee^{\eta_\alpha^{(t)}} + \ee^{\eta_e^{(t)}} }} \log N   \nonumber \\
&\stackrel{(i)}{\leqslant}  \frac{\left(C_y  +  C_x B_0 p\right)\log N + B_{3N} (\log N)^{E_1/2+1}}{1 + \ee^{\eta_e^{(t)}-\eta_\alpha^{(t)}}/n^{(t)}_{i \bullet}}  + \sqrt{\frac{ 1 }{n^{(t)}_{i \bullet}\ee^{-\eta_e^{(t)}} + \ee^{-\eta_{\alpha}^{(t)}} }} \log N   \nonumber \\
&\stackrel{(ii)}{\leqslant} \left(C_y  +  C_x B_0 p\right) \log N + \frac{B_{3N} (\log N)^{E_1/2+1}} {1+\ee^{-(A_1+E_1)\log\log N}/c} + (\log N)^{\min(A_1,E_1)/2+1} \nonumber \\
&\leqslant (\sfc_0 + \sfc_{1N} B_{3N}) (\log N)^{E_1/2+1} \stackrel{(iii)}{=} A_{3N} (\log N)^{E_1/2+1} ,
\end{align}
where $(i)$ follows from the definition of $\Gcal_{1N}$ and the induction assumption $\big|\beta_{q_j,k-1}^{(t)}\big| \leqslant B_{3N}(\log N)^{E_1/2+1}$, $(ii)$ follows because $n^{(t)}_{i \bullet}\leqslant c$ for all $i=1,\ldots,r$ and all $t$, $|\eta_e^{(t)}|\leqslant E_1 \log \log N$, $|\eta_\alpha^{(t)}|\leqslant A_1 \log \log N$ on $\Theta_N$, and $(iii)$ follows from \eqref{eq:A3B3}. Therefore, \eqref{eq:alphat.1} and \eqref{eq:alphat2} with the triangle inequality imply that $\PP \big( \big|\alpha_{s_i,k}^{(t)}\big| > A_{3N} (\log N)^{E_1/2+1}$, and $\Gcal_{1N} ~\big|~ \theta^{(t)}, \betab_n^{(t)}, \Yb_n^{(t)}\big) \leqslant \exp\left\{-(1/2)\log^2 N\right\}$. Since this upper bound is data-free, the law of iterated expectation implies \eqref{eq:Gtail.1} for $\alpha_{s_i,k}^{(t)}$. The inequality \eqref{eq:Gtail.2} for $\beta_{q_j,k}^{(t)}$ can be proved similarly.

From the definition of $A_{3N}$ and $B_{3N}$ in \eqref{eq:abc.lem4}, we can further upper bound them by
\begin{align*}
A_{3N} & \leqslant 2\sfc_0\left(\frac{1 }{(\log N)^{-(A_1+E_1)}/c}+ \frac{1}{(\log N)^{-(B_1+E_1)}/r} \right)  \leqslant 2\sfc_0 (r+c) (\log N)^{A_1+B_1+E_1},
\end{align*}
and similarly $B_{3N}\leqslant 2\sfc_0 (r+c) (\log N)^{A_1+B_1+E_1}$. Therefore, we can combine \eqref{eq:E1n.bound}, \eqref{eq:Gtail.1} and \eqref{eq:Gtail.2} and apply a simple union bound to obtain \eqref{eq:Gcal.tail}.
\end{proof}

Throughout the rest of the proof, we will always condition on the event $\Gcal_N$ defined in \eqref{eq:Gcal} which happens with large probability according to Lemma \ref{lem:upper.bound}. In particular, for the expectation of the latent variables $\vartheta$ conditional on $\theta,Y_n$, we will use $\overline \pi (\vartheta\mid \theta, Y_n) \propto \pi (\vartheta\mid \theta, Y_n) \mathbbm{1}(\Gcal_N)$ to denote its restricted density on the set $\Gcal_N$. For Algorithms \ref{algo:SGLD} and \ref{algo:pigeonhole SGLD}, we only need to consider their restricted versions on the set $\Gcal_N$ and finally combine the conclusions with the probabilistic statement in Lemma \ref{lem:upper.bound}.

\subsection{Technical Lemmas on the Gradients} \label{subsec:lem.grad}
We derive several technical lemmas on the bounds for the gradients. Since all the quantities of concern here are from the same iteration in the stochastic gradient MCMC algorithms, we will suppress the superscript $(t)$ which indicates the quantities in the $t$th iteration to ease the notation. For example, $Y_n^{(t)}, \theta^{(t)}, \vartheta^{(t)}, \Ecal^{(t)}$ will be written as $Y_n,\theta,\vartheta,\Ecal$, etc. We use $\vartheta$ to denote the vector of all latent variables of row random effect $\alpha$'s and column random effect $\beta$'s. For a subset $Y_n$, we define
\begin{align} \label{eq:g.func.def}
g_{\vartheta}(\theta, Y_n) & = \big(g_{\vartheta 1}(b,Y_n)^\top, g_{\vartheta 2}(\eta_{\alpha},Y_n), g_{\vartheta 3}(\eta_{\beta},Y_n), g_{\vartheta 4}(\eta_e,Y_n)\big)^\top, \\
g_{\vartheta 1}(b,Y_n) & = - \frac{1}{m}\sum^m_{k=1} \left[\frac{N}{n} \nabla_{b} \log p(Y_n \mid \theta,\vartheta_k) + \nabla_{b} \log \pi(b,\vartheta_k) \right] , \nonumber \\
& = - \frac{N}{mn}\sum_{k=1}^m \sum^r_{i=1}\sum^c_{j=1} x_{s_iq_j} (Y_{s_iq_j} - \alpha_{s_i,k} - \beta_{q_j,k} - x_{s_iq_j}^\top b)Z_{s_iq_j} \ee^{-\eta_e} \nonumber \\
g_{\vartheta 2}(\eta_{\alpha},Y_n) & = - \frac{1}{m}\sum^m_{k=1} \left[\frac{N}{n} \nabla_{\eta_{\alpha}} \log p(Y_n \mid \theta,\vartheta_k) + \frac{R}{r}\nabla_{\eta_{\alpha}} \log \pi(\vartheta_k \mid \eta_{\alpha}) + \nabla_{\eta_\alpha} \log \pi(\eta_\alpha)\right] \nonumber \\
& = (R/2 + \afrak_1) - \frac{1}{m}\sum_{k=1}^m \left(\frac{R}{r}\sum_{i=1}^r \alpha_{s_i,k}^2/2 + \bfrak_1 \right) \ee^{-\eta_{\alpha}} ,  \nonumber \\
g_{\vartheta 3}(\eta_{\beta},Y_n) & =  - \frac{1}{m}\sum^m_{k=1} \left[\frac{N}{n} \nabla_{\eta_{\beta}} \log p(Y_n \mid \theta,\vartheta_k) + \frac{C}{c}\nabla_{\eta_{\beta}} \log \pi(\vartheta_k \mid \eta_\beta) + \nabla_{\eta_\beta}\log \pi(\eta_\beta)\right] \nonumber \\
& = (C/2 + \afrak_2) - \frac{1}{m}\sum_{k=1}^m \left(\frac{C}{c}\sum_{j=1}^c \beta_{q_j,k}^2/2 + \bfrak_2  \right) \ee^{-\eta_{\beta}} ,  \nonumber \\
g_{\vartheta 4}(\eta_e,Y_n) & = - \frac{1}{m}\sum^m_{k=1} \left[\frac{N}{n} \nabla_{\eta_e} \log p(Y_n \mid \theta,\vartheta_k) + \nabla_{\eta_e} \log \pi(\eta_e,\vartheta_k) \right] \nonumber \\
& = (N/2 + \afrak_3) - \frac{1}{m}\sum_{k=1}^m \Bigg[\frac{N}{2n}\sum_{i=1}^r \sum_{j=1}^c (Y_{s_iq_j} - \alpha_{s_i,k} - \beta_{q_j,k} - x_{s_iq_j}^\top b)^2 Z_{s_iq_j} + \bfrak_3 \Bigg]\ee^{-\eta_e} . \nonumber
\end{align}
where $\{\vartheta_1,\ldots,\vartheta_m\}$ denote the length-$m$ Markov chain of latent variables $(\alphab_n,\betab_n)$ sampled from $\pi(\alphab_n, \betab_n \mid \theta, Y_n)$ in Step (b) of Algorithm \ref{algo:pigeonhole SGLD}.

\begin{lemma}\label{lea2}
Suppose that Assumptions \ref{assump:asymp}, \ref{assump:bound}, \ref{assump:eigen} and \ref{assump:initial} hold. There exists a constant
\begin{align} \label{eq:L.def}
L & = \big[C_xp\{C_y +4\sfc_0(r+c)+C_x\} + C_x^2B_0p^2 \big] + \left[\sfc_0^2(r+c)^2 + \bfrak_1 \right]   \nonumber \\
&\quad + \left[\sfc_0^2(r+c)^2 + \bfrak_2 \right] + \left[ \left\{C_y+C_xB_0p + 4\sfc_0(r+c) \right\}^2/2 + \bfrak_3\right] ,
\end{align}
with $\sfc_0=C_y+C_xB_0p+1$, such that for any subset $Y_n$, any $\theta, \theta' \in \Theta_N$ and all sufficiently large $N$, on the event $\Gcal_N$ as defined in \eqref{eq:Gcal}, the functions $g_{\vartheta}(\theta, Y_n)$ and $g_\vartheta(\theta', Y_n)$ in \eqref{eq:g.func.def} satisfy
\begin{align}
& \|g_{\vartheta}(\theta, Y_n) - g_{\vartheta}(\theta', Y_n)\|_2 \leqslant L \|\theta - \theta'\|_2 N (\log N)^{3A_1+3B_1+4E_1+2}, \text{ and }  \label{eq:g.diff1} \\
& {\EE}_{\pi(\vartheta \mid \theta, \Yb_n)} \left\{\left[g_{\vartheta}(\theta, Y_n) - g_{\vartheta}(\theta', Y_n)\right]\cdot \mathbbm{1}(\Gcal_N) \right\} \leqslant L \|\theta - \theta'\|_2 N (\log N)^{3A_1+3B_1+4E_1+2}. \label{eq:g.diff2}
\end{align}
\end{lemma}

\begin{proof}[Proof of Lemma \ref{lea2}]
Since $g_{\vartheta}(\theta, Y_n)$ is a $(p+3)$-dimensional differentiable function, $\nabla_{\theta}g_{\vartheta}(\theta, Y_n)$ is a $(p+3)\times (p+3)$ matrix. Since $\theta=(b^\top,\eta_{\alpha},\eta_{\beta},\eta_e)^\top$, we divide the rows and columns of the $(p+3)\times (p+3)$ matrix $\nabla_{\theta}g_{\vartheta}(\theta, Y_n)$ accordingly into blocks, such that the row and column 1 to $p$ correspond to $b$, and the $p+1,p+2,p+3$th row and column correspond to $\eta_{\alpha},\eta_{\beta},\eta_e$, respectively.  By the definition of $g_{\vartheta}(\theta, Y_n)$ in \eqref{eq:g.func.def} and the model specified in \eqref{eq:lme} and \eqref{eq:normalre}, it is straightforward to calculate that
\begin{align*}
& \nabla_{\theta}g_{\vartheta}(\theta, Y_n) = {\left[ \begin{array}{cccc}
\big(\nabla_{\theta}g_{\vartheta}(\theta, Y_n)\big)_{11} & 0 & 0 & \big(\nabla_{\theta}g_{\vartheta}(\theta, Y_n)\big)_{14} \\
0 & \big(\nabla_{\theta}g_{\vartheta}(\theta, Y_n)\big)_{22} & 0 & 0\\
0 & 0 & \big(\nabla_{\theta}g_{\vartheta}(\theta, Y_n)\big)_{33} &0 \\
\big(\nabla_{\theta}g_{\vartheta}(\theta, Y_n)\big)_{41}& 0 & 0 & \big(\nabla_{\theta}g_{\vartheta}(\theta, Y_n)\big)_{44}
\end{array}
\right ]}, \\
&\text{where} \\
& \big(\nabla_{\theta}g_{\vartheta}(\theta, Y_n)\big)_{11} = \frac{N}{n} \ee^{-\eta_e} \sum_{i=1}^r \sum_{j=1}^c Z_{s_iq_j} x_{s_iq_j}x_{s_iq_j}^\top , \\
& \big(\nabla_{\theta}g_{\vartheta}(\theta, Y_n)\big)_{14} = \big(\nabla_{\theta}g_{\vartheta}(\theta, Y_n)\big)_{41}^\top \nonumber \\
&\qquad \qquad \qquad \quad = \frac{N}{nm} \sum_{k=1}^m\sum_{i=1}^r\sum_{j=1}^c x_{s_iq_j} \big(Y_{s_iq_j} - \alpha_{s_i,k} - \beta_{q_j,k} - x_{s_iq_j}^\top b \big) Z_{ij} \ee^{-\eta_e}, \\
& \big(\nabla_{\theta}g_{\vartheta}(\theta, Y_n)\big)_{22} =  \frac{1}{m}\sum_{k=1}^m \left(\frac{R}{2r}\sum_{i=1}^r\alpha_{s_i,k}^2 + \bfrak_1 \right) \ee^{-\eta_{\alpha}}, \\
& \big(\nabla_{\theta}g_{\vartheta}(\theta, Y_n)\big)_{33} =  \frac{1}{m}\sum_{k=1}^m \left(\frac{C}{2c}\sum^c_{j=1}\beta_{q_j,k}^2 + \bfrak_2 \right) \ee^{-\eta_{\beta}}, \\
& \big(\nabla_{\theta}g_{\vartheta}(\theta, Y_n)\big)_{44}  =  \frac{1}{m}\sum_{k=1}^m \Bigg[\frac{N}{2n}\sum_{i=1}^r \sum_{j=1}^c (Y_{s_iq_j} - \alpha_{s_i,k} - \beta_{q_j,k} - x_{s_iq_j}^\top b)^2 Z_{s_iq_j} + \bfrak_3 \Bigg]\ee^{-\eta_e} .
\end{align*}
On the event $\Gcal_N$ defined in \eqref{eq:Gcal}, we have the following upper bound in the Frobenius norm for each term above:
\begin{align*}
\left\| \big(\nabla_{\theta}g_{\vartheta}(\theta, Y_n)\big)_{11} \right\|_F  &\leqslant N C_x^2p (\log N)^{E_1},  \\
\left\| \big(\nabla_{\theta}g_{\vartheta}(\theta, Y_n)\big)_{14} \right\|_F  &\leqslant \big[C_xp\{C_y\log N +4\sfc_0(r+c)(\log N)^{A_1+B_1+3E_1/2+1}\} + C_x^2B_0p^2\log N \big]  \\
&\quad \times (\log N)^{E_1} N \\
&\leqslant \big[C_xp\{C_y +4\sfc_0(r+c)\} + C_x^2B_0p^2 \big] N(\log N)^{A_1+B_1+5E_1/2+1} , \\
\left| \big(\nabla_{\theta}g_{\vartheta}(\theta, Y_n)\big)_{22} \right| & \leqslant \left(2R \sfc_0^2(r+c)^2(\log N)^{2A_1+2B_1+3E_1+2} + \bfrak_1 \right) (\log N)^{A_1} \\
&\leqslant \left[\sfc_0^2(r+c)^2 + \bfrak_1 \right] N (\log N)^{3A_1+2B_1+3E_1+2}, \\
\left| \big(\nabla_{\theta}g_{\vartheta}(\theta, Y_n)\big)_{33} \right| & \leqslant \left(2C \sfc_0^2(r+c)^2(\log N)^{2A_1+2B_1+3E_1+2} + \bfrak_2\right) (\log N)^{B_1} \\
& \leqslant \left[\sfc_0^2(r+c)^2 + \bfrak_2 \right] N (\log N)^{2A_1+3B_1+3E_1+2}   ,\\
\left| \big(\nabla_{\theta}g_{\vartheta}(\theta, Y_n)\big)_{44} \right| & \leqslant  \Big[ \left\{(C_y+C_xB_0p)\log N + 4\sfc_0(r+c)(\log N)^{A_1+B_1+3E_1/2+1} \right\}^2 N/2 \\
&\quad  + \bfrak_3\Big] (\log N)^{E_1} \\
& \leqslant \left[ \left\{C_y+C_xB_0p + 4\sfc_0(r+c) \right\}^2/2 + \bfrak_3\right] N (\log N)^{2A_1+2B_1+4E_1+2} .
\end{align*}
With $L$ defined in \eqref{eq:L.def}, we have that for any $\theta,\theta' \in \Theta_N$, on the event $\Gcal_N$,
\begin{align*}
& \|g_{\vartheta}(\theta, Y_n) - g_{\vartheta}(\theta', Y_n)\|_2 \leqslant \big\| \nabla_{\theta}g_{\vartheta}(\theta, Y_n) \big\|_2 \|\theta - \theta'\|_2 \\
& \leqslant  \big\| \nabla_{\theta}g_{\vartheta}(\theta, Y_n) \big\|_F \|\theta - \theta'\|_2 \leqslant L \|\theta - \theta'\|_2 N (\log N)^{3A_1+3B_1+4E_1+2}.
\end{align*}
This proves \eqref{eq:g.diff1}. Then we can simply take the posterior conditional expectation with respect to all the latent variables of $\alpha$'s and $\beta$'s (whose distribution is $\pi(\vartheta\mid \theta,\Yb_n)$) on the event $\Gcal_N$ to obtain \eqref{eq:g.diff2}.
\end{proof}

\vspace{3mm}

\begin{lemma} \label{leb1}
Suppose that Assumptions \ref{assump:asymp}, \ref{assump:bound}, \ref{assump:eigen} and \ref{assump:initial} hold. For any $ \theta \in \Theta_N$ and any subset $Y_n$, there exists a constant
\begin{align} \label{eq:M0}
M_0 & =  2C_xp[C_y+C_xB_0p + 4\sfc_0(r+c) ] + [1+2\afrak_1+2\sfc_0^2(r+c)^2+2\bfrak_1] \nonumber \\
& + [1+2\afrak_2+2\sfc_0^2(r+c)^2+2\bfrak_2] + [1+2\afrak_3+2\bfrak_2+ 2\{C_y+C_xB_0p + 4\sfc_0(r+c)\}^2] ,
\end{align}
with $\sfc_0=C_y+C_xB_0p+1$, such that on the event $\Gcal_N$, for all $\theta\in\Theta_N$ and all sufficiently large $N$,
\begin{align} \label{eq:lipschitz}
& \|g_{\vartheta}(\theta, Y_n)\|_2 \leqslant M_0 N (\log N)^{3A_1+3B_1+4E_1+2}  , \nonumber \\
& \|\nabla_{\theta} \log \pi(\theta\mid \vartheta, \Yb) \|_2 \leqslant M_0 N (\log N)^{3A_1+3B_1+4E_1+2} .
\end{align}
\end{lemma}

\begin{proof}[{Proof of Lemma \ref{leb1}}]
Based on the definition \eqref{eq:g.func.def} and the Cauchy-Schwarz inequality, we have that on the event $\Gcal_N$,
\begin{align*}
&\|g_{\vartheta}(\theta, Y_n)\|_2^2 \leqslant \left\| \frac{N}{mn}\sum_{k=1}^m \sum^r_{i=1}\sum^c_{j=1} x_{s_iq_j} (Y_{s_iq_j} - \alpha_{s_i,k} - \beta_{q_j,k} - x_{s_iq_j}^\top b)Z_{s_iq_j} \ee^{-\eta_e}\right\|_2^2  \nonumber \\
&\quad + \left\|  - (R/2 + \afrak_1) + \frac{1}{m}\sum_{k=1}^m \left(\frac{R}{r}\sum_{i=1}^r \alpha_{s_i,k}^2/2 + \bfrak_1 \right) \ee^{-\eta_{\alpha}} \right\|_2^2  \nonumber \\
&\quad + \left\| - (C/2 + \afrak_2) + \frac{1}{m}\sum_{k=1}^m \left(\frac{C}{c}\sum_{j=1}^c \beta_{q_j,k}^2/2 + \bfrak_2  \right) \ee^{-\eta_{\beta}}  \right\|_2^2 \nonumber \\
&\quad + \left\| - (N/2 + \afrak_3) + \frac{1}{m}\sum_{k=1}^m \Bigg[\frac{N}{2n}\sum_{i=1}^r \sum_{j=1}^c (Y_{s_iq_j} - \alpha_{s_i,k} - \beta_{q_j,k} - x_{s_iq_j}^\top b)^2 Z_{s_iq_j} + \bfrak_3 \Bigg]\ee^{-\eta_e} \right\|_2^2 \nonumber \\
&\leqslant \big\{N(\log N)^{E_1} C_xp\big[C_y\log N + 4\sfc_0(r+c) (\log N)^{A_1+B_1+3E_1/2+1} + C_xB_0p \log N\big] \big\}^2  \nonumber \\
&\quad + \big\{R/2 + \afrak_1 + \big[2R\sfc_0^2(r+c)^2 (\log N)^{2A_1+2B_1+3E_1+2} + \bfrak_1\big] (\log N)^{A_1}\big\}^2 \nonumber \\
&\quad + \big\{C/2 + \afrak_2 + \big[2C\sfc_0^2(r+c)^2 (\log N)^{2A_1+2B_1+3E_1+2} + \bfrak_2\big] (\log N)^{B_1}\big\}^2 \nonumber \\
&\quad + \big\{N/2 + \afrak_3 + \big[N\big\{C_y \log N + 4\sfc_0(r+c) (\log N)^{A_1+B_1+3E_1/2+1} + C_xB_0p \log N \big\}^2 \nonumber \\
&\quad + \bfrak_3\big](\log N)^{E_1} \big\}^2  \nonumber \\
&\leqslant M_0^2 N^2 (\log N)^{2(3A_1+3B_1+4E_1+2)},
\end{align*}
where the last inequality follows from the definition of $M_0$ in \eqref{eq:M0}. This proves the first relation in \eqref{eq:lipschitz}. The proof of the second relation in \eqref{eq:lipschitz} follows similarly since it is just a full data version of the first relation.
\end{proof}

\subsection{Technical Lemmas on SGLD for Non-Log-Concave Posterior} \label{subsec:lem.SGLD}

In this section, we prove several technical lemmas for showing the convergence of the SGLD for the non-log-concave posterior distribution in the crossed mixed effects model defined by \eqref{eq:lme} and \eqref{eq:normalre}. In particular, to prove the approximation from the output of the pigeonhole SGLD to the target posterior distribution, we introduce three auxiliary sequences of the projected SGLD $\big(\big\{\theta^{(t)}_{\text{Proj-SGLD}}\big\}^T_{t=1}\big)$, the $1/2$-lazy projected SGLD $\big(\big\{\theta^{(t)}_{\laz}\big\}^T_{t=1}\big)$, and the Metropolized SGLD $\big(\big\{\theta^{(t)}_{\text{MH}}\big\}^T_{t=1}\big)$, following the proof idea in \citet{zou2020faster}. These three auxiliary sequences are only utilized in theoretical analysis and not implemented in practice. Their convergence results will be presented in Lemma \ref{lem:lazy}, Lemma \ref{le1} and Lemma \ref{le234}, respectively, whose proof will depend on Lemma \ref{lea2} and Lemma \ref{leb1} in the previous section. For an overview, the approximations and their proofs are given in the following roadmap:
\begin{align*} 
&\text{PSGLD} \xrightarrow{\text{Lemma \ref{le1}}}
\text{projected SGLD} \xrightarrow{\text{Lemma \ref{lem:lazy}}} \text{$1/2$-lazy projected SGLD} \\
&\xrightarrow{\text{Lemma 6.4 of \citet{zou2020faster}}} \text{Metropolized SGLD} \xrightarrow{\text{Lemma \ref{le234}}} \text{Truncated posterior on } \Theta_N 
\end{align*}

Based on these auxiliary Markov processes, we develop convergence analysis below in Lemma \ref{le1}, Lemma \ref{lem:lazy} and Lemma \ref{le234} for the proof of Theorem \ref{th1}. We show that the total variation distance between the empirical distributions of the output from the PSGLD $\Pi_T$ (constrained to the parameter set $\Theta_N$ and the event $\Gcal_N$) and the projected SGLD $\Pi_T^{\text{Proj-SGLD}}$ can be made arbitrarily small in Lemma \ref{le1}. Then we show in Lemma \ref{lem:lazy} that with probability close to 1, the empirical distribution of the projected SGLD $\Pi_T^{\text{Proj-SGLD}}$ can be approximated by that of the $1/2$-lazy projected SGLD $\tilde \Pi_{T_{\laz}}^{\text{Proj-SGLD}}$ with a chain length $T_{\laz}\approx 2T$. Finally, Lemma \ref{le234} shows that the total variation distance between the empirical distribution of the output from the $1/2$-lazy projected SGLD $\tilde \Pi_{T_{\laz}}^{\text{Proj-SGLD}}$ and the posterior distribution $\Pi^*_N \propto \Pi(\ud \theta \mid \Yb) \mathbbm{1}(\theta\in \Theta_N)$ can be made arbitrarily small. Combining Lemmas \ref{le1}, \ref{lem:lazy} and \ref{le234}, we can show the convergence of the empirical distribution from the pigeonhole SGLD $\Pi_T$ to the target posterior distribution $\Pi^*_N$ by using the triangle inequality in Theorem \ref{th1}.

The projected SGLD adds an acceptance/rejection step at each iteration of the pigeonhole SGLD. The proposal from the pigeonhole SGLD will only be accepted if it falls in the set of $\Bcal(\theta^{(t)},\sfr) \cap \Theta_N$, i.e., in the projected SGLD algorithm, $\theta^{(t+1)} = \theta^{(t+1)}\mathbbm{1}\{\theta^{(t+1)} \in \Bcal(\theta^{(t)},\sfr) \cap \Theta_N \} + \theta^{(t)}\mathbbm{1}\{\theta^{(t+1)} \notin \Bcal(\theta^{(t)},\sfr) \cap \Theta_N \}$, where the radius $\sfr$ is given in \eqref{eq:sfr} below.

We introduce some additional notation. We will use $P(\cdot\mid \cdot)$ and $Q(\cdot\mid \cdot)$ to denote the transition distributions and $p(\cdot \mid \cdot)$ and $q(\cdot \mid \cdot)$ to denote the transition densities of Markov chains. To distinguish different parameter vectors, we will use $\ub,\vb,\wb$, which represent the parameter vector $\theta$ at different stages of the algorithm. In particular, we use $\ub = (u_b^\top, u_{\eta_{\alpha}},u_{\eta_{\beta}},u_{\eta_e} )^{\top}$ to denote the current parameter, $\vb =(v_b^\top, v_{\eta_{\alpha}},v_{\eta_{\beta}},v_{\eta_e} )^{\top}$ to denote the point after one step updating of the pigeonhole SGLD in Algorithm \ref{algo:pigeonhole SGLD}, and $\wb=(w_b^\top, w_{\eta_{\alpha}},w_{\eta_{\beta}},w_{\eta_e} )^{\top}$ to denote the point obtained after the acceptance/rejection step. The conditional distribution of $\vb$ given $\ub, \vartheta, Y_n$ is the normal distribution $N\big(\ub- (\Ecal/2)g_{\vartheta}(\ub, Y_n), ~ \Ecal\big)$  truncated to $\Theta_N$, whose density satisfies
\begin{align*}
p(\vb \mid \ub, Y_n) = \int_{\Gcal_N} p(\vb\mid \ub, \vartheta, Y_n)\overline \pi( \vartheta\mid \ub, Y_n)\ud \vartheta = \EE_{\overline \pi(\vartheta \mid \ub, Y_n)}[p(\vb\mid \ub, \vartheta, Y_n)],
\end{align*}
where $\Gcal_N$ is defined in \eqref{eq:Gcal}.

Now we also take into account the randomness in the selection of subset $Y_n$ (i.e., $\Yb_n$ from the full data $\Yb$). Let $\Scal =\{s_1,\ldots,s_r\}\otimes \{q_1,\ldots,q_c\}\subseteq \{1,\ldots,R\}\otimes \{1,\ldots,C\}$ denote the random index set associated with the subset data $\Yb_n$. The transition probability of the pigeonhole SGLD constrained to the space of $\Theta_N$ and the event $\Gcal_N$ defined in \eqref{eq:Gcal} is then $p(\vb\mid \ub)=  \EE_{\Scal}\left\{\EE_{\overline\pi(\vartheta \mid \ub, Y_n)}\left[p(\vb\mid \ub, \vartheta, Y_n)\right] \right\}$. The acceptance probability of the projected SGLD can be denoted by $p(\ub) = \PP_{\vb\sim P(\cdot\mid \ub)}\left( \vb \in \Bcal(\ub, \sfr) \cap \Theta_N \right)$. As a result, the full transition probability density of $\ub \rightarrow \wb$ is
\begin{align} \label{eq:proj.sgld.trans}
q(\wb \mid \ub)=[1-p(\ub)] \delta_{\ub}(\wb) + p(\wb \mid \ub) \cdot \mathbbm{1}[\wb \in \Bcal(\ub, \sfr) \cap \Theta_N],
\end{align}
where $\delta_{\ub}(\cdot)$ is the Dirac delta function at $\ub$.

For any $T\in \ZZ^+$, $\epsilon_{\max}>0$ and $\tau \in (0,1)$, we define
\begin{align} \label{eq:sfr}
\sfr & = 2\sqrt{{\epsilon}_{\max}} \big[ \sqrt{p+3} + \sqrt{2\{\log(8T/\tau)+(p+4)\log 2\}} \big] .
\end{align}

\begin{lemma} \label{le1}
Suppose that Assumptions \ref{assump:asymp}, \ref{assump:bound}, \ref{assump:eigen} and \ref{assump:initial} hold. Suppose that ${\epsilon}_{\max} \prec 1/[N^2(\log N)^{2(3A_1+3B_1+4E_1+2)}]$ as $N\to\infty$. For the distributions of the output from the pigeonhole SGLD $\Pi_T$ (constrained to the parameter set $\Theta_N$ and the event $\Gcal_N$) and those from the projected SGLD $\Pi_T^{\emph{Proj-SGLD}}$, for all sufficiently large $N$, it holds that
\begin{align*}
\left\|\Pi_T - \Pi_T^{\emph{Proj-SGLD}}\right\|_{\TV} \leqslant \frac{\tau}{8}.
\end{align*}
\end{lemma}

\begin{proof}[Proof of Lemma \ref{le1}]
We proceed using similar arguments to the proof of Lemma 6.1 in \cite{zou2020faster}. Let $\theta^{[T]}=\big\{\theta^{(t)}\big\}^{T}_{t=0}$ and $\theta_{\textup{Proj-SGLD}}^{[T]}=\big\{\theta^{(t)}_{\textup{Proj-SGLD}}\big\}^{T}_{t=0}$ denote the whole output vectors of the pigeonhole SGLD and the projected SGLD, respectively. The proof of Lemma 6.1 in \cite{zou2020faster} shows that for any $\tau \in (0,1)$, and any set $\Acal \subseteq \Theta_N$, if it holds that $\PP \left(\theta_{\textup{Proj-SGLD}}^{[T]} \ne \theta^{[T]}\right) \leqslant \tau/8$, then by the definition of total variation distance, we have that
\begin{align*}
\norm{\Pi_T-\Pi^{\text{Proj-SGLD}}_T}_{\TV} & = \sup_{\Acal \in \Theta}\left|\Pi_T(\Acal)- \Pi^{\text{Proj-SGLD}}_T(\Acal)\right| \\
&\leqslant \EE\left[\mathbbm{1}\left(\theta^{[T]} \ne \theta_{\textup{Proj-SGLD}}^{[T]}\right)\right] = \PP \left(\theta_{\textup{Proj-SGLD}}^{[T]} \ne \theta^{[T]}\right)  \leqslant \frac{\tau}{8}.
\end{align*}
Therefore, the main idea of proof is to show that the projected SGLD generates the same samples as those of the pigeonhole SGLD with probability at least $1-\tau/8$.

We show that uniformly for all $t=1,\ldots,T$, $\big\|\theta^{(t)}-\theta^{(t-1)}\big\|_2 \leqslant \sfr$ with probability at least $1-\tau/8$. From the pigeonhole SGLD updating equation in \eqref{eq:PSGLD} and the assumption that we truncate all $\theta^{(t)}$ to the sieve $\Theta_N$ (only for the theory in Section \ref{sec:theorem}), the updating equation for $\theta^{(t+1)}$ can be equivalently written as
\begin{align} \label{eq:theta.t+1.eq1}
&\theta^{(t+1)} = \theta^{(t)}  + \frac{\Ecal}{2} g_{\vartheta^{(t)}}(\theta^{(t)},Y_n^{(t)}) + \tilde \psi^{(t)} , \\
\text{where } & \tilde \psi^{(t)} = \psi^{(t)}\cdot \mathbbm{1}\left(\theta^{(t)}  + \frac{\Ecal}{2} g_{\vartheta^{(t)}}(\theta^{(t)},Y_n^{(t)}) +  \psi^{(t)}\in \Theta_N\right), \psi^{(t)}\sim N(0,\Ecal). \nonumber 
\end{align}
As such, the probability density function of the truncated normal random vector $\tilde \psi^{(t)}$, denoted by $f_{\tilde \psi^{(t)}}$, is given by
\begin{align}\label{eq:density.tilde.psi.1}
f_{\tilde \psi^{(t)}}(x) &= \frac{\mathbbm{1}\left(\theta^{(t)}   + \frac{\Ecal}{2} g_{\vartheta^{(t)}}(\theta^{(t)},Y_n^{(t)}) +  x \in \Theta_N\right)}{\PP(\theta^{(t)}   + \frac{\Ecal}{2} g_{\vartheta^{(t)}}(\theta^{(t)},Y_n^{(t)}) +  \psi^{(t)}\in \Theta_N)} \nonumber \\
&\quad \times \frac{1}{(2\pi)^{(p+3)/2}\dett(\Ecal)^{1/2}}\exp\left\{-\frac{1}{2}x^\top \Ecal^{-1} x\right\} .
\end{align}

We derive a lower bound for the probability $\PP(\theta^{(t)}   + \frac{\Ecal}{2} g_{\vartheta^{(t)}}(\theta^{(t)},Y_n^{(t)}) +  \psi^{(t)}\in \Theta_N)$ for $\psi^{(t)}\sim N(0,\Ecal)$. Recall from the definition of $\Theta_N$ in the main text that $\Theta_N$ is the rectangular region 
\begin{align} \label{eq:ThetaN.def2}
\Theta_N &= \Big\{\theta=(b^\top,\eta_{\alpha},\eta_{\beta},\eta_e)^\top \in \RR^{p+3}:~ \|b\|_{\infty}\leqslant B_0\log N, |\eta_{\alpha}| \leqslant A_1 \log\log N, \nonumber \\
&\qquad |\eta_{\beta}| \leqslant B_1 \log\log N, ~ |\eta_e| \leqslant E_1 \log\log N \Big\} .
\end{align}
We first check how large $\frac{\Ecal}{2} g_{\vartheta^{(t)}}(\theta^{(t)},Y_n^{(t)})$ is. Let $\sfs_{n,t}=\left\|\frac{\Ecal}{2} g_{\vartheta^{(t)}}(\theta^{(t)},Y_n^{(t)})\right\|_2$. By Lemma \ref{leb1}, 
\begin{align} \label{eq:grad.ub1}
& \frac{\sfs_{n,t}}{\sqrt{\epsilon_{\max}}} \leqslant \frac{\sqrt{\epsilon_{\max}}}{2}M_0 N(\log N)^{3A_1+3B_1+4E_1+2} = o(1),
\end{align}
as $N\to\infty$, given that ${\epsilon}_{\max} \prec 1/[N^2(\log N)^{2(3A_1+3B_1+4E_1+2)}]$.

To minimize the probability $\PP(\theta^{(t)}   + \frac{\Ecal}{2} g_{\vartheta^{(t)}}(\theta^{(t)},Y_n^{(t)}) +  \psi^{(t)}\in \Theta_N)$ for $\psi^{(t)}\sim N(0,\Ecal)$, we look for the worst case of placing the point $\theta^{(t)}$ inside the rectangle set $\Theta_N$, such that this probability is as small as possible. Because the $p+3$ components of $\psi^{(t)}\sim N(0,\Ecal)$ are independent, we only need to find the worst case for each marginal normal probability given that $\Theta_N$ is a rectangular region. For $j=1,\ldots,p+3$, let $\theta^{(t)}_j$ and $\sfg_j$ be the $j$th component of $\theta^{(t)}$ and $\frac{\Ecal}{2} g_{\vartheta^{(t)}}(\theta^{(t)},Y_n^{(t)})$, respectively.
Then we have that for all sufficiently large $N$,
\begin{align}\label{eq:psi.prob.lb1}
&\quad~ \PP\left(\theta^{(t)}   + \frac{\Ecal}{2} g_{\vartheta^{(t)}}(\theta^{(t)},Y_n^{(t)}) +  \psi^{(t)}\in \Theta_N \right) \nonumber\\
&\overset{(i)}{=} \prod_{j=1}^p \PP\left( -B_0\log N - \theta^{(t)}_j - \sfg_j \leqslant \psi^{(t)}_j \leqslant B_0\log N - \theta^{(t)}_j - \sfg_j\right)  \nonumber \\
&\qquad \times \PP\left( -A_1\log\log N - \theta^{(t)}_{p+1} - \sfg_{p+1} \leqslant \psi^{(t)}_{p+1}\leqslant A_1\log\log N - \theta^{(t)}_{p+1} - \sfg_{p+1}  \right)  \nonumber \\
&\qquad \times \PP\left( -B_1\log\log N - \theta^{(t)}_{p+2} - \sfg_{p+2} \leqslant \psi^{(t)}_{p+2}\leqslant B_1\log\log N - \theta^{(t)}_{p+2} - \sfg_{p+2}  \right)  \nonumber \\
&\qquad \times \PP\left( -E_1\log\log N - \theta^{(t)}_{p+3} - \sfg_{p+3} \leqslant \psi^{(t)}_{p+3}\leqslant E_1\log\log N - \theta^{(t)}_{p+3} - \sfg_{p+3}  \right)  \nonumber \\
&\overset{(ii)}{\geqslant} \prod_{j=1}^p \PP\left( -\theta^{(t)}_j -[B_0\log N -  \sfs_{n,t}]  \leqslant \psi^{(t)}_j \leqslant -\theta^{(t)}_j + [B_0\log N -  \sfs_{n,t}] \right)  \nonumber \\
&\qquad \times \PP\left( - \theta^{(t)}_{p+1} - [A_1\log\log N - \sfs_{n,t}]  \leqslant \psi^{(t)}_{p+1}\leqslant - \theta^{(t)}_{p+1} + [A_1\log\log N - \sfs_{n,t}]  \right)  \nonumber \\
&\qquad \times \PP\left( - \theta^{(t)}_{p+2} - [B_1\log\log N - \sfs_{n,t}] \leqslant \psi^{(t)}_{p+2}\leqslant - \theta^{(t)}_{p+2} + [B_1\log\log N - \sfs_{n,t}] \right)  \nonumber \\
&\qquad \times \PP\left( - \theta^{(t)}_{p+3}  - [E_1\log\log N - \sfs_{n,t}]  \leqslant \psi^{(t)}_{p+3}\leqslant - \theta^{(t)}_{p+3} + [E_1\log\log N - \sfs_{n,t}]  \right) ,
\end{align}
where $(i)$ follows from the independence among the $p+3$ components of $\psi^{(t)}\sim N(0,\Ecal)$ and the rectangular shape of $\Theta_N$, and $(ii)$ follows from the fact that $|\sfg_j|\leqslant \sfs_{n,t}$ for all $j=1,\ldots,p+3$ and $\sfs_{n,t}=o(1)$ by \eqref{eq:grad.ub1}. Now we only need to choose $\theta^{(t)} =(\theta^{(t)}_1,\ldots,\theta^{(t)}_{p+3})\in \Theta_N$ such that the right-hand side of \eqref{eq:psi.prob.lb1} is minimized. Given that $\sfs_{n,t}=o(1)$ as $N\to\infty$ and each $\psi^{(t)}_j$ for $j=1,\ldots,p+3$ is a normal random variable centered at zero, it is straightforward to see that the terms $B_0\log N -  \sfs_{n,t}$, $A_1\log\log N - \sfs_{n,t}$, $B_1\log\log N - \sfs_{n,t}$ and $E_1\log\log N - \sfs_{n,t}$ are all positive for sufficiently large $N$, and that the worst case happens when $\theta^{(t)}$ is placed at one of the vertices of $\Theta_N$, such that each probability on the right-hand side of \eqref{eq:psi.prob.lb1} is minimized. Given the symmetric shape of $\Theta_N$, without loss of generality, we can take $\theta^{(t)}_j=-B_0\log N$ for $j=1,\ldots,p$, $\theta^{(t)}_{p+1}=-A_1\log\log N$, $\theta^{(t)}_{p+2}=-B_1\log\log N$, and $\theta^{(t)}_{p+3}=-E_1\log\log N$, such that from \eqref{eq:psi.prob.lb1}, 
\begin{align}\label{eq:psi.prob.lb2}
&\quad~ \PP\left(\theta^{(t)}   + \frac{\Ecal}{2} g_{\vartheta^{(t)}}(\theta^{(t)},Y_n^{(t)}) +  \psi^{(t)}\in \Theta_N \right) \nonumber\\
&\geqslant \prod_{j=1}^p \PP\left( \sfs_{n,t} \leqslant \psi^{(t)}_j \leqslant 2B_0\log N - \sfs_{n,t}\right)   \times \PP\left( \sfs_{n,t} \leqslant \psi^{(t)}_{p+1}\leqslant 2A_1\log\log N - \sfs_{n,t}  \right)   \nonumber \\
&\qquad \times \PP\left(  \sfs_{n,t} \leqslant \psi^{(t)}_{p+2}\leqslant 2B_1\log\log N - \sfs_{n,t}  \right) \times \PP\left( \sfs_{n,t} \leqslant \psi^{(t)}_{p+3}\leqslant 2E_1\log\log N - \sfs_{n,t}  \right) \nonumber \\
&\overset{(i)}{\geqslant} \left[\Phi\left(\frac{2B_0\log N - \sfs_{n,t}}{\sqrt{\epsilon_{\max}}}\right) - \Phi\left(\frac{\sfs_{n,t}}{\sqrt{\epsilon_{\min}}}\right) \right]^p  \nonumber \\
&\qquad~~ \times   \left[\Phi\left(\frac{2A_1\log\log N - \sfs_{n,t}}{\sqrt{\epsilon_{\max}}}\right) - \Phi\left(\frac{\sfs_{n,t}}{\sqrt{\epsilon_{\min}}}\right) \right] \nonumber \\
&\qquad ~~\times  \left[\Phi\left(\frac{2B_1\log\log N - \sfs_{n,t}}{\sqrt{\epsilon_{\max}}}\right) - \Phi\left(\frac{\sfs_{n,t}}{\sqrt{\epsilon_{\min}}}\right) \right] \nonumber \\
&\qquad~~ \times  \left[\Phi\left(\frac{2E_1\log\log N - \sfs_{n,t}}{\sqrt{\epsilon_{\max}}}\right) - \Phi\left(\frac{\sfs_{n,t}}{\sqrt{\epsilon_{\min}}}\right) \right]\nonumber \\
&\overset{(ii)}{\geqslant} (1-o(1))\cdot 1/2^{p+3} \geqslant 1/2^{p+4},
\end{align}
where in $(i)$, $\Phi(\cdot)$ denotes the cumulative distribution function of $N(0,1)$, and $(i)$ follows because $\epsilon_{\min} \asymp \epsilon_{\max}$ by Assumption \ref{assump:initial}, and in each marginal probability, we make the standard deviation in the first $\Phi(\cdot)$ term as large as possible and make the standard derivation in the second $\Phi(\cdot)$ term as small as possible, such that the right-hand side of $(i)$ is a lower bound. The inequality $(ii)$ in \eqref{eq:psi.prob.lb2} follows from the following facts: By Assumption \ref{assump:initial} and the condition ${\epsilon}_{\max} \prec 1/[N^2(\log N)^{2(3A_1+3B_1+4E_1+2)}]$, as $N\to\infty$, $\epsilon_{\min} \asymp \epsilon_{\max} =o(1)$, so $B_0\log N/\sqrt{\epsilon_{\max}}\to +\infty$, $A_1\log\log N/\sqrt{\epsilon_{\max}}\to +\infty$, $B_1\log\log N/\sqrt{\epsilon_{\max}}\to +\infty$, $E_1\log\log N/\sqrt{\epsilon_{\max}}\to +\infty$; furthermore, $0\leqslant \sfs_{n,t}/\sqrt{\epsilon_{\min}} \leqslant \sqrt{\overline c_{\epsilon}} \sfs_{n,t}/\sqrt{\epsilon_{\max}} \to 0$ and $\sfs_{n,t}\to 0$ according to \eqref{eq:grad.ub1}. These relations imply that as $N\to\infty$, $\Phi\left(\frac{2B_0\log N - \sfs_{n,t}}{\sqrt{\epsilon_{\max}}}\right)  \to 1$, $\Phi\left(\frac{2A_1\log\log N - \sfs_{n,t}}{\sqrt{\epsilon_{\max}}}\right)\to 1$, $\Phi\left(\frac{2B_1\log\log N - \sfs_{n,t}}{\sqrt{\epsilon_{\max}}}\right) \to 1$, $\Phi\left(\frac{2E_1\log\log N - \sfs_{n,t}}{\sqrt{\epsilon_{\max}}}\right)\to 1$, and $\Phi\left(\frac{\sfs_{n,t}}{\sqrt{\epsilon_{\min}}}\right)\to 1/2$.

Let $\overline\psi^{(t)}$ be a random vector following $N(0,\epsilon_{\max}I_{p+3})$. Using the lower bound in \eqref{eq:psi.prob.lb2} together with the density \eqref{eq:density.tilde.psi.1}, we have the following inequality for the truncated random variable $\tilde \psi^{(t)}$ in \eqref{eq:theta.t+1.eq1}: for any $z >0$ and all sufficiently large $N$,
\begin{align} \label{eq.tilde.psi.norm}
\PP\left(\left\|\tilde \psi^{(t)}\right\|_2\geqslant z \right)
& =\int_{\|x\|_2\geqslant z} f_{\tilde \psi^{(t)}}(x) \ud x \nonumber \\
&\overset{(i)}{\leqslant} 2^{p+4} \int_{\|x\|_2\geqslant z} \frac{1}{(2\pi)^{(p+3)/2}\dett(\Ecal)^{1/2}}\exp\left\{-\frac{1}{2}x^\top \Ecal^{-1} x\right\} \ud x  \nonumber \\
&\overset{(ii)}{\leqslant} 2^{p+4} \int_{\|x\|_2\geqslant z} \frac{1}{(2\pi \epsilon_{\max})^{(p+3)/2}}\exp\left\{-\frac{1}{2\epsilon_{\max}}x^\top x\right\} \ud x  \nonumber \\
&= 2^{p+4} \PP\left(\left\|\overline\psi^{(t)}\right\|_2\geqslant z\right),
\end{align}
where $(i)$ follows from \eqref{eq:psi.prob.lb2} and ignoring the indicator function in \eqref{eq:density.tilde.psi.1} to make the integral larger, and $(ii)$ follows because the random normal vector with a larger variance has a larger probability outside radius $z$.

Therefore, from the updating equation \eqref{eq:theta.t+1.eq1}, we have that
\begin{align} \label{le1.21}
&\quad \PP\left(\theta^{(t+1)}\notin \Bcal(\theta^{(t)},\sfr)\right)
= \PP\left(\big\|\theta^{(t+1)}-\theta^{(t)}\big\|_2  > \sfr \right) \nonumber \\
&\leqslant  \PP\left(\left\|\tilde\psi^{(t)}\right\|_2 > \sfr-\frac{{\epsilon}_{\max}}{2}\big\|g_{\vartheta^{(t)}}(\theta^{(t)},Y_n^{(t)})\big\|_2 \right) \nonumber \\
&\overset{(i)}{\leqslant} 2^{p+4} \PP\left(\left\|\overline\psi^{(t)}\right\|_2\geqslant \sfr-\frac{{\epsilon}_{\max}}{2}\big\|g_{\vartheta^{(t)}}(\theta^{(t)},Y_n^{(t)})\big\|_2 \right) \nonumber \\
&\overset{(ii)}{=} 2^{p+4} \PP\left(\left\|\frac{1}{\sqrt{\epsilon_{\max}}}\overline\psi^{(t)}\right\|_2\geqslant \frac{1}{\sqrt{\epsilon_{\max}}} \left[\sfr-\frac{{\epsilon}_{\max}}{2}M_0 N(\log N)^{3A_1+3B_1+4E_1+2} \right]\right),
\end{align}
where $(i)$ follows from \eqref{eq.tilde.psi.norm} and $(ii)$ follows from Lemma \ref{leb1}. On the left-hand side of the last expression of \eqref{le1.21}, $\overline\psi^{(t)}/\sqrt{\epsilon_{\max}}\sim N(0,I_{p+3})$ so $\|\overline\psi^{(t)}/\sqrt{\epsilon_{\max}}\|_2^2 \sim \chi^2_{p+3}$, the chi-square distribution with $p+3$ degrees of freedom. On the right-hand side of the expression, since $\sfr = 2\sqrt{ {\epsilon}_{\max}}\big[ \sqrt{p+3} + \sqrt{2\{\log(8T/\tau)+(p+4)\log 2\}} \big]$ from the definition in \eqref{eq:sfr}, and  ${\epsilon}_{\max} \prec 1/N^2$, we have that for sufficiently large $N$, 
\begin{align*}
&\quad \big[\sfr -{\epsilon}_{\max} M_0 N(\log N)^{3A_1+3B_1+4E_1+2}/2\big]/\sqrt{{\epsilon}_{\max}}\\
&\geqslant \sqrt{p+3} +\sqrt{2\{\log (8T/\tau)+(p+4)\log 2\}}.
\end{align*}
Thus, by the tail bound of chi-square distribution (for example, Lemma 1 of \citealt{LauMas00}), \eqref{le1.21} implies that
\begin{align*}
&\quad~\PP\left(\theta^{(t+1)}\notin \Bcal(\theta^{(t)},\sfr)\right) \\
&\leqslant 2^{p+4} \PP_{W\sim\chi^2_{p+3}} \left( \sqrt{W} \geqslant \sqrt{p+3} + \sqrt{2[\log (8T/\tau)+(p+4)\log 2]} \right) \\
&\leqslant 2^{p+4} \exp\left\{-\log(8T/\tau)-(p+4)\log 2\right\} = \frac{\tau}{8T},
\end{align*}
which implies that $\PP\left(\theta^{(t+1)} \in \Bcal(\theta^{(t)},\sfr)\right) \geqslant 1 - \tau/(8T)$ for each $t$. A union bound over all $t=0,\ldots,T-1$ leads to $\PP\left(\theta^{(t+1)} \in \Bcal(\theta^{(t)},\sfr), \text{ for all } t=0,\ldots,T-1 \right) \geqslant 1 - \tau/8$.

We conclude that the projected SGLD generates the same output as that of the pigeonhole SGLD with a probability as least $1-\tau/8$. Therefore, the total variation distance between the distributions of these two outputs does not exceed $\tau/8$. This completes the proof of Lemma \ref{le1}.
\end{proof}

Similar to \citet{zou2020faster}, we define a $1/2$-lazy version of the projected SGLD Markov process above with the following transition distribution
\begin{align} \label{eq:Tu}
& \Tcal_{\ub}(\wb)=\frac{1}{2} \delta_{\ub}(\wb)+\frac{1}{2} q(\wb \mid \ub) .
\end{align}
First, we notice that this $1/2$-lazy Markov process with the transition kernel $\Tcal_{\ub}(\wb)$ as given in \eqref{eq:Tu} has the same stationary distribution as the projected SGLD with transition kernel $q(\wb \mid \ub)$ given in \eqref{eq:proj.sgld.trans}. This is because if $\pi^*(\cdot)$ is the density of the stationary distribution of the projected SGLD with transition kernel $q(\wb \mid \ub)$, then by definition, for any $\wb\in \Theta_N$, $\int_{\Theta_N} \pi^*(\ub)q(\wb \mid \ub) \ud \ub = \pi^*(\wb) $, which implies that
\begin{align*}
\int_{\Theta_N} \pi^*(\ub) \Tcal_{\ub}(\wb) \ud \ub &= \int_{\Theta_N} \pi^*(\ub) \left\{\frac{1}{2} \delta_{\ub}(\wb)+\frac{1}{2} q(\wb \mid \ub)\right\} \ud \ub \\
&= \frac{1}{2}\pi^*(\wb)  + \frac{1}{2} \int_{\Theta_N} \pi^*(\ub)q(\wb \mid \ub) \ud \ub \\
&= \frac{1}{2}\pi^*(\wb)  + \frac{1}{2}\pi^*(\wb)  = \pi^*(\wb),
\end{align*}
i.e., $\pi^*(\cdot)$ is also the density of the stationary distribution of the $1/2$-lazy version. 

Next, we prove that each Markov chain $\Ccal(T)=\left\{\theta^{(1)},\ldots,\theta^{(T)}\right\}$ drawn from the projected PSGLD can be well approximated in total variation norm with high probability by some chain $\tilde \Ccal(T_{\laz}) = \left\{\tilde \theta^{(1)},\ldots,\tilde \theta^{(T_{\laz})}\right\}$ with the transition kernel $\Tcal_{\ub}(\wb)$ as given in \eqref{eq:Tu}, for $T_{\laz}\approx 2T$. 

For the ease of presentation, we assume that the index of iteration starts at 1. With the initial value $\tilde \theta^{(1)}$, consider the Markov chain $\tilde \theta^{(1)},\tilde \theta^{(2)},\ldots$ generated from the transition kernel $\Tcal_{\ub}(\wb)$ given in \eqref{eq:Tu}. For $t=1,2,\ldots$, define the state variable $\gamma^{(t)}$ as follows: $\gamma^{(t)}=0$ if $\tilde \theta^{(t)} = \tilde \theta^{(t-1)}$ with probability $1/2$, and $\gamma^{(t)}=1$ if $\tilde \theta^{(t)}\sim q(\cdot \mid \tilde \theta^{(t-1)})$ with probability $1/2$. Define the injection $\Pcal:\tilde \Ccal(T_{\laz}) \mapsto \Ccal(T)$ as follows: for $t=1,\ldots,T_{\laz}$, if $\gamma^{(t)}=0$, then remove the element $\tilde\theta^{(t)}$ from the sequence $\tilde \Ccal(T_{\laz})$; the remaining elements in $\tilde \Ccal(T_{\laz})$ are ordered from $1$ to $T$ as the new sequence $\Ccal(T)=\left\{\theta^{(1)},\ldots,\theta^{(T)}\right\}$ with $T\equiv \sum_{t=1}^{T_{\laz}} \gamma^{(t)}$ and $\theta^{(1)}=\tilde \theta^{(1)}$. In other words, the injection $\Pcal$ maps each lazy Markov chain to its non-lazy version. We notice that although we have removed the ``replicates" in the chain $\tilde \Ccal(T_{\laz})$ due to the ``lazy'' $1/2$-probability of staying at the current state, the remaining chain $\Ccal(T)$ may still contain duplicates due to the PSGLD transition kernel $q(\cdot \mid \cdot)$ defined in \eqref{eq:proj.sgld.trans}. 

We further define the random lengths of $\gamma^{(t)}=0$ after each occurrence $\gamma^{(t)}=1$ as $n_1,n_2,\ldots,n_T$. In other words, the chain $\tilde \Ccal(T_{\laz})= \left\{\tilde \theta^{(1)},\ldots,\tilde \theta^{(T_{\laz})}\right\}$ can be equivalently written as 
$$\theta^{(1)},\underbrace{\theta^{(1)},\ldots,\theta^{(1)}}_{n_1},
\theta^{(2)},\underbrace{\theta^{(2)},\ldots,\theta^{(2)}}_{n_2},\ldots \ldots,\theta^{(T)}, \underbrace{\theta^{(T)},\ldots,\theta^{(T)}}_{n_T}.$$
It is clear that by the definition of $\Tcal_{\ub}(\wb)$ in \eqref{eq:Tu}, $n_1,\ldots,n_T$ are independent random variables and for each $t=1,\ldots,T$, $\PP(n_t=s)=1/2^{s+1}$ for $s=0,1,2,\ldots$. Furthermore, $T_{\laz}-T = \sum_{t=1}^T n_t$. The next lemma shows that for each large $T$ and PSGLD chain $\Ccal(T)$, we can reversely find a chain $\tilde \Ccal(T_{\laz})$, such that the empirical distributions based on the parameter values in $\Ccal(T)$ and $\tilde \Ccal(T_{\laz})$ are close in total variation distance with high probability. 

\begin{lemma} \label{lem:lazy}
Then for all sufficiently large $T$, for any $\tau\in (0,1)$, 
$$|T_{\laz}-2T|\leqslant \frac{\tau}{8}T, \text{ and}\quad\left\| \tilde \Pi^{\textup{Proj-SGLD}}_{T_{\laz}} - \Pi^{\textup{Proj-SGLD}}_T \right\|_{\TV} \leqslant \frac{\tau}{8} , $$
with probability at least $1-4\exp\left(-\sqrt{T} \tau/8\right)$, where $\tilde \Pi^{\textup{Proj-SGLD}}_{T_{\laz}}$ is the empirical distribution with the chain length $T_{\laz}$ from the $1/2$-lazy projected SGLD.
\end{lemma}

\begin{proof}[Proof of Lemma \ref{lem:lazy}]
The empirical distribution based on the draws in $\Ccal(T)$ is $\Pi_T = T^{-1} \sum_{t=1}^T \delta_{\theta^{(t)}}$, where $\delta_{\cdot}$ denotes the Dirac measure. Similarly, the empirical distribution based on the draws in $\tilde \Ccal(T_{\laz})$ is 
$$\tilde \Pi^{\textup{Proj-SGLD}}_{T_{\laz}} = T_{\laz}^{-1} \sum_{t=1}^{T_{\laz}} \delta_{\tilde \theta^{(t)}} = T_{\laz}^{-1} \sum_{t=1}^T (n_t+1) \delta_{ \theta^{(t)}},$$
where the latter expression is due to collapsing the duplicates at those steps with $\gamma^{(t)}=0$. The total variation distance between them is then
\begin{align} \label{eq:tv1.lazy}
\left\| \tilde \Pi^{\textup{Proj-SGLD}}_{T_{\laz}} - \Pi^{\textup{Proj-SGLD}}_T \right\|_{\TV} &= \frac{1}{2} \sum_{t=1}^T \left| \frac{n_t+1}{T_{\laz}} - \frac{1}{T} \right| .
\end{align}
For two sequences of nonnegative numbers $\{a_t\}_{t=1}^{T}$ and $\{b_t\}_{t=1}^{T}$, we have that
\begin{align*}
&\sum_{t=1}^T \left|\frac{a_t}{\sum_{t=1}^T a_t} - \frac{b_t}{\sum_{t=1}^T b_t}\right| = \frac{\sum_{t=1}^T \left|a_t \sum_{t=1}^T b_t - b_t \sum_{t=1}^T a_t\right|}{\left(\sum_{t=1}^T a_t\right)\left(\sum_{t=1}^T b_t\right)} \\
\leqslant{}& \frac{\sum_{t=1}^T \left|b_t-a_t\right| \sum_{t=1}^T a_t + \sum_{t=1}^T a_t \sum_{t=1}^T \left| a_t- b_t\right|}{\left(\sum_{t=1}^T a_t\right)\left(\sum_{t=1}^T b_t\right)} \\
={}& \frac{2\sum_{t=1}^T \left|a_t-b_t\right|}{\sum_{t=1}^T b_t}.
\end{align*}
Therefore, if we set $a_t=n_t+1$ (with $\sum_{t=1}^T a_t = T_{\laz}$) and $b_t=2$ (with $\sum_{t=1}^T b_t = 2T$), then from \eqref{eq:tv1.lazy} we obtain that
\begin{align} \label{eq:tv2.lazy}
\left\| \tilde \Pi^{\textup{Proj-SGLD}}_{T_{\laz}} - \Pi^{\textup{Proj-SGLD}}_T \right\|_{\TV} &\leqslant \frac{\sum_{t=1}^T|n_t-1|}{2T}.
\end{align}
Now we derive a concentration bound for $\sum_{t=1}^T|n_t-1|/(2T)$. Since $\PP(n_t=s)=1/2^{s+1}$ for $s=0,1,2,\ldots$, obviously $n_t$'s are sub-exponential random variables with mean $\EE(n_t)=1$, and we can use the Chernoff bound to control the tail probability. Specifically, for any $c\in (-T\log 2/2,T\log 2/2)$, direct calculation gives 
\begin{align*}
\EE\left\{\frac{c}{T}(n_t-1)\right\} = \frac{1}{\ee^{c/T}(2-\ee^{c/T})} \leqslant \exp(2c^2/T^2),
\end{align*}
where the last step follows from $\exp(2x^2) - 1/[\ee^{x}(2-\ee^{x})] \geqslant 0 $ for all $|x|\leqslant \log 2/2$. Therefore, by the Markov inequality, we have that for any given $\tau\in (0,1)$, 
\begin{align*}
&\quad~ \PP\left(\frac{\sum_{t=1}^T(n_t-1)}{2T} \geqslant \frac{\tau}{8} \right) \\
& \leqslant \exp(-c\tau/4) \EE\left\{\frac{c}{T} \sum_{t=1}^T(n_t-1)\right\} = \exp(-c\tau/4) \prod_{t=1}^T \EE\left\{\frac{c}{T}(n_t-1)\right\} \\
&\leqslant \exp\left(\frac{2c^2}{T} - \frac{c\tau}{4}\right)  .
\end{align*}
For sufficiently large $T$, we choose $c=\sqrt{T}/2$ which satisfies $c\in (-T\log 2/2,T\log 2/2)$, such that the upper bound above becomes
\begin{align} \label{eq:tv3.lazy}
& \PP\left(\frac{\sum_{t=1}^T(n_t-1)}{2T} \geqslant \frac{\tau}{8} \right) \leqslant 2\exp\left(-\sqrt{T} \tau/8\right) .
\end{align}
Similarly, for the left side inequality, 
\begin{align*}
&\quad~ \PP\left(\frac{\sum_{t=1}^T(n_t-1)}{2T} \leqslant -\frac{\tau}{8} \right) \\
& \leqslant \exp(c\tau/4) \EE\left\{\frac{c}{T} \sum_{t=1}^T(1-n_t)\right\} = \exp(c\tau/4) \prod_{t=1}^T \EE\left\{\frac{-c}{T}(n_t-1)\right\} \\
&\leqslant \exp\left(\frac{2c^2}{T} + \frac{c\tau}{4}\right)  .
\end{align*}
For sufficiently large $T$, we choose $c=-\sqrt{T}/2$ which satisfies $c\in (-T\log 2/2,T\log 2/2)$, such that the upper bound above becomes
\begin{align} \label{eq:tv4.lazy}
& \PP\left(\frac{\sum_{t=1}^T(n_t-1)}{2T} \leqslant -\frac{\tau}{8} \right) \leqslant 2\exp\left(-\sqrt{T} \tau/8\right) .
\end{align}
Finally we combine \eqref{eq:tv3.lazy} and \eqref{eq:tv4.lazy} to obtain that
\begin{align} \label{eq:tv5.lazy}
& \PP\left(\frac{\sum_{t=1}^T|n_t-1|}{2T} \geqslant \frac{\tau}{8} \right) \leqslant 4\exp\left(-\sqrt{T} \tau/8\right) ,
\end{align}
which also implies that
\begin{align*} 
& \PP\left(|T_{\laz}-2T| \geqslant \frac{\tau}{8}T \right) \leqslant 4\exp\left(-\sqrt{T} \tau/8\right) .
\end{align*}
The conclusion on total variation distance follows from \eqref{eq:tv2.lazy} and \eqref{eq:tv5.lazy}.
\end{proof}

Given the conclusion of Lemma \ref{lem:lazy}, in all the following lemmas and proofs, we will always assume that $T_{\laz}$ and $T$ are of the same order, i.e., $T_{\laz}/T \asymp 1$, and we will momentarily treat $T_{\laz}$ as deterministic instead of random, until the proof of Theorem \ref{th1} in Section \ref{subsec:proof.thm}.

\vspace{6mm}

To show the convergence of the Markov process of $1/2$-lazy projected SGLD with the transition distribution $\Tcal_{\ub}(\wb)$, we follow the idea of \citet{Zhangetal17} and \citet{zou2020faster} to utilize the Metropolized SGLD, which is constructed by adding a correction step into the transition distribution $\Tcal_{\ub}(\cdot)$. A point $\wb$ generated by the algorithm from the starting point $\ub$ is accepted with the probability
\begin{align} \label{eq:alpha.metro}
\alpha_{\ub}(\wb)=\min \left\{1,~ \frac{\Tcal_{\wb}(\ub)}{\Tcal_{\ub}(\wb)} \cdot \exp \left[\log \pi(\wb \mid \Yb)- \log \pi(\ub \mid \Yb)\right]\right\},
\end{align}
where $\pi(\theta \mid \Yb)$ is the true posterior based on the full data. The transition distribution of this Markov process is
\begin{align} \label{eq:Tu.star}
& \Tcal_{\ub}^{\star}(\wb)=\left(1-\alpha_{\ub}(\wb)\right) \delta_{\wb}(\ub)+\alpha_{\ub}(\wb) \Tcal_{\ub}(\wb),
\end{align}
where $\alpha_{\ub}(\wb)$ is defined in \eqref{eq:alpha.metro}.

\vspace{3mm}

In the next series of lemmas, we establish the convergence of the output of the $1/2$-lazy projected SGLD to the target posterior distribution in total variation distance with some upper-bounded error converging to zero as $N,T_{\laz}\to\infty$. The final convergence result is shown in Lemma \ref{le234}, whose proof depends on a few technical lemmas, Lemma \ref{le2}, Lemma \ref{le3}, and Lemma \ref{le4}.

\begin{lemma}\label{le2}
Suppose that Assumptions \ref{assump:asymp}, \ref{assump:bound}, \ref{assump:eigen} and \ref{assump:initial} hold. For any given $\tau \in(0,1)$, define $\delta=\delta(L, M_0, N,{\epsilon}_{\max}, \sfr)$ as
\begin{align} \label{eq:delta}
\delta = & \sfr^2 \left[LN(\log N)^{3A_1+3B_1+4E_1+2} + M_0^2 N^2 (\log N)^{2(3A_1+3B_1+4E_1+2)}\right]  \nonumber \\
& + \frac{\epsilon_{\max}}{4} M_0^2 N^2(\log N)^{2(3A_1+3B_1+4E_1+2)}  + \frac{\epsilon_{\max}^2}{4} M_0^4 N^4(\log N)^{4(3A_1+3B_1+4E_1+2)} ,
\end{align}

where $L,M_0$ are as defined in Lemmas \ref{lea2} and \ref{leb1}, and $\sfr$ is defined in \eqref{eq:sfr} (both $\epsilon_{\max}$ and $\sfr$ depend on $\tau$). Then for any set $\Acal \subseteq \Theta_N$ and any point $\ub \in \Theta_N$, on the event $\Gcal_N$ defined in \eqref{eq:Gcal}, for all sufficiently large $N$,
\begin{align} \label{le2.2}
& (1-\delta)\Tcal_{\ub}^{\star}(\Acal) \leqslant \Tcal_{\ub}(\Acal)\leqslant(1+\delta)\Tcal_{\ub}^{\star}(\Acal),
\end{align}
where $\Tcal_{\ub}$ and $\Tcal_{\ub}^{\star}$ are defined in \eqref{eq:Tu} and \eqref{eq:Tu.star}, respectively. Furthermore, for any point $\ub \in \Theta_N$ and $\wb \in \Theta_N \cap \Bcal(\ub, \sfr) \backslash\{\ub\}$, we have $\alpha_{\ub}(\wb) \geqslant 1-\delta / 2$.
\end{lemma}

\begin{proof}[Proof of Lemma \ref{le2}]
Our proof of Lemma \ref{le2} is similar to that of Lemma 6.2 in Section B.2 of \cite{zou2020faster}.  The proof is divided into two cases, $\ub\notin \Acal$ and $\ub\in \Acal$. In the first case, i.e., when $\ub\notin \Acal$, we have from \eqref{eq:Tu.star} that
\begin{align}\label{le2.1}
\Tcal_{\ub}^{\star}(\mathcal{A})=\int_{\mathcal{A}} \Tcal_{\ub}^{\star}(\wb) \ud  \wb=\int_{\mathcal{A}} \alpha_{\ub}(\wb) \Tcal_{\ub}(\wb) \ud  \wb.
\end{align}
By the definition of the projected SGLD and its 1/2-lazy version in \eqref{eq:Tu}, we know that the next iteration under $\Tcal_{\ub}^{\star}$ satisfies $\wb \in \Theta_N \cap \Bcal(\ub, \sfr) $. By \eqref{le2.1}, $\Tcal_{\ub}^{\star}(\mathcal{A}) \leqslant \Tcal_{\ub}(\mathcal{A})$ since $\alpha_{\ub}(\wb) \leqslant 1$. Thus, the first inequality of \eqref{le2.2} holds. To prove the second inequality of \eqref{le2.2} holds, it is sufficient to show that $\alpha_{\ub}(\wb) \geqslant 1-\delta / 2$. According to the definition of $\alpha_{\ub}(\wb)$ in \eqref{eq:alpha.metro}, this is equivalent to proving that $N_1/D_1 \geqslant 1-\delta/2$, where
\begin{align} \label{eq:N1D1}
N_1 & = \exp\left\{\log \pi(\wb \mid \Yb)-\log \pi(\ub \mid \Yb)\right\} \cdot \EE_{\Scal} \EE_{\overline\pi(\vartheta|\wb,Y_n)} [p(\ub\mid \wb,\vartheta,Y_n)] , \nonumber \\
D_1 & = \EE_{\Scal'} \EE_{\overline\pi(\vartheta'|\ub,Y_n)} [p(\wb\mid \ub,\vartheta',Y_n)] ,
\end{align}
$\Scal, \Scal' \subseteq \{1,\ldots,R\}\otimes \{1,\ldots,C\}$ are the two independent randomly selected index sets, $Y_n$ and $Y_n'$ are the corresponding vectorized subsets of the full data $\Yb$, and $\vartheta,\vartheta'$ are the corresponding Monte Carlo draws of the latent variables.

We notice that from Lemma \ref{leb1}, on the event $\Gcal_N$, for any parameter $\ub\in\Theta_N$ and any subset data $Y_n$, $\|g_{\vartheta}(\ub,Y_n)\|_2\leqslant M_0 N(\log N)^{3A_1+3B_1+4E_1+2}$ for all large $N$. Since $g_{\vartheta}(\ub,Y_n)$ defined in \eqref{eq:g.func.def} is the subset-based unbiased estimator of $\nabla_{\theta} \log \pi(\ub \mid \Yb)$, we apply the Hoeffding's lemma to obtain that for any vector $\ab \in \RR^{p+3}$,
\begin{align}\label{eq:hoeffding}
& \EE_{\Scal} \EE_{\overline\pi(\vartheta|\ub,Y_n)} \exp \left\{\ab^\top \left[g_{\vartheta}(\ub,Y_n) + \nabla_{\theta} \log \pi(\ub \mid \Yb)\right]\right\} \nonumber \\
& \leqslant \exp\left\{M_0^2 N^2(\log N)^{6A_1+6B_1+8E_1+4} \|\ab\|_2^2 \right\}.
\end{align}

By the Jensen's inequality,
\begin{align}\label{eq:N1.1}
&N_1 \geqslant \exp\left\{\log \pi(\wb \mid \Yb) -\log \pi(\ub \mid \Yb) + \EE_{\Scal} \EE_{\overline\pi(\vartheta|\wb,Y_n)} [\log p(\ub\mid \wb,\vartheta,Y_n)] \right\} \nonumber \\
&=  (2\pi)^{-(p+3)/2} [\dett(\Ecal)]^{1/2} \exp\left\{\log \pi(\wb \mid \Yb) -\log \pi(\ub \mid \Yb) \right\} \nonumber \\
&~~ \times \exp\left\{- \frac{1}{2} \EE_{\Scal} \EE_{\overline\pi(\vartheta|\wb,Y_n)} \Big( \ub - \wb + \frac{\Ecal}{2} g_{\vartheta}(\wb,Y_n)  \Big)^\top \Ecal^{-1} \Big( \ub - \wb + \frac{\Ecal}{2} g_{\vartheta}(\wb,Y_n)  \Big)  \right\} \nonumber \\
&\geqslant (2\pi)^{-(p+3)/2} [\dett(\Ecal)]^{1/2} \exp\left\{\log \pi(\wb \mid \Yb) -\log \pi(\ub \mid \Yb) \right\} \times \nonumber \\
&\quad \exp\Big\{- \frac{ (\ub - \wb)^\top \Ecal^{-1} (\ub - \wb)}{2}  - \frac{(\ub - \wb)^\top  \EE_{\Scal} \EE_{\overline\pi(\vartheta|\wb,Y_n)} g_{\vartheta}(\wb,Y_n)}{2} \nonumber \\
&\quad - \frac{1}{8} \EE_{\Scal} \EE_{\overline\pi(\vartheta|\wb,Y_n)} \big[g_{\vartheta}(\wb,Y_n)^\top \Ecal g_{\vartheta}(\wb,Y_n) \big]  \Big\} \nonumber \\
&\stackrel{(i)}{\geqslant} (2\pi)^{-(p+3)/2} [\dett(\Ecal)]^{1/2} \exp\left\{\log \pi(\wb \mid \Yb) -\log \pi(\ub \mid \Yb) \right\} \times \nonumber \\
&\quad \exp\Big\{- \frac{ (\ub - \wb)^\top \Ecal^{-1} (\ub - \wb)}{2}  + \frac{(\ub - \wb)^\top  \nabla_{\theta} \log \pi(\wb \mid \Yb)}{2} \nonumber \\
&\quad -  \frac{\epsilon_{\max}}{8}  M_0^2 N^2(\log N)^{2(3A_1+3B_1+4E_1+2)}
\Big\},
\end{align}
where $(i)$ follows Lemma \ref{leb1}.

For $D_1$, we have that
\begin{align} \label{eq:D1.1}
&D_1 =  (2\pi)^{-(p+3)/2} [\dett(\Ecal)]^{1/2} \nonumber \\
&~~\times  \EE_{\Scal'} \EE_{\overline\pi(\vartheta'|\ub,Y_n')} \exp\left\{- \frac{1}{2} \Big( \wb - \ub + \frac{\Ecal}{2} g_{\vartheta'}(\ub,Y_n')  \Big)^\top \Ecal^{-1} \Big( \wb - \ub + \frac{\Ecal}{2} g_{\vartheta'}(\ub,Y_n')  \Big)  \right\} \nonumber \\
&\leqslant (2\pi)^{-(p+3)/2} [\dett(\Ecal)]^{1/2} \exp\left\{- \frac{ (\ub - \wb)^\top \Ecal^{-1} (\ub - \wb)}{2} + \frac{(\wb - \ub )^\top  \nabla_{\theta} \log \pi(\ub \mid \Yb)}{2} \right\} \nonumber \\
&~~\times  \EE_{\Scal'} \EE_{\overline\pi(\vartheta'|\ub,Y_n')} \exp\Bigg\{  -\frac{(\wb - \ub )^\top \left[g_{\vartheta'}(\ub,Y_n') +\nabla_{\theta} \log \pi(\ub \mid \Yb)\right]}{2} \nonumber \\
&~~ - \frac{1}{8} g_{\vartheta'}(\ub,Y_n')  ^\top \Ecal g_{\vartheta'}(\ub,Y_n')    \Bigg\} \nonumber \\
&\leqslant (2\pi)^{-(p+3)/2} [\dett(\Ecal)]^{1/2} \exp\Bigg\{- \frac{ (\ub - \wb)^\top \Ecal^{-1} (\ub - \wb)}{2} + \frac{(\wb - \ub )^\top  \nabla_{\theta} \log \pi(\ub \mid \Yb)}{2} \Bigg\} \nonumber \\
&~~\times  \EE_{\Scal'} \EE_{\overline\pi(\vartheta'|\ub,Y_n')} \exp\Bigg\{ - \frac{(\wb - \ub )^\top \left[g_{\vartheta'}(\ub,Y_n') + \nabla_{\theta} \log \pi(\ub \mid \Yb)\right]}{2} \nonumber \\
&~~ + \frac{1}{4} \left[g_{\vartheta'}(\ub,Y_n') +  \nabla_{\theta} \log \pi(\ub \mid \Yb)\right]^\top \Ecal \nabla_{\theta} \log \pi(\ub \mid \Yb) \nonumber \\
&~~ - \frac{1}{8} \nabla_{\theta} \log \pi(\ub \mid \Yb)^\top \Ecal \nabla_{\theta} \log \pi(\ub \mid \Yb)  \Bigg\} \nonumber \\
&\stackrel{(i)}{\leqslant}  (2\pi)^{-(p+3)/2} [\dett(\Ecal)]^{1/2} \exp\Bigg\{- \frac{ (\ub - \wb)^\top \Ecal^{-1} (\ub - \wb)}{2} + \frac{(\wb - \ub )^\top  \nabla_{\theta} \log \pi(\ub \mid \Yb)}{2}  \nonumber \\
&~~ - \frac{1}{8} \nabla_{\theta} \log \pi(\ub \mid \Yb)^\top \Ecal \nabla_{\theta} \log \pi(\ub \mid \Yb)  \nonumber \\
&~~ + \frac{1}{4} M_0^2 N^2(\log N)^{2(3A_1+3B_1+4E_1+2)} \left\|\wb - \ub + (\Ecal/4) \nabla_{\theta} \log \pi(\ub \mid \Yb)  \right\|_2^2 \Bigg\} \nonumber \\
&\stackrel{(ii)}{\leqslant} (2\pi)^{-(p+3)/2} [\dett(\Ecal)]^{1/2} \exp\Bigg\{- \frac{ (\ub - \wb)^\top \Ecal^{-1} (\ub - \wb)}{2} + \frac{(\wb - \ub )^\top  \nabla_{\theta} \log \pi(\ub \mid \Yb)}{2}  \nonumber \\
&~~ - \frac{1}{8} \nabla_{\theta} \log \pi(\ub \mid \Yb)^\top \Ecal \nabla_{\theta} \log \pi(\ub \mid \Yb)  \nonumber \\
&~~ + \frac{1}{2} M_0^2 N^2(\log N)^{2(3A_1+3B_1+4E_1+2)} \left(\|\wb - \ub\|_2^2 + \frac{\epsilon_{\max}^2}{4} M_0^2 N^2(\log N)^{2(3A_1+3B_1+4E_1+2)} \right) \Bigg\} ,
\end{align}
where $(i)$ follows from \eqref{eq:hoeffding} and Lemma \ref{leb1}, and $(ii)$ follows from the inequality $\|\ab_1+\ab_2\|_2^2\leqslant 2(\|\ab_1\|_2^2+\|\ab_2\|_2^2)$ and Lemma \ref{leb1}.

Therefore, we combine \eqref{eq:N1.1} and \eqref{eq:D1.1} to obtain that
\begin{align}\label{eq:N1D1.2}
\frac{N_1}{D_1} &\geqslant \exp\Bigg\{\log \pi(\wb \mid \Yb) - \log \pi(\ub \mid \Yb)  \nonumber \\
&\quad  - \frac{1}{2} (\wb-\ub)^\top \left[\nabla_{\theta} \log \pi(\ub \mid \Yb) + \nabla_{\theta} \log \pi(\wb \mid \Yb)\right] \nonumber \\
&\quad  + \frac{1}{8} \nabla_{\theta} \log \pi(\ub \mid \Yb)^\top \Ecal \nabla_{\theta} \log \pi(\ub \mid \Yb)
- \frac{\epsilon_{\max}}{8} M_0^2 N^2(\log N)^{2(3A_1+3B_1+4E_1+2)} \nonumber \\
&\quad  - \frac{1}{2} M_0^2 N^2(\log N)^{2(3A_1+3B_1+4E_1+2)}  \nonumber \\
&\quad~~ \times \left(\|\wb - \ub\|_2^2 + \frac{\epsilon_{\max}^2}{4} M_0^2 N^2(\log N)^{2(3A_1+3B_1+4E_1+2)} \right) \Bigg\} .
\end{align}
For any $\ub,\wb \in \Theta_N$, on the event $\Gcal_N$, by Lemma \ref{lea2}, we have that
\begin{align*}
&\quad~ \left|\log \pi(\wb \mid \Yb) - \log \pi(\ub \mid \Yb) - (\wb - \ub)^\top \nabla_{\theta} \log \pi(\ub \mid \Yb) \right| \\
&\leqslant  \frac{LN(\log N)^{3A_1+3B_1+4E_1+2}}{2} \|\wb-\ub\|_2^2 , \\
&\quad \left|\log \pi(\ub \mid \Yb) - \log \pi(\wb \mid \Yb) - (\ub - \wb)^\top \nabla_{\theta} \log \pi(\wb \mid \Yb) \right| \\
&\leqslant \frac{LN(\log N)^{3A_1+3B_1+4E_1+2}}{2} \|\ub-\wb\|_2^2 .
\end{align*}
By adding and averaging the two inequalities, we obtain that

\begin{align} \label{eq:uw.1}
&\quad \left|\log \pi(\wb \mid \Yb) - \log \pi(\ub \mid \Yb) - \frac{1}{2} (\wb - \ub)^\top \left[\nabla_{\theta} \log \pi(\ub \mid \Yb) + \nabla_{\theta} \log \pi(\wb \mid \Yb) \right]\right| \nonumber \\
&\leqslant  \frac{LN(\log N)^{3A_1+3B_1+4E_1+2}}{2} \|\ub-\wb\|_2^2 .
\end{align}

We combine \eqref{eq:N1D1.2} and \eqref{eq:uw.1} to obtain that on the event $\Gcal_N$, when $\|\ub-\wb\|_2\leqslant \sfr$,
\begin{align} \label{eq:N1D1.3}
\frac{N_1}{D_1} &\geqslant \exp\Bigg\{- \frac{LN(\log N)^{3A_1+3B_1+4E_1+2} + M_0^2 N^2 (\log N)^{2(3A_1+3B_1+4E_1+2)}}{2} \sfr^2  \nonumber \\
&- \frac{\epsilon_{\max}}{8}  M_0^2 N^2(\log N)^{2(3A_1+3B_1+4E_1+2)}  - \frac{1}{8} M_0^4 N^4(\log N)^{4(3A_1+3B_1+4E_1+2)} \epsilon_{\max}^2 \Bigg\}  \nonumber \\
& = \exp(-\delta/2) \stackrel{(i)}{\geqslant} 1 - \frac{\delta}{2},
\end{align}
where $(i)$ is from the definition of $\delta$ in \eqref{eq:delta} and the inequality $\exp(-x)\geqslant 1-x$ for all $x\in \RR$. This completes the proof for the case of $\ub \notin \Acal$.

The other case of $\ub \in \Acal$ can be proved similarly, using the same proof given in Section B.2 of \cite{zou2020faster}. Thus we complete the proof of Lemma \ref{le2}.
\end{proof}

With the transition distribution of the $1/2$-lazy projected SGLD $\Tcal_{\ub}(\cdot)$ $\delta-$close to that of the Metropolized SGLD $\Tcal_{\ub}^*(\cdot)$, Lemma 6.4 of \cite{zou2020faster} shows that the distribution of the output of the projected SGLD is close to the true posterior distribution $\Pi^*_N$ in total variation distance, with the approximation error quantified by $\delta$ and the conductance of $\Tcal_{\ub}^{\star}(\cdot)$.
\begin{lemma} \label{le3}
Suppose that Assumptions \ref{assump:asymp}, \ref{assump:bound}, \ref{assump:eigen} and \ref{assump:initial} hold and $T_{\laz}/T \asymp 1$. If $\Tcal_{\ub}(\cdot)$ and $\Tcal_{\ub}^{\star}(\cdot)$ satisfy \eqref{le2.2} with a number $\delta \leqslant \min\{1-\sqrt{2}/2, \phi/16\}$, then for any $\lambda-$warm start initial distribution with respect to $\Pi_N^*$, it holds that
\begin{align*}
\left\| \tilde\Pi^{\textup{Proj-SGLD}}_{T_{\laz}} - \Pi_N^* \right\|_{\TV} \leqslant \lambda \left(1-\phi^2/8\right)^{T_{\laz}} + 16\delta/\phi,
\end{align*}
where $\phi$ is the conductance of $\Tcal_{\ub}^{\star}(\cdot)$, defined as
\begin{align} \label{eq:phi}
\phi = \inf_{\Acal : \Acal \subseteq \Theta_N, \Pi_N^*(\Acal) \in (0,1)}\frac{\int_{\Acal}\Tcal_{\ub}^{\star} (\Theta_N \backslash \Acal) \Pi_N^*(\ud \ub)}{\min \left\{\Pi_N^*(\Acal), \Pi_N^*(\Theta_N \backslash \Acal)\right\} } .
\end{align}
\end{lemma}

\begin{proof}[Proof of Lemma \ref{le3}]
The proof is the same as that of Lemma 6.4 in \citet{zou2020faster}.
\end{proof}

To exactly quantify the total variation distance between the output of the $1/2$-lazy projected SGLD $\tilde \Pi^{\textup{Proj-SGLD}}_{T_{\laz}}$ and the target posterior distribution $\Pi_N^*$, we will further give a lower bound of the conductance $\phi$ in Lemma \ref{le4}. Before that, we present two more technical lemmas.
\begin{lemma}\label{b.4}
(Lemma 3.1 in \cite{lee2018convergence}). Let $\Tcal_{\ub}^{\star}(\cdot)$ be a time-reversible Markov chain on $\Theta_N$ with the stationary distribution $\Pi_N^*$. For any given $\Delta >0$, suppose for any $\ub, \vb \in \Theta_N$ with $\|\ub- \vb\|_2 \leqslant \Delta$, we have $\|\Tcal_{\ub}^{\star}(\cdot) - \Tcal_{\vb}^{\star}(\cdot)\|_{\TV} \leqslant 0.99$, then the conductance of $\Tcal_{\ub}^{\star}(\cdot)$ satisfies $\phi \geqslant C_3 \rho\Delta$ for some absolute constant $C_3>0$, where $\rho$ is the Cheeger constant of $\Pi_N^*$.
\end{lemma}

From \eqref{eq:proj.sgld.trans} and \eqref{eq:Tu.star}, for all $\wb \in \Bcal(\ub,\sfr ) \cap \Theta_N$, the transition probability of the Metropolized SGLD $\Tcal_{\ub}^{\star}(\wb)$ is
\begin{align*}
\Tcal_{\ub}^{\star}(\wb) = \frac{2-p(\ub)+p(\ub)[1-\alpha_{\ub}(\wb)]}{2} \delta_{\ub}(\wb) + \frac{\alpha_{\ub}(\wb)}{2}p(\wb|\ub)\cdot \mathbbm{1} \left(\wb \in \Bcal(\ub, \sfr) \cap \Theta_N \right).
\end{align*}

\begin{lemma}\label{lem:b.5}
Suppose that Assumptions \ref{assump:asymp}, \ref{assump:bound}, \ref{assump:eigen} and \ref{assump:initial} hold, and that $T_{\laz}/T \asymp 1$. If for the $\tau \in (0,1)$ in \eqref{eq:sfr},
\begin{align} \label{eq:emax.cond1}
\epsilon_{\max}\prec \min\left\{1/[N^2(\log N)^{2(3A_1+3B_1+4E_1+2)}], 1/ \log (T/\tau) \right\},
\end{align}
as $N,T\to\infty$, then for any $\ub \in \Theta_N$ and for all sufficiently large $N,T$, on the event $\Gcal_N$ defined in \eqref{eq:Gcal}, the acceptance probability $p(\ub)=\PP_{\vb\sim P(\cdot\mid \ub)}\left( \vb \in \Bcal(\ub, \sfr) \cap \Theta_N \right) $ satisfies $p(\ub) \geqslant 0.4 $.
\end{lemma}

\begin{proof}[Proof of Lemma \ref{lem:b.5}]
Let $\tilde p(\wb\mid \ub,\vartheta,Y_n)$ be the density of $N(\ub-(\Ecal/2)g_{\vartheta}(\ub,Y_n), \Ecal)$. Then by our definition, the SGLD transition density constrained to the parameter set $\Theta_N$ is $p(\wb\mid \ub,\vartheta,Y_n) = \tilde p(\wb\mid \ub,\vartheta,Y_n) / \int_{\Theta_N} \tilde p(\wb\mid \ub,\vartheta,Y_n) \ud \wb$.

By Lemma \ref{leb1}, for all $\theta\in \Theta_N$, $(Y_n, \vartheta) \in \Gcal_N$, we have that for all sufficiently large $N$, $\|g_{\vartheta}(\theta,Y_n)\|_2\leqslant M_0 N (\log N)^{3A_1+3B_1+4E_1+2}$. Therefore, if ${\epsilon}_{\max} \prec 1/[N^2(\log N)^{2(3A_1+3B_1+4E_1+2)}]$, then  $({\epsilon}_{\max}/2)\|g_{\vartheta}(\ub,Y_n)\|_2\prec 1$ as $N\to\infty$. According to the definition of $\sfr$ in \eqref{eq:sfr}, as $T\to\infty$, we have $\sfr-({\epsilon}_{\max}/2)\|g_{\vartheta}(\ub,Y_n)\|_2 \geqslant C_{p1}\sqrt{{\epsilon}_{\max}}$ for a constant $C_{p1}>0$ that depends only on $p$ whose value will be chosen below. Hence,
\begin{align}\label{b.5i1}
&\quad \int_{\Bcal(\ub,\sfr)^c}  \tilde p(\wb \mid \ub,\vartheta, Y_n) \ud \wb  \nonumber \\
& = \PP\left(\|\wb-\ub\|_2 - \left\|\frac{\Ecal}{2}g_{\vartheta}(\ub,Y_n)\right\|_2 \geqslant \sfr- \left\|\frac{\Ecal}{2}g_{\vartheta}(\ub, Y_n)\right\|_2\right) \nonumber\\
& \leqslant  \PP\left(\left\|\wb-\ub+\frac{\Ecal}{2}g_{\vartheta}(\ub,Y_n)\right\|^2_2 \geqslant \left(\sfr- \left\|\frac{\Ecal}{2}g_{\vartheta}(\ub, Y_n)\right\|_2\right)^2\right) \nonumber\\
& \leqslant  \PP_{W \sim \chi_{p+3}^{2}}\left(W \geqslant \frac{\left[\sfr-({\epsilon}_{\max}/2)\|g_{\vartheta}(\ub ,Y_n)\|_2\right]^2}{2{\epsilon}_{\max}}\right) \nonumber \\
& \leqslant \PP_{W\sim \chi_{p+3}^{2}}\big( W \geqslant C_{p1}^2/2 \big)  < 2^{-(p+10)},
\end{align}
where in the last step, we choose $C_{p1}$ such that $\PP_{W\sim \chi_{p+3}^{2}}\big( W \geqslant C_{p1}^2/2 \big)  < 2^{-(p+10)}$.

Since $\sfr\prec 1$ as $N,T\to\infty$ and $\Theta_N$ is a rectangle shaped compact set, for any $\ub\in \Theta_N$, we can always find a point $\vb \in \Theta_N$ such that $\|\vb-\ub\|_2\leqslant \sfr^2$ and at least $1/2^{p+3}$ of the ball $\Bcal(\vb,\sfr)$ is inside the set $\Theta_N$. Let $p'(\cdot \mid \vb)$ be the density of $N(\vb, \Ecal)$. Then we have that
\begin{align} \label{eq:I21}
& \int_{\Theta_N} p'(\wb \mid \vb) \ud \wb
\geqslant \frac{1}{2^{p+3}}\int_{\Bcal(\vb, \sfr)}  p'(\wb \mid \vb) \ud \wb
 \geqslant \frac{1}{2^{p+3}}\PP_{W \sim \chi_{p+3}^2}\bigg(W \leqslant \frac{\sfr^2}{2{\epsilon}_{\max}}\bigg) \nonumber \\
&\geqslant \frac{1}{2^{p+3}}\PP_{W \sim \chi_{p+3}^2}\bigg(W \leqslant \frac{4[\sqrt{p+3}+\sqrt{2\{\log(8T/\tau)+(p+4)
\log 2\}}]^2 {\epsilon}_{\max}}{2{\epsilon}_{\max}}\bigg) \nonumber \\
&\geqslant 2^{-(p+4)},
\end{align}
as $T\to\infty$. By the Pinsker's inequality, we have that
\begin{align} \label{eq:I22}
&\quad \bigg|\int_{\Theta_N}  \tilde p(\wb \mid \ub,\vartheta, Y_n) \ud  \wb - \int_{\Theta_N} p'(\wb \mid \vb) \ud  \wb \bigg| \nonumber \\
&\leqslant \sqrt{2 D_{\text{KL}}(N(\ub-(\Ecal/2)g_{\vartheta}(\ub,Y_n),\Ecal), N(\vb,\Ecal))} \nonumber \\
& \leqslant  \frac{\|\vb-[\ub-(\Ecal/2)g_{\vartheta}(\ub,Y_n)]\|_2}{\sqrt{{\epsilon}_{\min}}} \leqslant  \frac{\|\vb-\ub\|_2}{\sqrt{{\epsilon}_{\min}}}+\frac{{\epsilon}_{\max}}{2\sqrt{{\epsilon}_{\min}}} \|g_{\vartheta}(\ub,Y_n))\|_2 \nonumber \\
& \stackrel{(i)}{\leqslant}   4 \sqrt{\overline c_{\epsilon}} \sqrt{\epsilon_{\max}} \big[\sqrt{p+3} + \sqrt{2\{\log(8T/\tau)+(p+4)
\log 2\}}\big]^2 \nonumber \\
&\quad + \frac{\sqrt{\overline c_{\epsilon}}}{2} \sqrt{\epsilon_{\max}} M_0 N (\log N)^{3A_1+3B_1+4E_1+2} \nonumber  \\
& \stackrel{(ii)}{\leqslant} 2^{-(p+5)},
\end{align}
where $D_{\text{KL}}(N(\ub-(\Ecal/2)g_{\vartheta}(\ub,Y_n),\Ecal), N(\vb,\Ecal))$ denotes the Kullback-Leibler divergence from $N(\ub-(\Ecal/2)g_{\vartheta}(\ub,Y_n),\Ecal)$ to $N(\vb,\Ecal)$. The inequality $(i)$ follows from our choice of $\vb$, Assumption \ref{assump:initial} and Lemma \ref{leb1}, and the inequality $(ii)$ follows from the relation \\ $\epsilon_{\max}\prec 1/[N^2(\log N)^{2(3A_1+3B_1+4E_1+2)}]$ and $\epsilon_{\max}\prec 1/ \log (T/\tau)$.

\eqref{eq:I21} and \eqref{eq:I22} together imply that as $N,T\to\infty$,
\begin{align} \label{eq:I23}
&\quad \int_{\Theta_N} \tilde p(\wb\mid \ub,\vartheta,Y_n) \ud \wb \nonumber \\
&\geqslant \int_{\Theta_N} p'(\wb \mid \vb) \ud  \wb - \left|\int_{\Theta_N} \tilde p(\wb\mid \ub,\vartheta,Y_n) \ud \wb - \int_{\Theta_N} p'(\wb \mid \vb) \ud  \wb \right| \nonumber \\
&\geqslant 2^{-(p+4)} - 2^{-(p+5)}  = 2^{-(p+5)} .
\end{align}
We combine \eqref{b.5i1} and \eqref{eq:I23} to obtain that as $N,T\to\infty$,
\begin{align*}
& \int_{\Bcal(\ub,r)\cap \Theta_N} p(\wb\mid \ub,\vartheta,Y_n) \ud \wb  = 1 - \frac{\int_{\Bcal(\ub,r)^c\cap \Theta_N} \tilde p(\wb\mid \ub,\vartheta,Y_n) \ud \wb}{\int_{\Theta_N} \tilde p(\wb\mid \ub,\vartheta,Y_n) \ud \wb} \nonumber \\
&\geqslant 1 - \frac{2^{-(p+10)}}{2^{-(p+5)}} = 1 - 2^{-5} > 0.4.
\end{align*}
Therefore, by definition,
\begin{align*}
p(\ub) = \EE_{\Scal}\left[\EE_{\overline \pi(\vartheta\mid\ub,Y_n)}\left\{\int_{\Bcal(\ub,\sfr) \cap \Theta_N}p(\wb |\ub, \vartheta,Y_n)\ud \wb\right\}\right] > 0.4,
\end{align*}
where $\Scal \subseteq \{1,\ldots,R\} \otimes \{1,\ldots,C\}$ denotes the index set of the random subset $\Yb_n$ in the full dataset $\Yb$. This completes the proof of Lemma \ref{lem:b.5}.
\end{proof}
\vspace{3mm}

\begin{lemma}\label{le4}
Suppose that Assumptions \ref{assump:asymp}, \ref{assump:bound}, \ref{assump:eigen} and \ref{assump:initial} hold. If the step size ${\epsilon}_{\max}$ satisfies the condition \eqref{eq:emax.cond1}, then there exists a constant $C_4>0$ such that the conductance $\phi$ defined in \eqref{eq:phi} satisfies
\begin{align*}
& \phi \geqslant C_4\rho\sqrt{{\epsilon}_{\max}} ,
\end{align*}
where $\rho$ is the Cheeger constant of the target posterior distribution $\Pi_N^*$.
\end{lemma}

\begin{proof}[Proof of Lemma \ref{le4}]
From Lemma \ref{le2}, we have that for any $\ub \in \Theta_N$ and $\wb \in \Theta_N \cap \Bcal(\ub, \sfr) \backslash\{\ub\}$,  $\alpha_{\ub}(\wb) \geqslant 1-\delta / 2$ for all sufficiently large $N$ on the event $\Gcal_N$, with $\delta$ given in \eqref{eq:delta}. As $\epsilon_{\max}$ satisfies \eqref{eq:emax.cond1}, we have $\delta\to 0$ as $N,T\to\infty$. By following the same proof of Lemma 6.5 in \cite{zou2020faster}, with Lemma \ref{lem:b.5}, we can derive from the definition \eqref{eq:Tu.star} that for any $\ub,\vb\in \Theta_N$, on the event $\Gcal_N$,
\begin{align} \label{eq:Tstar.uv1}
&\left\|\Tcal_{\ub}^{\star}(\cdot)-\Tcal_{\vb}^{\star}(\cdot)\right\|_{\TV}\leqslant 0.8 + 0.6 \delta + \frac{1}{2}\|P(\cdot \mid \ub)-P(\cdot \mid \vb)\|_{\TV} \nonumber \\
&~~ + \frac{1}{2}\max \bigg\{\int_{\wb \in \Theta_N \backslash  \Bcal(\ub, \sfr)} p(\wb \mid \ub) \ud  \wb, \int_{\wb \in \Theta_N \backslash  \Bcal(\vb, \sfr) } p(\wb \mid \ub) \ud  \wb\bigg\} ,
\end{align}
where $P(\cdot \mid \ub),P(\cdot \mid \vb)$ denote the distributions with transition densities $p(\cdot \mid \ub),p(\cdot \mid \vb)$.

We first bound the term of $\|P(\cdot \mid \ub)-P(\cdot \mid \vb)\|_{\TV}$ in \eqref{eq:Tstar.uv1}. Since $\tilde p(\cdot \mid \ub,\vartheta,Y_n)$ is the density of $N\big(\ub-(\Ecal/2)g_{\vartheta}(\ub,Y_n), \Ecal\big)$ and $ p(\cdot \mid \ub,\vartheta,Y_n)$ is the density of this normal truncated to $\Theta_N$, we have that on the event $\Gcal_N$, for all sufficiently large $N$ and $T$,
\begin{align} \label{eq:Tstar.uv2}
&\quad~ \|P(\cdot \mid \ub)-P(\cdot \mid \vb)\|_{\TV} \nonumber \\
& = \sup_{\Acal \subseteq \Theta_N} \left| \EE_{\Scal} \EE_{\overline \pi(\vartheta|\ub,Y_n)} \left[\int_{\Acal} p(\wb \mid \ub,\vartheta,Y_n) \ud \wb - \int_{\Acal} p(\wb \mid \vb,\vartheta,Y_n) \ud \wb \right] \right| \nonumber \\
&\leqslant \frac{\EE_{\Scal} \EE_{\overline \pi(\vartheta|\ub,Y_n)} \left\|N\big(\ub-(\Ecal/2)g_{\vartheta}(\ub,Y_n), \Ecal\big) -  N\big(\vb-(\Ecal/2)g_{\vartheta}(\vb,Y_n), \Ecal\big) \right\|_{\TV}}{\int_{\Theta_N} \tilde p(\wb\mid \ub,\vartheta,Y_n) \ud \wb} \nonumber \\
&\stackrel{(i)}{\leqslant} 2^{p+5} \EE_{\Scal} \EE_{\overline \pi(\vartheta|\ub,Y_n)} \sqrt{2D_{\text{KL}}\big( N\big(\ub-(\Ecal/2)g_{\vartheta}(\ub,Y_n), \Ecal\big) ,  N\big(\vb-(\Ecal/2)g_{\vartheta}(\vb,Y_n), \Ecal\big)\big)} \nonumber \\
&\leqslant 2^{p+5} \EE_{\Scal} \EE_{\overline \pi(\vartheta|\ub,Y_n)} \frac{\|\ub-\vb\|_2 + \|(\Ecal/2)[g_{\vartheta}(\ub,Y_n)-g_{\vartheta}(\vb,Y_n)]\|_2}{\sqrt{\epsilon_{\min}}}  \nonumber \\
&\stackrel{(ii)}{\leqslant} 2^{p+5} \left[\frac{\|\ub-\vb\|_2}{\sqrt{\epsilon_{\min}}} + \frac{\epsilon_{\max}\cdot L\|\ub-\vb\|_2 N(\log N)^{3A_1+3B_1+4E_1+2}}{2\sqrt{\epsilon_{\min}}} \right]  \nonumber \\
&\stackrel{(iii)}{\leqslant} (2^{p+5}+L) \sqrt{\overline c_{\epsilon}} \cdot \frac{\|\ub-\vb\|_2}{\sqrt{\epsilon_{\max}}} ,
\end{align}
where $(i)$ follows from the Pinsker's inequality and the lower bound in \eqref{eq:I23}, $(ii)$ follows from Lemma \ref{lea2}, and $(iii)$ follows from Assumption \ref{assump:initial} and the condition that \\
$\epsilon_{\max} \prec 1/[N^2(\log N)^{2(3A_1+B_1+4E_1+2)}]$. Therefore, if we take $\Delta= \sqrt{\epsilon_{\max}}/[10^3(2^{p+5}+L)\sqrt{\overline c_{\epsilon}}]$, then \eqref{eq:Tstar.uv2} leads to $\|P(\cdot \mid \ub)-P(\cdot \mid \vb)\|_{\TV}\leqslant 0.001$ for any $\|\ub-\vb\|_2\leqslant \Delta$.

Next, we bound the last term in \eqref{eq:Tstar.uv1}. We use the lower bound in \eqref{eq:I23} again to obtain that on the event $\Gcal_N$,
\begin{align} \label{eq:2max.1}
\int_{\Theta_N \backslash \Bcal(\ub,\sfr)}  p(\wb \mid \ub,\vartheta, Y_n) \ud  \wb
&\leqslant 2^{p+5} \int_{\Theta_N \backslash \Bcal(\ub,\sfr)} \tilde p(\wb \mid \ub,\vartheta, Y_n) \ud  \wb \nonumber \\
&\leqslant 2^{p+5} \mathbb{P}_{W \sim \chi_{p+3}^{2}} \left(W \geqslant \frac{[\sfr-({\epsilon}_{\max}/2)\|g_{\vartheta}(\ub,Y_n)\|_2]^2}{2{\epsilon}_{\max}} \right) , \textup{ and }  \nonumber \\
\int_{\Theta_N \backslash \Bcal(\vb,\sfr)}  p(\wb \mid \ub,\vartheta, Y_n) \ud  \wb
&\leqslant 2^{p+5} \int_{\Theta_N \backslash \Bcal(\vb,\sfr)} \tilde p(\wb \mid \ub,\vartheta, Y_n) \ud  \wb \nonumber \\
&\leqslant \mathbb{P}_{W \sim \chi_{p+3}^{2}} \left(W \geqslant \frac{[\sfr-({\epsilon}_{\max}/2)\|g_{\vartheta}(\ub,Y_n)\|_2 -\|\ub-\vb\|_2]^2}{2{\epsilon}_{\max}} \right).
\end{align}
Given our choice of $\Delta= \sqrt{\epsilon_{\max}}/[10^3(2^{p+5}+L)\sqrt{\overline c_{\epsilon}}]$ and $\sfr$ defined in \eqref{eq:sfr}, $\Delta \prec \sfr$ as $T\to\infty$ for any $\tau\in (0,1)$. By Lemma \ref{leb1}, $\|g_{\vartheta}(\ub,Y_n)\|_2 \leqslant M_0N(\log N)^{3A_1+3B_1+4E_1+2}$ for all $\ub\in \Theta_N$ on the event $\Gcal_N$. Given the condition $\epsilon_{\max} \prec 1/[N^2(\log N)^{2(3A_1+3B_1+4E_1+2)}]$, $({\epsilon}_{\max}/2)\|g_{\vartheta}(\ub,Y_n)\|_2 \prec \sfr$ as $N,T\to\infty$ on the event $\Gcal_N$. Therefore, we can choose a large constant $C_{p2}$ such that $\mathbb{P}_{W \sim \chi_{p+3}^{2}} (W \geqslant C_{p2}) \leqslant 0.001$ and for all sufficiently large $N,T$, on the event $\Gcal_N$,
\begin{align} \label{eq:2max.2}
&\frac{[\sfr-({\epsilon}_{\max}/2)\|g_s(\ub,\vartheta,Y_n)\|_2]^2}{2{\epsilon}_{\max}} \geqslant C_{p2}, \text{ and }  \nonumber \\
& \frac{[\sfr-({\epsilon}_{\max}/2)\|g_s(\ub,\vartheta,Y_n)\|_2 -\|\ub-\vb\|_2]^2}{2{\epsilon}_{\max}} \geqslant C_{p2}.
\end{align}
Therefore, by combining \eqref{eq:2max.1} and \eqref{eq:2max.2} and taking the expectation $\EE_{\Scal}\EE_{\overline \pi(\vartheta \mid \ub, Y_n)}$ in \eqref{eq:2max.1}, we have that on the event $\Gcal_N$, as $N,T\to\infty$,
\begin{align} \label{eq:2max.3}
\max \left\{\int_{\wb \in \Theta_N \backslash \Bcal(\ub,\sfr)} p(\wb \mid \ub) \ud  \wb, ~ \int_{\wb \in \Theta_N \backslash \Bcal(\vb,\sfr)} p(\wb \mid \ub) \ud  \wb\right\}  \leqslant 0.001.
\end{align}

We combine \eqref{eq:Tstar.uv1}, \eqref{eq:Tstar.uv2}, and \eqref{eq:2max.2} to conclude that on the event $\Gcal_N$, as $N,T\to\infty$,
\begin{align} \label{eq:Tstar.uv4}
\|\Tcal_{\ub}^{\star}(\cdot) -\Tcal^*_{\vb}(\cdot) \|_{\TV} \leqslant  0.801 + 0.6\delta \leqslant 0.99,
\end{align}
where the last inequality holds since $\delta$ defined in \eqref{eq:delta} satisfies $\delta\to 0$ as $N,T\to\infty$ given the condition of $\epsilon_{\max}$ in \eqref{eq:emax.cond1}.

The relation \eqref{eq:Tstar.uv4} with $\Delta= \sqrt{\epsilon_{\max}}/[10^3(2^{p+5}+L)\sqrt{\overline c_{\epsilon}}]$ implies that the condition of Lemma \ref{b.4} is satisfied. Therefore, the conductance of $\Tcal_{\ub}^{\star}(\cdot)$ has the lower bound $\phi \geqslant C_4 \rho \sqrt{{\epsilon}_{\max}}$, where $C_4>0$ is a constant and $\rho$ is the Cheeger constant of the target posterior distribution $\Pi_N^*$. This completes the proof.
\end{proof}

\vspace{3mm}

\begin{lemma}\label{le234}
Suppose that Assumptions \ref{assump:asymp}, \ref{assump:bound}, \ref{assump:eigen} and \ref{assump:initial} hold. Suppose that $T_{\laz}/T \asymp 1$ and $\log T \asymp \log N$ as $N,T,T_{\laz}\to\infty$. Suppose that for a constant $\zeta>0$, $\epsilon_{\max}$ and $\tau$ in \eqref{eq:sfr} satisfy
\begin{align} \label{eq:emax.cond2}
& {\epsilon}_{\max} \asymp \min(\rho^2,1) N^{-4(1+\zeta)} , \quad \tau = N^{-\zeta},
\end{align}
where $\rho$ is the Cheeger constant of the posterior distribution $\Pi_N^*$. For any $\lambda$-warm start initial distribution $\nu_0$ with respect to $\Pi_N^*$, the total variation distance between the distribution of the output of the projected SGLD $\tilde\Pi^{\textup{Proj-SGLD}}_{T_{\laz}}$ and the truncated posterior distribution $\Pi^*_N$ satisfies that on the event $\Gcal_N$, as $N,T_{\laz}\to\infty$,
\begin{align}\label{le234f}
\left\|\tilde\Pi_{T_{\laz}}^{\textup{Proj-SGLD}}-\Pi_N^*\right\|_{\TV} \leqslant  \lambda \left(1- C_1 \rho^2 \epsilon_{\max}\right)^{T_{\laz}} + \tilde C_2 N^{-\zeta} ,
\end{align}
for some positive constants $C_1, \tilde C_2$.
\end{lemma}

\begin{proof}[Proof of Lemma \ref{le234}]
The proof of Lemma \ref{le234} is by combining Lemma \ref{le2}, Lemma \ref{le3}, and Lemma \ref{le4}. Lemma \ref{le2} shows that the transition distribution of the $/12$-lazy projected SGLD $\Tcal_{\ub}(\cdot)$ is $\delta-$close to that of the Metropolized SGLD $\Tcal_{\ub}^{\star}(\cdot)$, where $\delta$ is defined in \eqref{eq:delta}. We first verify the condition of Lemma \ref{le3} which is $\delta \leqslant \min\{1-\sqrt{2}/2,\phi/16\}$. We notice that by the definition of $\sfr$ in \eqref{eq:sfr} and $\delta$ in \eqref{eq:delta}, if we set $\epsilon_{\max},\tau$ as in the condition \eqref{eq:emax.cond2}, $T_{\laz}/T \asymp 1$ and $\log T \asymp \log N$ as $N,T\to\infty$, then $\log (T/\tau) \asymp \log N$ and it is straightforward to verify that
\begin{align} \label{eq:delta.bound1}
\delta &\preceq \epsilon_{\max} N^2 (\log N)^{2(3A_1+3B_1+4E_1+2)+1}   + \epsilon_{\max}^2 N^4 (\log N)^{4(3A_1+3B_1+4E_1+2)} \nonumber \\
&\preceq \epsilon_{\max} N^2 (\log N)^{2(3A_1+3B_1+4E_1+2)+1}.
\end{align}
Hence, by Lemma \ref{le4} and \eqref{eq:delta.bound1}, as $N\to\infty$,
\begin{align}\label{eq:delta.bound2}
\frac{16\delta}{\phi} & \preceq \frac{ N^{-2(1+\zeta)}N^2(\log N)^{2(3A_1+3B_1+4E_1+2)+1} \rho\sqrt{\epsilon_{\max}} }{\rho\sqrt{\epsilon_{\max}}} \prec N^{-\zeta} ,
\end{align}
By \eqref{eq:emax.cond2} and Lemma \ref{le4}, $\delta\prec C_4 \rho \sqrt{\epsilon_{\max}}/16 \leqslant \phi/16$ as $N,T\to\infty$. Meanwhile, since $\rho\preceq 1$, $\delta \prec \rho \sqrt{\epsilon_{\max}} \leqslant 1-\sqrt{2}/2$ as $N\to\infty$ is trivially satisfied. Therefore, we conclude from Lemma \ref{le3} that for any $\lambda-$warm start $\nu_0$ with respect to $\Pi_N^*$ and any given $\tau \in (0,1)$, for ${\epsilon}_{\max}$ satisfying \eqref{eq:emax.cond2} and for all sufficiently large $N,T_{\laz}$,
\begin{align}
&\quad~ \left\|\tilde\Pi_{T_{\laz}}^{\textup{Proj-SGLD}}-\Pi^*_N\right\|_{\TV} \leqslant \lambda \left(1-\phi^2/8\right)^{T_{\laz}} + 16\delta/\phi \nonumber \\
&\stackrel{(i)}{\leqslant} \lambda \left(1- C_4^2 \rho^2 \epsilon_{\max}/8\right)^{T_{\laz}} + 16\delta/\phi \stackrel{(ii)}{\leqslant} \lambda \left(1- C_1 \rho^2 \epsilon_{\max} \right)^{T_{\laz}} + \tilde C_2 N^{-\zeta} , \nonumber
\end{align}
for some positive constants $C_1,\tilde C_2$, where $(i)$ follows from Lemma \ref{le4}, and $(ii)$ follows from \eqref{eq:delta.bound2} with $C_1=8C_4^{-2}$. This has proved \eqref{le234f}.
\end{proof}

\vspace{3mm}

\subsection{Proof of Theorem \ref{th1}} \label{subsec:proof.thm}

\begin{proof}[Proof of Theorem \ref{th1}]
$ $ \newline

\noindent \textit{Proof of Part (i).} The proof is by combining Lemmas \ref{le1}, \ref{lem:lazy} and \ref{le234}. If the maximum step size $\epsilon_{\max}$ and $\tau$ satisfy \eqref{eq:emax.cond2} and hence $\log (T/\tau) \asymp \log N$ as $N,T\to\infty$, then by Lemmas \ref{le1}, \ref{lem:lazy} and \ref{le234},
\begin{align} 
\left\|\Pi_T-\Pi_N^*\right\|_{\TV} & \leqslant \left\|\Pi_T^{\textup{Proj-SGLD}} - \Pi_T\right\|_{\TV} + \left\| \tilde\Pi_{T_{\laz}}^{\textup{Proj-SGLD}} - \Pi_T^{\textup{Proj-SGLD}}\right\|_{\TV} \nonumber \\
&\quad + \left\|\tilde\Pi_{T_{\laz}}^{\textup{Proj-SGLD}}-\Pi_N^*\right\|_{\TV} \nonumber \\
&\leqslant \frac{N^{-\zeta}}{8} +  \frac{N^{-\zeta}}{8} + \lambda \left(1- C_1 \rho^2 \epsilon_{\max}\right)^{T_{\laz}} + \tilde C_2 N^{-\zeta}  \nonumber \\
&\overset{(i)}{\leqslant} \lambda  \left(1- C_1 \rho^2 \epsilon_{\max}\right)^{(2-N^{-\zeta}/8)T} + C_2 N^{-\zeta}\nonumber \\
&\leqslant \lambda  \left(1- C_1 \rho^2 \epsilon_{\max}\right)^{T} + C_2 N^{-\zeta}, \nonumber 
\end{align}
for $C_2=\tilde C_2+1/4$ with probability at least $1-4\exp(-\sqrt{T}N^{-\zeta}/8)$, where $(i)$ follows from Lemma \ref{lem:lazy} that $T_{\laz}\geqslant (2-\tau/8)T$ with high probability.
\vspace{5mm}

\noindent \textit{Proof of Part (ii)} 
If we further choose $T$ and $\tau$ as
\begin{align} \label{eq:T.tau.cond}
& T = 8C_4^{-2} \zeta \rho^{-4} N^{4(1+\zeta)}\log N, \quad \tau=N^{-\zeta},
\end{align}
then by the inequality $(1-1/x)^x\leqslant \exp(-1)$ for all $x>0$, we have that for all sufficiently large $N$,
\begin{align*}
&\lambda \left(1- C_4^2 \rho^2 \epsilon_{\max}/8 \right)^{T}  \leqslant \lambda \left(1- C_4^2 \rho^2 \epsilon_{\max}/8 \right)^{8C_4^{-2} \zeta  \rho^{-4} N^{4(1+\zeta)}\log N}  \\
&\leqslant \lambda \exp(-\zeta \log N) = \lambda N^{-\zeta}.
\end{align*}
then by the conclusion of Part (i), 
\begin{align} \label{eq:2pi.diff2}
\left\|\Pi_T-\Pi_N^*\right\|_{\TV} & = O\left(N^{-\zeta}\right),
\end{align}
on the event $\Gcal_N \cap \Hcal_N$, where $\Hcal_N$ is defined as the event that Lemma \ref{lem:lazy} happens and $\PP(\Hcal_N^c)\leqslant 4\exp(-\sqrt{T}N^{-\zeta}\tau/8) =4\exp(-C_4^{-2}\zeta\rho^{-4}N^{4+2\zeta}
\log N/2)$.

When $m \leqslant N^{\varsigma}$ and $\rho\succeq N^{-c_{\nu}}$ for some positive constants $\varsigma, c_{\nu}$, we have from Lemma \ref{lem:upper.bound} that the probability of $\Gcal_N^c$ is at most
\begin{align*} 
& \PP (\Gcal_N^c) \preceq \left[ 8C_4^{-2} \zeta(r + c) N^{4(1+\zeta)+\varsigma+4c_{\nu}} + \underline c^{-1} N \right] \exp\left\{-(1/2)\log^2 N\right\} \\
&\leqslant \exp\left\{-(1/4)\log^2 N\right\} ,
\end{align*}
for all sufficiently large $N$. This implies that 
\begin{align*}
\PP ((\Gcal_N\cap\Hcal_n)^c) &\leqslant \PP(\Gcal_N^c) + \PP(\Hcal_N^c) \\
&\leqslant \exp\left\{-(1/4)\log^2 N\right\} + 4\exp(-C_4^{-2}\zeta\rho^{-4}N^{4+2\zeta}
\log N/2), 
\end{align*}
which is summable. By the Borel-Cantelli lemma, the relation \eqref{eq:2pi.diff2} holds almost surely as $N\to\infty$.

\vspace{3mm}

\noindent \textit{Proof of Part (iii).}  For $t=1,\ldots,T$, let $\nu_t=f\sharp \Pi_t$ and $\nu^*=f\sharp \Pi^*_N$ be the push-forward measures of $\Pi_T$ and $\Pi_N^*$, i.e., for any Lebesgue measurable set $\Acal\subseteq [-C_f,C_f]$, $\nu_t(\Acal)=\Pi_t(f^{-1}(\Acal))$ and  $\nu_N^*(\Acal)=\Pi_N^*(f^{-1}(\Acal))$. Let $\Gamma(\nu_t, \nu^*)$ denote the set of all probability measures on $[-C_f,C_f]\times [-C_f,C_f]$ with marginals $\nu_t$ and $\nu_N^*$ respectively. Then for any $\gamma \in \Gamma(\nu_t, \nu_N^*)$, we have
\begin{align}\label{part2}
& \left| \int_{\Theta_N} f(\theta) \Pi_t(\ud \theta) - \int_{\Theta_N} f(\theta') \Pi_N^*(\ud \theta') \right| = \left| \int_{[-C_f,C_f]} x \nu_t(\ud x) - \int_{[-C_f,C_f]}x' \nu_N^*(\ud x') \right| \nonumber \\
&=  \left| \int_{[-C_f,C_f] \times [-C_f,C_f]}(x-x') \ud \gamma(\nu_t,\nu^*_N) \right| \leqslant  \int_{[-C_f,C_f] \times [-C_f,C_f]}\left|x-x'\right| \ud \gamma(\nu_t,\nu_N^*),
\end{align}
where $x$ and $x'$ are two random variables with marginal distributions $\Pi_t$ and $\Pi_N^*$. Let $W_1(\Pi_t, \Pi_N^*)=\inf_{\gamma\in \Gamma(\Pi_t,\Pi_N^*)}\int_{[-C_f,C_f] \times [-C_f,C_f]}\left|x-x'\right| \ud \gamma(\nu_t,\nu_N^*)$ denote the Wasserstein-$1$ distance between $\nu_t$ and $\nu_N^*$. Suppose the infimum is taken at a joint measure $\gamma_1$. Since \eqref{part2} holds for any $\gamma \in \Gamma(\nu_t, \nu_N^*)$, we have that
\begin{align}\label{p22}
&\quad~ \left| \int_{\Theta_N}f(\theta)\Pi_t(\ud \theta) - \int_{\Theta_N}f(\theta')\Pi_N^*(\ud \theta') \right| = \left| \int_{[-C_f,C_f]\times[-C_f,C_f]}(x-x')\ud \gamma(\nu_t,\nu_N^*) \right| \nonumber \\
&\leqslant  \int_{[-C_f,C_f]\times[-C_f,C_f]}\left|x-x'\right| \ud \gamma_1(\nu_t,\nu_N^*)  = \inf_{\gamma \in \Gamma(\nu_t,\nu_N^*)} \int_{[-C_f,C_f]\times[-C_f,C_f]}\left|x-x'\right| \ud \gamma(\nu_t,\nu_N^*) \nonumber\\
&= W_1(\nu_t, \nu_N^*) \stackrel{(i)}{\leqslant} 2C_f\left\|\nu_t - \nu_N^*\right\|_{\TV}  = 2C_f\sup_{\Acal\subseteq [-C_f,C_f]} \left|\Pi_t(f^{-1}(\Acal))-\Pi_N^*(f^{-1}(\Acal))\right| \nonumber \\
&\leqslant 2C_f\sup_{\Acal' \subseteq \Theta_N} \left|\Pi_t(\Acal')-\Pi_N^*(\Acal')\right| = 2C_f \left\|\Pi_t-\Pi_N^*\right\|_{\TV} ,
\end{align}
where $(i)$ follows from the inequality between the Wasserstein-$1$ distance and the total variation distance; see for example, Theorem 6.15 in \cite{Vil08}. Since $\left\|\Pi_t-\Pi_N^*\right\|_{\TV}\to 0$ as $t\to\infty$ and $N\to\infty$ as shown in Part (i), \eqref{p22} implies that $\left| \int_{\Theta_N}f(\theta)\Pi_t(\ud \theta) - \int_{\Theta_N}f(\theta')\Pi_N^*(\ud \theta') \right| \rightarrow 0$ as  $t\to\infty$ and $N\to\infty$.

By the central limit theorem of the Markov chain, we have that
\begin{align}\label{p23}
T^{-1} \sum^T_{t=1} f\left(\theta^{(t)}\right) - T^{-1} \sum^T_{t=1}\int_{\Theta_N}f\left(\theta'\right)\Pi_t(\ud \theta') = O_p\left(T^{-1/2}\right).
\end{align}

Thus, for any $ \varepsilon \in (0,1)$, there exists $T_1\in \ZZ^+$, such that for all $T \geqslant T_1$, it holds that
\begin{align*}
\PP\left( \left| T^{-1} \sum^T_{t=1} f\left(\theta^{(t)}\right) - T^{-1} \sum^T_{t=1}\int_{\Theta_N}f\left(\theta'\right)\Pi_t(\ud \theta')\right|  \geqslant T^{-1/2}\right) < \varepsilon .
\end{align*}
Therefore, for the difference in Part (ii), we have that
\begin{align} \label{p24}
&\quad~ \left| T^{-1} \sum^T_{t=1} f\left(\theta^{(t)}\right) - \int_{\Theta_N} f(\theta) \Pi_N^*(\ud \theta)  \right|  \nonumber \\
& \leqslant  \left| T^{-1} \sum^T_{t=1} f\left(\theta^{(t)}\right) - T^{-1} \sum^T_{t=1}\int_{\Theta_N}f\left(\theta'\right)\Pi_t\left(\ud \theta'\right)\right| \nonumber \\
& \quad + T^{-1} \sum^T_{t=1} \left|\int_{\Theta_N}f\left(\theta'\right) \Pi_t\left(\ud \theta'\right) - \int_{\Theta_N}f(\theta) \Pi_N^*(\ud \theta) \right|  \nonumber \\
& \leqslant  \left| T^{-1} \sum^T_{t=1} f\left(\theta^{(t)}\right) - T^{-1} \sum^T_{t=1}\int_{\Theta_N}f\left(\theta'\right)\Pi_t\left(\ud \theta'\right)\right| + T^{-1} \sum^T_{t=1} 2C_f \left\| \Pi_t - \Pi_N^*\right\|_{\TV},
\end{align}
where the last step follows from \eqref{p22}.

From the conclusion of Part (i), the second term on the right-hand side of \eqref{p24} goes to zero as $T\to\infty$ and $N\to\infty$ since it is a Cesaro average of \eqref{p22}. Therefore, for any $\xi >0$, there exists $T_2\in \ZZ^+$ and $T_2>\max(T_1,4/\xi^2)$, such that for all $ T>  T_2$, $T^{-1/2}<\xi/2$ and $ T^{-1} \sum^T_{t=1} 2C_f \left\| \Pi_t - \Pi_N^*\right\|_{\TV} <\xi/2$. Therefore, from \eqref{p24}, we have that for all $T>T_2$,
\begin{align*}
&\quad~ \PP\left(\left| T^{-1} \sum^T_{t=1}f\left(\theta^{(t)}\right) - \int_{\Theta_N}f(\theta) \Pi_N^*(\ud \theta) \right| \geqslant \xi \right)  \\
& \leqslant \PP\left( \left| T^{-1} \sum^T_{t=1} f\left(\theta^{(t)}\right) - T^{-1} \sum^T_{t=1}\int_{\Theta_N}f\left(\theta'\right)\Pi_t(\ud \theta')\right|  \geqslant T^{-1/2}\right) \nonumber\\
&\qquad + \PP\left( T^{-1} \sum^T_{t=1} 2C_f \left\| \Pi_t - \Pi_N^*\right\|_{\TV} \geqslant \xi/2 \right)  \\
&< \varepsilon + 0 = \varepsilon.
\end{align*}
This completes the proof of Part (ii).
\end{proof}

\bibliographystyle{plainnat}
\bibliography{papers}

\end{document}